%% file: checs50-aa.tex
\documentclass{aa}

\usepackage{graphicx}
\usepackage{txfonts}

\usepackage{graphicx}	%
\usepackage{amsmath}	%
\usepackage{amssymb}	%

\usepackage{comment}
\usepackage{subfig}
\usepackage[T1]{fontenc}
\usepackage{ae,aecompl}
\usepackage{graphicx} %
\usepackage{amsmath}  %
\usepackage{amssymb}    %
\usepackage{lscape}
\usepackage{rotating}
\usepackage{newtxtext,newtxmath}
\graphicspath{{./images/}}
\input{newcom}

\newcommand{\muXh}{\hat{\mu_{X}}}
\newcommand{\muCh}{\hat{\mu_{C}}}
\newcommand{\muX}{\mu_{X}}
\newcommand{\muC}{\mu_{C}}
\newcommand{\sigx}{\sigma_{X}}
\newcommand{\sigc}{\sigma_{C}}
\newcommand{\kapx}{\kappa_{X}}
\newcommand{\kapc}{\kappa_{C}}
\newcommand{\delx}{\delta_{X}}
\newcommand{\delc}{\delta_{C}}

\newcommand{\Ngal}{\text{Ngal}}
\newcommand{\MfX}{M_{500,X}}
\newcommand{\MfC}{M_{500,C}}
\newcommand{\MXMC}{M_{500,X}/M_{500,C}}

\begin{document}

\title{\textit{Chandra} follow-up of the Hectospec Cluster Survey: comparison of caustic and hydrostatic masses and constraints on the hydrostatic bias}
\titlerunning{Comparison of X-ray and caustic masses of galaxy clusters}
\authorrunning{C.~H.~A.~Logan et al.}

\author{Crispin~H.~A.~Logan\inst{1}\fnmsep\inst{2},
  Ben~J.~Maughan\inst{1}\thanks{email: ben.maughan@bristol.ac.uk},
  Antonaldo~Diaferio\inst{3}\fnmsep\inst{4},
  Ryan~T.~Duffy\inst{1},
  Margaret~J.~Geller\inst{5},
  Kenneth~Rines\inst{6} \and
  Jubee~Sohn\inst{5}
}

\institute{HH Wills Physics Laboratory, University of Bristol, Tyndall
  Avenue, Bristol, BS8 1TL, UK
  \and
  European Space Agency (ESA), European Space Astronomy Centre (ESAC),
  Camino Bajo del Castillo s/n, 28692 Villanueva de la Ca\~{n}ada,
  Madrid, Spain
  \and
  Universit\`a di Torino, Dipartimento di Fisica, via P. Giuria 1,
  I-10125, Torino, Italy
  \and
  Istituto Nazionale di Fisica Nucleare (INFN), sezione di Torino, Via
  P. Giuria 1, I-10125 Torino, Italy
  \and
  Smithsonian Astrophysical Observatory, 60 Garden St., Cambridge, MA 02138
  \and
  Department of Physics \& Astronomy, Western Washington University,
  Bellingham, WA 98225, USA
}

\date{Received 17/02/2022; accepted 06/07/2022}

\abstract
{Clusters of galaxies are powerful probes with which to study
  cosmology and astrophysics. However, for many applications, an
  accurate measurement of a cluster's mass is essential. A
  systematic underestimate of hydrostatic masses from X-ray
  observations (the so-called hydrostatic bias) may be responsible
  for tension between the results of different cosmological
  measurements.}
{We compare X-ray hydrostatic masses with masses estimated using
  the caustic method (based on galaxy velocities) in order to
  explore the systematic uncertainties of both methods and place
  new constraints on the level of hydrostatic bias.}
{Hydrostatic and caustic mass profiles were determined
  independently for a sample of 44 clusters based on
  \textit{Chandra} observations of clusters from the Hectospec
  Cluster Survey. This is the largest systematic comparison of its
  kind. Masses were compared at a standardised radius ($R_{500}$)
  using a model that includes possible bias and scatter in both
  mass estimates. The systematics affecting both mass
  determination methods were explored in detail.}
{The hydrostatic masses were found to be systematically higher than
  caustic masses on average, and we found evidence that the caustic
  method increasingly underestimates the mass when fewer galaxies
  are used to measure the caustics. We limit our analysis to the 14
  clusters with the best-sampled caustics where this bias is
  minimised ($\ge210$ galaxies), and find that the average ratio of
  hydrostatic-to-caustic mass at $R_{500}$ is
  $\MXMC=1.12^{+0.11}_{-0.10}$.}
{We interpret this result as a constraint on the level of
  hydrostatic bias, favouring small or zero levels of hydrostatic
  bias (less than $20\%$ at the $3\sigma$ level). However, we find
  that systematic uncertainties associated with both mass estimation
  methods remain at the $10-15\%$ level, which would permit
  significantly larger levels of hydrostatic bias.}


\maketitle

\section{Introduction} \label{section.intro}

Galaxy clusters are the largest gravitationally bound objects in the
Universe, and thus they are used to study the growth of cosmic
structures and as cosmological probes. Tight constraints on
cosmological parameters can be determined from measurements of the
mass function of clusters, or their baryon fraction, as a function of
redshift \citep[see e.g.][for a review]{Allen:2011a}. However, the
accuracy of these constraints crucially depends on the accuracy with
which cluster masses can be measured. For example, the exponential
drop in the number density of clusters at high masses means that their
mass function is a sensitive cosmological probe, but an error of
$\sim$ 10\% on the cluster mass can lead to a factor of two difference
in the expected space density\footnote{For this approximation we used
  \texttt{hmfcalc} \citep{Murray:2013a}.}. The measurement of galaxy
cluster masses is challenging, as most of the cluster mass is in the
form of dark matter. Mass estimation techniques thus probe the masses
of clusters indirectly via the effect of the gravitational potential
of the cluster on its intracluster medium (ICM), its member galaxies,
or images of background galaxies.

Optical data were used to give the first mass estimates of galaxy
clusters. \citet{Zwicky:1933a} measured the line-of-sight velocities
of a number of cluster galaxies in the nearby Coma cluster, and, by
computing the velocity dispersion, used the virial theorem to estimate
a mass for the cluster. In the late 1960s and 1970s, the first X-ray
satellites were launched, and it was in the 1990s that the first X-ray
samples of hundreds of clusters were produced
\citep{Ebeling:1998a,Ebeling:2001a,Bohringer:2000a,Bohringer:2001a}
from which X-ray hydrostatic masses could be calculated (e.g.
\citealp{Reiprich:2002a}). The launch of \emph{XMM-Newton} and
\emph{Chandra} in 1999 has led to even more precise X-ray masses being
measured for large samples of clusters (e.g. \citealp{Martino:2014a}).
The first work using weak gravitational lensing (WL) to map the dark
matter distribution, undertaken by \citet{Tyson:1990a}, was followed
by a number of other papers (e.g.
\citealp{Blandford:1991a,Miralda-Escude:1991a,Kaiser:1992a}), and in
the last 3 decades it has become increasingly common to use WL as a
mass estimation technique (see \citealp{ume20} for a review).

For cosmological studies, the calibration of masses is critical to
understand the biases that may be inherent to different mass
estimation techniques. Cosmology requires large samples of clusters
with masses which in general must be estimated from simple mass
proxies that must be calibrated with more direct, higher fidelity mass
measurements. Thus, the accuracy of the mass scale from those
measurements is crucial, and biases in the mass scale can manifest as
biases in the derived cosmological parameters. For example, there is
currently tension between the cosmological constraints ($\Omega_{m}$
and $\sigma_{8}$) from the \emph{Planck} cluster number counts
\citep{Planck-Collaboration:2016d} and the \emph{Planck} cosmic
microwave background (CMB) results \citep{Planck-Collaboration:2016a}.
The \textit{Planck} cluster masses were calibrated using hydrostatic
masses, and the tension in the results could be due to a negative bias
(i.e. the true mass is underestimated) in the hydrostatic masses,
often referred to as the hydrostatic bias.

More recent analyses have found lower tension between the values of
$\sigma_8$ obtained from the \textit{Planck} CMB and from cluster
number counts measured with SZE or X-ray cluster surveys
\citep[e.g.][]{Douspis:2019a,boc19,pla20,gar21,sal21}. The improved
agreement comes from updates to the CMB analysis and the mass
calibration of the clusters, with all analyses favouring a
non-negligible hydrostatic bias. However the magnitude of the
hydrostatic bias is not agreed upon, and remains a critical point for
cluster studies.

Some level of hydrostatic bias is expected a-priori, due to the
assumption that the only pressure support in the ICM is the thermal
pressure measured in X-rays. If any other sources of pressure (e.g.
from turbulence, bulk motions or clumping of the ICM, or from cosmic
rays) are also present, this leads to the X-ray hydrostatic mass
method underestimating the true cluster mass
\citep{Nagai:2007a,Lau:2009a,Nelson:2014a}.

The level of hydrostatic bias is expected to be higher in unrelaxed
clusters compared with relaxed clusters, as unrelaxed systems are
expected to have more non-thermal pressure support from bulk motions
and turbulence of the cluster gas. However, all clusters, regardless
of dynamical state, will experience some amount of non-thermal
pressure support, as gas and substructures are always infalling onto
the cluster as they undergo constant growth, leading to non-thermal
pressure due to the residual motion of the ICM and the turbulence
created. Constraining turbulence in the ICM requires high resolution
X-ray spectrometry, and the only mission thus far capable of this has
been \textit{Hitomi}. Before the mission failed, \textit{Hitomi}
observed the Perseus cluster, finding that in the core of the Perseus
cluster the pressure support from turbulence was 4\% of the
thermodynamic pressure \citep{Hitomi-Collaboration:2016a}, which on
its own would not lead to a large value of hydrostatic bias (though we
note that the Perseus cluster is a dynamically relaxed cluster and, of
course, not representative of the cluster population as a whole).

The hydrostatic bias may also be a function of radius, as the
infalling gas may lead to stronger gas motions and therefore
significant non-thermal pressure support at the outskirts of clusters.
\citet{Bonamente:2013a} and \citet{Fusco-Femiano:2018a} both find
evidence for significant non-thermal pressure at the cluster outskirts
(for A1835 and A2142 respectively), and work using hydrodynamical
simulations suggests that the non-thermal pressure due to the bulk gas
motion increases with radius
\citep{Nagai:2007a,Lau:2009a,Vazza:2009a,Vazza:2018a,Battaglia:2012a,Rasia:2012a}.
Specifically, \citet{Nagai:2007a} and \citet{Rasia:2012a} both find
that at low ($<$ 0.2 $R_{500}$) and high radius ($>$ $R_{500}$) the
bias increases\footnote{We follow the convention that e.g. $R_{500}$
  refers to the radius within which the mean density of the clusters
  is 500 times the critical density of the Universe at the cluster's
  redshift.}. \citet{Nagai:2007a} find the bias to increase from
$<$10\% at $R_{500}$ to $\sim$ 30\% at twice this radius. Studies that
have used observational data agree with this general trend
\citep{Siegel:2018a,Eckert:2019a}.

To get a grasp on the biases of the cluster mass estimation methods,
comparisons between methods themselves are undertaken, and simulated
data are also used to compare the known mass from the simulation to
the mass that is recovered if observational methods are applied to
synthetic observations of the simulated cluster. Comparisons of X-ray
hydrostatic masses with WL masses can yield insights into the
magnitude of the hydrostatic bias, as WL masses are not sensitive to
the state of the ICM (in contrast to X-ray hydrostatic masses).
However, WL masses are sensitive to mass along the line of sight to
the observed cluster, which can lead to scatter of $\sim$ 20 - 30\%
and positive bias of up to $\sim$ 20\%
\citep{Hoekstra:2001a,Hoekstra:2003a,Hoekstra:2011a,Becker:2011a,Hwang:2014a},
which needs to be understood and taken into account in these
comparisons. A further complication is that the uncertainty on WL
masses increases with decreasing cluster mass and decreasing cluster
redshift \citep{Hoekstra:2003a}, and even for a cluster with mass
5$\times$10$^{14}$ $\text{M}_{\odot}$ at a redshift of 0.3, the one
sigma uncertainty can be as large as 40\%. Some comparisons between WL
masses and X-ray hydrostatic masses suggest that X-ray hydrostatic
masses underestimate the true mass by $\sim$ 20 - 30\%
\citep{von-der-Linden:2014a,Donahue:2014a,Sereno:2015a,Hoekstra:2015a}.
Work using hydrodynamical simulations also suggest a significant
hydrostatic bias of $\sim$10 - 30\%
\citep{Rasia:2006a,Rasia:2012a,Nagai:2007a,Lau:2009a,Nelson:2014a}.
However \citet{Gruen:2014a}, \citet{Israel:2014a},
\citet{Applegate:2016a} and \citet{Smith:2016a} find no significant
evidence for hydrostatic bias when comparing X-ray hydrostatic masses
to WL masses.

Comparisons between X-ray and WL masses from observations also support the idea that disturbed clusters will have a larger hydrostatic bias than relaxed clusters \citep{Zhang:2010a,Mahdavi:2013a}. \citet{Biffi:2016a} also find a difference in hydrostatic bias between relaxed and disturbed clusters using simulations (though find no difference between cool core and non cool core systems).

Comparisons between X-ray and caustic masses are much rarer. Caustic
masses \citep{Diaferio:1997a,Diaferio:1999a} are a useful alternative
to WL masses for testing the accuracy of hydrostatic masses. Like WL
masses they are not sensitive to the state of the ICM, and biases in
the caustic method are, in principle, well understood from simulations
\citep{Serra:2011a}. One drawback of using caustic masses for
comparison is that they have a large scatter of $\sim$ 30\%
\citep{Serra:2011a}, which comes predominantly from the assumption of
spherical symmetry (and thus the viewing angle of the cluster). Thus,
for a meaningful comparison, a large cluster sample is needed (several
tens of clusters). Another drawback of the comparison is that the
caustic mass method was conceived to measure the mass in the outer
region of clusters, from 1 - 3 $R_{200}$, where the X-ray hydrostatic
mass method cannot be applied because the X-ray surface brightness is
too low to return a reliable signal at these radii. However,
\citet{Serra:2011a} use simulations to estimate the bias in the
caustic method at lower radii, such that a meaningful comparison
between the two cluster mass estimators can still be made.

\citet{Diaferio:2005a} was the first work to compare X-ray and caustic
masses, for a sample of three clusters, finding that for two clusters
the mass profiles agreed well, and for the other cluster they did not,
as the cluster was clearly out of equilibrium. Much larger cluster
samples (from observational data) were used by \citet{Maughan:2016a},
\citet{Andreon:2017a} and \citet{Foex:2017a} to infer caustic masses
to be $\sim$20\% lower (at $R_{500}$), $\sim$15\% larger (at
$R_{500}$) and $\sim$30\% larger (at $R_{200}$) than the hydrostatic
masses respectively. \citet{Armitage:2019a}\footnote{We note that the
  caustic method used in \citet{Armitage:2019a} is from
  \citet{Gifford:2013b}. Their method, unlike the other X-ray caustic
  comparison papers mentioned which all use the \citet{Diaferio:1999a}
  method, heavily relies on the assumption that the cluster has an NFW
  density profile; in addition, they are interested in the mass within
  $R_{200}$, which is not the target of the \citet{Diaferio:1999a}
  caustic method.} use simulation data to compare caustic and X-ray
hydrostatic masses (at $R_{200}$) and find a similar mass ratio to
\citet{Foex:2017a}. \citet{Ettori:2019a}, however, find their caustic
masses to be underestimated by $\sim$ 40\% (at $R_{200}$) compared to
their hydrostatic masses (both computed using observational
data). These contradictory results clearly show that the situation is
complex and unsettled.

There have also been comparisons of caustic masses with weak lensing -
\citet{Diaferio:2005a} do so for a sample of three clusters, and
\citet{Geller:2013a} do so for a sample of 19 clusters. Both works
find the mass methods to agree (independent of the dynamical state of
the cluster) at around the virial radius to within $\sim$ 30\%,
consistent within the errors associated with each mass method.

In this paper, we aim to compare hydrostatic and caustic masses for
the largest sample so far used for these purposes, in order to
investigate the accuracy of the methods and provide new constraints on
the hydrostatic bias that are independent of those based on WL masses.
We improve on our earlier work \citep{Maughan:2016a} by expanding the
sample from 16 to 44 clusters. We also include a more comprehensive
analysis of the systematics, and our main conclusions are based on
a subset of 14 clusters for which the systematics are minimised.

In this paper we use the following notation to represent the masses of
the clusters. $M_X$ and $M_C$ denote the masses measured with the
hydrostatic and caustic methods, respectively. When these masses are
specifically measured within $\rf$, we denote them as $\MfX$ and
$\MfC$. Unless otherwise stated, these are always measured within the
$\rf$ determined from the hydrostatic analysis.

The structure of the paper is as follows. In \S \ref{section.sample}
we discuss the sample selection and data preparation. \S
\ref{section.analysis} details the data analysis and modelling. In \S
\ref{sec.results} we present initial results for the full sample, and
then our main results from the subset found to have the most reliable
mass estimates. We discuss our results in \S
\ref{section.discuss_results_mxmc}, including a review of the
systematics affecting the mass estimates, and the implications for the
hydrostatic bias, and summarise our findings in \S
\ref{section.summary}. Throughout this paper we assume a WMAP9
cosmology of $H_0 = 69.7$ km s$^{-1}$Mpc$^{-1}$,
$ \Omega_{\Lambda} = 0.718$, and $\Omega_{m} = 0.282$
\citep{Hinshaw:2013a}. All measurement uncertainties correspond to
68\% confidence level, unless explicitly stated otherwise. When
referring to base 10 logarithm we use `$\log$' or `$\log_{10}$', and
use `$\ln$' if referring to the natural logarithm.

\section{Cluster sample} \label{section.sample}

For our analysis, we used the X-ray flux-limited subset of the
Hectospec Cluster Survey (HeCS; \citealp{Rines:2013a}), a
spectroscopic survey of X-ray selected clusters. The HeCS sample was
constructed by matching clusters selected using the \textit{ROSAT}
All-Sky Survey (RASS; \citealp{Voges:1999a}) with the imaging
footprint of the SDSS Data Release 6 \citep{Adelman-McCarthy:2008a}.
The SDSS data were used to select candidate cluster member galaxies
for spectroscopic follow-up, which was performed with MMT/Hectospec
\citep{Fabricant:2005a}.

The X-ray flux limit applied to the HeCS sample in order to create the
flux-limited HeCS sub-sample was $5 \times 10^{-12} \flux$ (in the
$0.1-2.4\keV$ band), which excluded A750, A2187, A2396, A2631, and
A2645 from the full HeCS sample. We then excluded three clusters from
the flux-limited sample that were found to be clearly dominated by AGN
upon inspection of available X-ray observations (these were A689 and
A1366 based on \textit{Chandra} data, and A2055 based on
\textit{XMM-Newton} data). More specifically, A1366 and A2055 have
BLLacs in the brightest cluster galaxy (BCG), and A689 has a BLLac
very close to the BCG \citep{Giles:2012a}. This resulted in a
flux-limited sample of 50 clusters. We then obtained \textit{Chandra}
follow-up observations of those clusters lacking \textit{Chandra}
data.

To construct the final sample, we then excluded four clusters
due to flaring in the available \textit{Chandra} observations (A267,
A667, A1361, and A1902), one cluster as it was part of a double
cluster (A1758), and one cluster as it was off-chip in the
\emph{Chandra} observation available (A646 - the observation was of a
radio galaxy in the cluster, but around 7$'$ from the cluster core).
The final sample then consists of 44 clusters, which are
  summarised in Table \ref{table.samplesummary}. All clusters are in
the redshift range $0.1 < $ z $< 0.3$. We call our sample the CHeCS
(\textit{Chandra} observations of Hectospec Cluster Survey clusters)
sample. We note that the masses from \citet{Rines:2013a} were
converted to be consistent with the value of $H_{0}$ we use in this
paper.

\begin{center}
  \begin{table*}
    \scalebox{0.90}{
      \begin{tabular}{llcccccccccccc}
        \hline
        Cluster & Short name &       RA    &  DEC & z & ObsID & clean time & Ngal \\
                &            &       deg   &  deg &   &       & (ks)       &   \\
        \hline
        ZwCl 0755.8+5408 & Zw1478  &119.9190   & 53.9990  & 0.1027 & 18248                          & 20                     & 82 \\
        Abell 0655       & A655    &  126.3610 & 47.1320  & 0.1271 & 15159                          & 8                        &  315 \\
        Abell 0697       & A697    & 130.7362  & 36.3625  & 0.2812 & 4217                           & 20                     &  185 \\
        MS 0906.5+1110   & MS0906  & 137.2832  & 10.9925  & 0.1767 & 924                            & 30                     &  101 \\
        Abell 0773       & A773    & 139.4624  & 51.7248  & 0.2173 & 5006,533,3588                  & 15,10,9            & 173  \\
        Abell 0795       & A795    & 141.0240  & 14.1680  & 0.1374 & 11734$^{\star}$                & 29                     & 179 \\
        ZwCl 0949.6+5207 & Zw2701  & 148.1980  & 51.8910  & 0.2160 & 12903$^{\star}$,3195$^{\star}$ & 93,22                & 93 \\
        Abell 0963       & A963    &  154.2600 & 39.0484  & 0.2041 & 903$^{\star}$                  & 34                     & 211\\
        Abell 0980       & A980    &  155.6275 & 50.1017  & 0.1555 & 15105                          & 14                     &  222 \\
        ZwCl 1021.0+0426 & Zw3146  & 155.9117  & 04.1865  & 0.2894 & 909,9371                       & 43,33                & 106 \\
        Abell 0990       & A990    & 155.9120  & 49.1450  & 0.1416 & 15114                          & 10                      &  91  \\
        ZwCl 1023.3+1257 & Zw3179  & 156.4840  & 12.6910  & 0.1422 & 13375                          & 9                      &  69 \\
        Abell 1033       & A1033   & 157.9320  & 35.0580  & 0.1220 & 15614,15084                    & 33,29                & 191 \\
        Abell 1068       & A1068   & 160.1870  & 39.9510  & 0.1386 & 1652$^{\star}$                 & 26                     & 129 \\
        Abell 1132       & A1132   & 164.6160  & 56.7820  & 0.1351 & 19770,13376                    & 20,9                 &  160 \\
        Abell 1201       & A1201   & 168.2287  & 13.4448  & 0.1671 & 9616                           & 47                     & 165 \\
        Abell 1204       & A1204   & 168.3324  & 17.5937  & 0.1706 & 2205                           & 24                     & 92 \\
        Abell 1235       & A1235   & 170.8040  & 19.6160  & 0.1030 & 18247                          & 18                     & 131 \\
        Abell 1246       & A1246   & 170.9912  & 21.4903  & 0.1921 & 11770                          & 5                      & 226 \\
        Abell 1302       & A1302   & 173.3070  & 66.3990  & 0.1152 & 18245                          & 19                     & 162 \\
        Abell 1413       & A1413   & 178.8260  & 23.4080  & 0.1412 & 5003                           & 75                     & 116 \\
        Abell 1423       & A1423   & 179.3420  & 33.6320  & 0.2142 & 11724,538                      & 26,10                 & 230 \\
        Abell 1437       & A1437   & 180.1040  & 03.3490  & 0.1333 & 15306,15188                    & 10,9                  & 194 \\
        Abell 1553       & A1553   & 187.6959  & 10.5606  & 0.1668 & 12254                          & 9                      & 171 \\
        Abell 1682       & A1682   & 196.7278  & 46.5560  & 0.2272 & 11725                          & 20                     & 151 \\
        Abell 1689       & A1689   & 197.8750  & - 1.3353 & 0.1842 & 6930,7289,5004                 & 76,75,20           & 210 \\
        Abell 1763       & A1763   & 203.8257  & 40.9970  & 0.2312 & 3591                           & 20                     & 237 \\
        Abell 1835       & A1835   & 210.2595  & 02.8801  & 0.2506 & 6880,6881,7370                 & 118,36,40          & 219 \\
        Abell 1918       & A1918   & 216.3420  & 63.1830  & 0.1388 & 18249                          & 21                     & 80 \\
        Abell 1914       & A1914   & 216.5068  & 37.8271  & 0.1660 & 20026,18252,20023,20025,20024  & 29,27,25,22,17 & 255 \\
        Abell 1930       & A1930   & 218.1200  & 31.6330  & 0.1308 & 11733$^{\star}$                & 35                     & 76\\
        Abell 1978       & A1978   & 222.7750  & 14.6110  & 0.1459 & 18250                          & 20                     & 63 \\
        Abell 2009       & A2009   & 225.0850  & 21.3620  & 0.1522 & 10438                          & 20                     & 195 \\
        RXCJ 1504.1-0248 & RXJ1504 & 226.0321  & - 2.8050 & 0.2168 & 17670,17197,17669,4935         & 51,30,29,12      & 120 \\
        Abell 2034       & A2034   & 227.5450  & 33.5060  & 0.1132 & 12886,12885                    & 91,81                & 182 \\
        Abell 2050       & A2050   & 229.0680  & 00.0890  & 0.1191 & 18251                          & 15                     & 106 \\
        Abell 2069       & A2069   & 231.0410  & 29.9210  & 0.1139 & 4965                           & 39                     & 441 \\
        Abell 2111       & A2111   & 234.9337  & 34.4156  & 0.2291 & 11726, 544                     & 21,10                & 208 \\
        Abell 2219       & A2219   & 250.0892  & 46.7058  & 0.2257 & 14356,14431,14355,14451        & 49,39,30,20      & 461 \\
        ZwCl 1717.9+5636 & Zw8197  & 259.5480  & 56.6710  & 0.1132 & 18246                          & 19                     & 76 \\
        Abell 2259       & A2259   & 260.0370  & 27.6702  & 0.1605 & 3245                           & 10                     & 165 \\
        RXCJ 1720+2638   & RXJ1720 & 260.0370  & 26.6350  & 0.1604 & 4361,1453                      & 14,8                 &  376 \\
        Abell 2261       & A2261   & 260.6129  & 32.1338  & 0.2242 & 5007                           & 24                     & 209 \\
        RXCJ 2129+0005   & RXJ2129 & 322.4186  & 00.0973  & 0.2339 & 9370,552                       & 30,10                & 325 \\
        \hline
      \end{tabular}
    }
    \caption{\label{table.samplesummary} Summary of sample. Column 1
      gives the cluster name with a shorter alternative name given in
      column 2, columns 3 and 4 give the RA and DEC of the cluster
      respectively, column 5 gives the spectroscopic redshift of the
      cluster, column 6 gives the \textit{Chandra} ObsID(s), column 7
      gives the length of the observation after lightcurve cleaning,
      column 8 gives the number of galaxies classed as member galaxies
      of each cluster in the caustic method \citep{Serra:2013a}, as
      defined in \citet{Rines:2013a} (equivalent to the number of
      galaxies within the caustics). Starred ObsIDs are ACIS-S
      otherwise all are ACIS-I}
  \end{table*}
\end{center}

\section{Analysis} \label{section.analysis}

\subsection{X-ray data} \label{subsection.xray_analysis} The
\textit{Chandra} data analysis follows closely that presented in
\citet{Giles:2015a}, which in turn is based on the analysis presented
in \citet{mau08a} and \citet{Maughan:2012a}. We summarise the main
steps here, and highlight any differences in the updated analysis used
in this paper. Any peculiarities of the analysis of individual
clusters are noted in Appendix \ref{subsection.individualnotes}.

All of the clusters in our sample were analysed with the
\textsc{ciao}\footnote{See \url{http://cxc.harvard.edu/ciao}.} 4.10
software package and \textsc{caldb}\footnote{See
  \url{http://cxc.harvard.edu/caldb}.} version 4.8.1
\citep{Fruscione:2006a}. Projected emissivity profiles of the ICM were
obtained from the surface brightness profile, measured in circular
annuli centred on the centroid of the X-ray emission. The centroid
location was measured in an X-ray image produced in the $0.7-2\keV$
band, and was determined by iteration using a circular region of
$150\arcs$ and then $50\arcs$ initially centred on the X-ray peak
\citep{mau08a}. Projected temperatures profiles were obtained by
extracting spectra in circular annuli also centred on the X-ray
centroid. The spectra were fit with an absorbed APEC plasma model
\citep{Smith:2001a} in XSPEC \citep{Arnaud:1996a}. For the spectral
analysis, instead of using the $\chi^2$ statistic as in
\citet{Maughan:2016a}, we used the C-statistic, which assumes
Poissonian statistics, and the spectra were grouped to contain at
least five counts per energy bin. This minimal binning is required to
avoid a bias in the XSPEC implementation of the C-statistic when there
are bins in the spectrum with zero or very few counts \citep[see][and
the discussion in Appendix B in the XSPEC manual]{wil05}. We exclude
the outermost temperature measurement in the profile if the measured
temperature in the final bin is greater than 15 keV, has errors
$>$50\%, or including it leads to an unphysical increase in the best
fit temperature profile. %

A further minor difference from our previous analysis is that the
absorbing column used in the spectral fits are fixed at the
\textsc{nhtot} value from \citet{Willingale:2013a}, instead of the
absorbing column from \citet{Dickey:1990a}. The updated \textsc{nhtot}
value now contains absorption contributions from molecular hydrogen,
not previously accounted for. We note this update does not
significantly affect our results. %

In order to calculate hydrostatic mass profiles, we use the models for
the 3D temperature and gas density profiles as presented in
\citet{Vikhlinin:2006a} and fit them to the observed projected
temperature profile and surface brightness profile respectively (this
mass fitting method is sometimes referred to as the forward-fitting
mass method e.g. \citealp{Ettori:2013a} and more details of our
implementation are given in Appendix
\ref{app.section.vikh_3D_models}). In this work we fit the 3D
temperature and gas density models simultaneously, in contrast to the
X-ray mass method used in \citet{Maughan:2016a}. This fitting is done
using a Markov Chain Monte Carlo (MCMC) method using \texttt{emcee}
\citep{Foreman-Mackey:2013a}, and the priors used on the 3D gas
density and temperature models are presented in Table
\ref{table.priors}. Uncertainties on the mass profile are obtained
straightforwardly from the MCMC chains. We also compute the mean
$R_{500}$, and mean hydrostatic mass at $R_{500}$ from these chains
(presented in Table \ref{table.massresults}). Gas mass profiles can
also be obtained from the 3D gas density fit, with uncertainties
obtained from the MCMC chains in the same way. This means that the
uncertainties derived on the gas mass within $\rf$ self-consistently
include the uncertainties on $\rf$.

\begin{center}
  \begin{table*}
    \scalebox{1}{
      \begin{tabular}{lcccccccccccc}
        \hline
        Density profile \\
        \hline
        $n_0$ &       $r_{\text{c}}$    &  $r_{\text{s}}$ & $\alpha$ & $\beta$ & $\epsilon$ & $n_{02}$ & $r_{\text{c2}}$ & $\beta_2$  \\
        cm$^{-3}$ & kpc   & kpc  &          &         &             & cm$^{-3}$ & kpc    &              \\
        \hline
        ($10^{-4}$, $10^{-1}$) & (5, 800) & (100, 4000) & (0, 3) & (0.3, 1.5) & (0, 5) & (0, $10^{-1}$) & (1, 70) & (0.1, 5) \\
        \hline
        Temperature profile \\
        \hline
        $T_{\text{0}}$ &  $r_{\text{cool}}$ & $a_{\text{cool}}$ & $T_{\text{min}}$ & $r_{\text{t}}$ & a & b & c  \\
        keV     &   kpc          &                   &         keV    & kpc  &   &    & \\
        \hline
        (0.5, 18) & (10, 500) & (0, 3) & (0.1, 6 or $T_{\text{0}}$) & (100, 500) & (-0.5, 0.5) & (0, 5) & (0, 1) \\
        \hline
      \end{tabular}
    }
    \caption{\label{table.priors} Summary of priors for the 3D gas density and 3D temperature models (see \S\ref{app.section.vikh_3D_models} for equations) used in the X-ray hydrostatic mass method described in \S \ref{subsection.xray_analysis}. All are flat uniform priors (in linear space) with lower and upper bounds in parentheses.}
  \end{table*}
\end{center}

To investigate the influence of the clusters' dynamical states on our
results, we split our sample into relaxed cool core (RCC) and
non-relaxed cool core (NRCC) clusters, using the same three
diagnostics as in \citet{Maughan:2016a}. To be classed as RCC, a
cluster has to have a low central cooling time ($<$ 7.7 Gyr), a highly
peaked central density profile (logarithmic slope $>$ 0.7 within 0.048
$R_{500}$), and a low centroid shift ($<$ 0.009). This definition of
what constitutes a RCC cluster is a conservative one: for our sample
of 44 clusters, 10 are classified as RCC (see Table
\ref{table.massresults}).

This work is an extension of \citet{Maughan:2016a} from 16 to 44
clusters. As noted \S \ref{section.sample}, we have dropped A267 and
A2631 from the present analysis, leaving 14 clusters in common. For
these clusters, we compared the hydrostatic masses at $R_{500}$
between the two analyses, finding a weighted average ratio of the new
masses to those from \citet{Maughan:2016a} of 0.96$\pm$0.05. This
demonstrates that the improvements in our analysis, and changes to the
\textit{Chandra} calibration have not significantly affected our
hydrostatic mass estimates.

\subsection{Caustic masses} \label{subsection.causticmasses}
In this work, we use the caustic masses from HeCS, as presented in
\citealp{Rines:2013a}. We summarise the key aspects of the analysis in
the following. The caustic mass method uses spectroscopic redshifts
for a number of member galaxies in a cluster (on average $\sim$180 for
our sample, see column 7 Table \ref{table.samplesummary}), to
determine their line-of-sight velocity (relative to the median of the
distribution of the line-of-sight velocities of the galaxies of the
main group on the binary tree \citealp{Diaferio:1999a}), which when
combined with their projected distances from the cluster centre, can
be used to estimate a mass profile of a cluster.

By plotting the line-of-sight velocity versus the
projected distance from the cluster centre for each of the
member galaxies (referred to as a redshift diagram), an overpopulated
region of this parameter space is clearly seen (see e.g.
\citealp{Rines:2013a}), and is bounded by a `trumpet shape'. The
boundaries between galaxies within this trumpet shape and outside it
are known as caustics. The galaxies within this overpopulated
parameter space are assumed to be bound by the cluster's gravitational
potential, and their velocities are assumed to be less than the escape
velocity of the cluster. The amplitude, ${\cal A}(r)$, of this trumpet
shape is related to the cluster's escape velocity and the velocity
anisotropy parameter $\beta(r)$, and decreases as a function of
projected distance from the cluster centre. \citet{Diaferio:1997a}
show that this caustic amplitude, ${\cal A}(r)$, can be related to the
cluster mass within a radius, $r$:
\begin{equation} \label{equation.causticmass}
  GM(<r) =  \mathcal{F}_\beta \int_{0}^{r} {\cal A}^2(r)\,dr \, .
\end{equation}
Equation \eqref{equation.causticmass} simplifies the correct equation
$GM(<r)=\int_0^r{\cal F}_\beta(r){\cal A}^2(r){\mathrm d}r$ by
replacing the function ${\cal F}_\beta(r)$ with the constant filling
factor ${\cal F}_\beta$. The function ${\cal F}_\beta(r)$ combines the
cluster density profile and the anisotropy of the velocity field, both
of which vary as a function of $r$. Beyond $\sim 0.5R_{200}$,
${\cal F}_\beta(r)$ is roughly constant with radius, so assuming a
constant filling factor ${\cal F}_\beta$ should not lead to
significant overestimation or underestimation of mass at larger radii
\citep{Diaferio:1999a}. However, within $\sim 0.5R_{200}$,
approximating ${\cal F}_\beta(r)$ to a constant might lead to an
overestimate of the cluster mass by $\sim 15\%$ \citep{Serra:2011a}.

We note that we use $\mathcal{F}_\beta=0.5$ for the filling factor,
which is appropriate for the algorithm of \citet{Diaferio:1999a} used to
calculate the caustic masses presented in \citet{Rines:2013a}.

The uncertainties on the caustic masses represent the statistic
precision of the caustic mass estimate due to the uncertainties in the
location of the caustics. They do not represent the accuracy with
which the caustic mass measures the true mass, which is $\approx30\%$
\citep{Serra:2011a}, driven largely by the effects of projection and
the viewing angle of the cluster (this is accounted for in our model
below by an intrinsic scatter between caustic and true masses).

The uncertainty on the caustic location increases with decreasing
ratio $\rho$ between the number of galaxies within the caustics and
the number of galaxies in the redshift diagram. On average,
this recipe is a robust estimate of the statistical uncertainty
\citep{Serra:2011a}. However, in sparsely-sampled redshift
diagrams, the caustic technique might underestimate the statistical
uncertainty, because the ratio $\rho$ might remain relatively large,
despite the low number of galaxies in the diagram. We verified that
these cases have a negligible effect on our results by repeating our
analysis with a uniform $50\%$ error on all caustic masses. In this
test, the intrinsic scatter term in our model is effectively removed
and we prevent any points with underestimated errors from driving the
results.

The number of galaxies within the caustics, Ngal, can also have a
systematic effect on the mass derived using the caustic method. The
simulations of \citet{Serra:2011a} suggest an underestimate of the
mass of $\sim10\%$ at $R_{500}$ for $\Ngal\sim100$, with the masses
being unbiased for $200\lta\Ngal\lta1000$. Any underestimate of mass
due to low values of Ngal is expected to approximately cancel with the
overestimate of mass at $R_{500}$ due to the assumption of constant
filling factor. However, as discussed below (\S
\ref{subsection.ngal}), our results suggest that the magnitude of the
mass underestimate for low values of Ngal may be larger than
predicted, and our main conclusions are thus based on the clusters
with the largest values of Ngal.

\subsection{Modelling the mass biases} \label{subsection.modellingbias}

\begin{figure}
  \centering
  \includegraphics[width = 80mm]{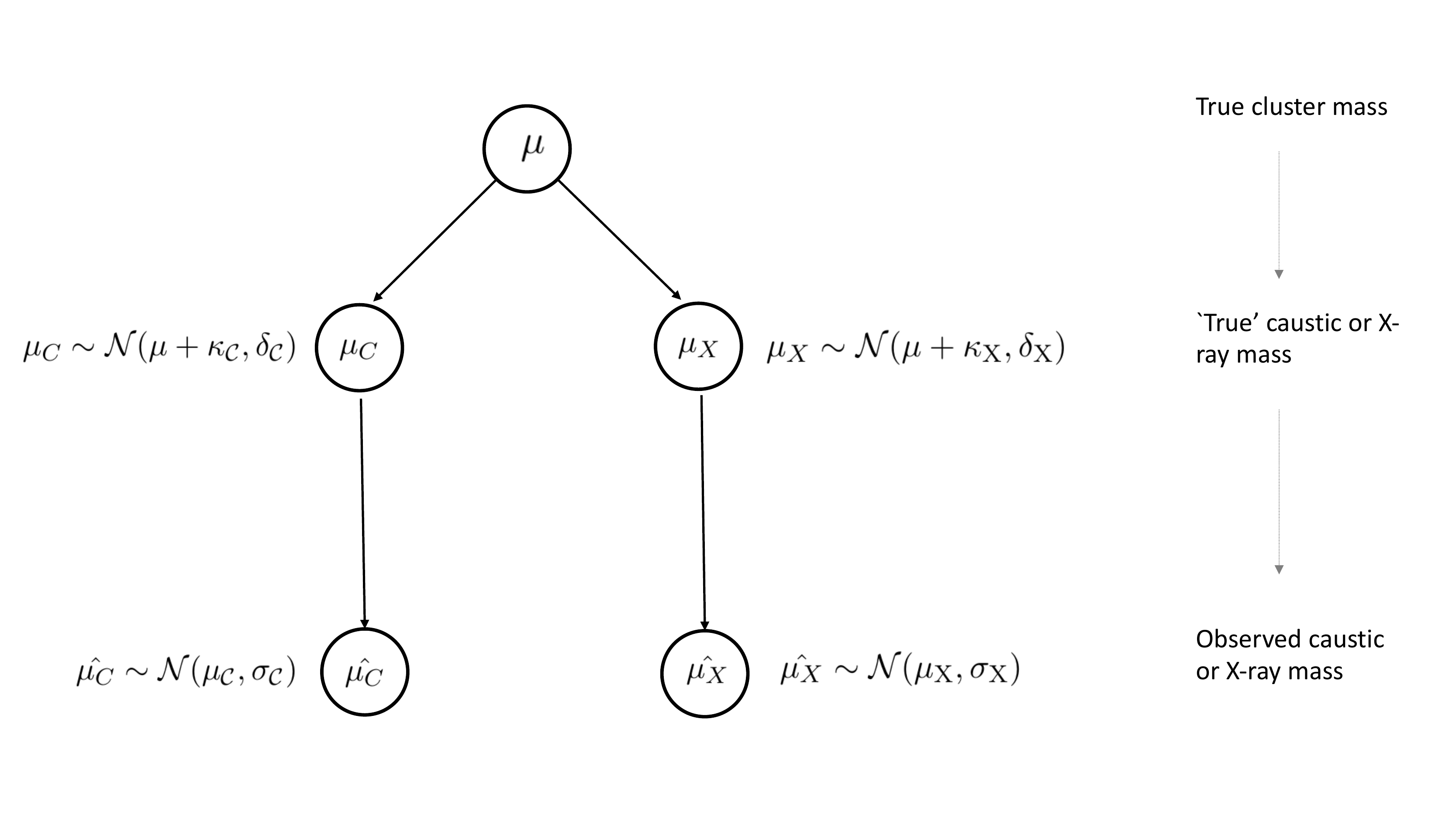}
  \caption{Graphical model summarising the mass modelling framework that we present in \S \ref{subsection.modellingbias}. The symbols are also described in that section; in summary, the symbol $\mu$ represents mass, $\kappa$ is the bias term, $\delta$ is the scatter term and $\sigma$ is the measurement error term (all in log space), and the $C$ and $X$ subscripts denote whether the quantity refers to that from the hydrostatic or caustic method. The $\hat{}$ refers to an observed quantity. $\cal{N}$ refers to a normal distribution, and `$\sim$' denotes `is distributed as'.}
  \label{figure.graphical_plot_mxmc_bayes_analysis}
\end{figure}

In order to constrain the bias and scatter between the hydrostatic and
caustic mass measurement techniques, we use a Bayesian framework (of
which a graphical representation is shown in Figure
\ref{figure.graphical_plot_mxmc_bayes_analysis}). The modelling
framework is the same as in \citet{Maughan:2016a}, but we summarise
the key points here. In the following, we use $\mu$ to denote the base
10 logarithm of mass.

A given cluster with a `true' mass, $\mu$ also has a `true' caustic mass, $\mu_C$, and `true' hydrostatic mass, $\mu_X$, which are related to the true mass, $\mu$, as follows:
\begin{align}
  \muX \sim \cal{N}(\mu + \rm \kapx,\rm \delx) \\
  \muC \sim \cal{N}(\mu + \rm \kapc,\rm \delc)
\end{align}
where $\kapx$ and $\kapc$ parametrise the bias between the true mass and the `true' hydrostatic and caustic mass, respectively, and $\delx$ and $\delc$ parametrise the intrinsic scatter between the true mass and the `true' hydrostatic and caustic mass, respectively. $\cal{N}$ refers to a normal distribution, and ``$\sim$" is equivalent to ``is distributed as".

These `true' hydrostatic and caustic masses are related to the observed hydrostatic and caustic masses ($\hat{\mu_X}$ and $\hat{\mu_C}$, respectively), as follows:
\begin{align}
  \muXh \sim \cal{N}(\rm \muX,\rm \sigx) \\
  \muCh \sim \cal{N}(\rm \muC,\rm \sigc)
\end{align}
where $\sigx$ and $\sigc$ represent the standard deviation of the
lognormal error for the observed hydrostatic and caustic masses,
respectively. Note that our model does not take into account selection
effects that could arise due to the X-ray flux limit. However, we show
in \S \ref{sec:impact-select-effect} that this has a negligible impact
on our results.

We used the same (loose) priors as in \citet{Maughan:2016a} on the cluster masses, bias and scatter terms. Specifically, for the cluster masses in log space, $\mu$, we applied a uniform prior from 12 to 17. For the bias terms in log space, $\kapx$ and $\kapc$, we applied normal priors with mean 0 and a standard deviation of 1. For the intrinsic scatter terms in log space, $\delx$ and $\delc$, we applied normal priors (truncated at zero) with mean 0.09 and a standard deviation of 2.2, corresponding to a weak prior with a mean of $\sim$20\% in normal space.

Using the model described above, we used the observational data that we have for each cluster ($\muXh$, $\sigx$, $\muCh$, $\sigc$) to constrain the 'true' hydrostatic and caustic mass terms ($\muX$, $\muC$), the scatter terms ($\delx$ and $\delc$) and bias terms ($\kapx$ and $\kapc$) for all of the clusters in our sample. Both the scatter terms and the bias terms are degenerate, however, the intrinsic scatter between the hydrostatic and caustic mass measurements
\begin{align}
  \delta = \sqrt{\delx^2+\delc^2}
\end{align}
and the mean bias between the hydrostatic and caustic mass measurements
\begin{align} \label{eqn.mx_mc_ratio}
  \kappa = \kapx - \kapc = \muX - \muC = \log_{10}\left(\frac{M_{X}}{M_{C}}\right)
\end{align}
can be constrained by the data.

When reporting the values of $\kappa$ and $\delta$ in Tables
\ref{table.resultssummary44RCC} and \ref{table.resultssummary44Ngal},
we report the median value of $\delta$, with errors given as the
difference between the median and the 16th and 84th percentiles. For
$\kappa$, we report the mean value. As
$\kappa=\log_{10}(M_{X}/M_{C})$, and $\kappa$ (a quantity in log
space) is normally distributed, the posterior of $M_{X}/M_{C}$ (a
quantity in linear space) is lognormally distributed. Therefore, in
Tables \ref{table.resultssummary44RCC} and
\ref{table.resultssummary44Ngal} we summarise the posterior of
$M_{X}/M_{C}$ by reporting its median and errors in the same way that
we did for $\delta$. In these tables, the $\kappa$ values are given in
base 10 log space, and the $\delta$ values have been converted to
percentage scatters.

We also note that in Table \ref{table.massresults} and in all figures
showing mass errors, we report the mean and standard
deviation of the  hydrostatic and caustic mass measurements in \textit{linear
  space}, which is the form in which masses are often reported in
other work. However, we find that the uncertainties on the mass
measurements are in fact better described using lognormal errors, and
so choose to model the biases associated with the caustic and X-ray
hydrostatic mass measurement methods in log space.

We use the probabilistic programming language
\textsc{stan}\footnote{\url{http://mc-stan.org}} to implement the
model described above, specifically using the No-U-Turn Sampler
\citep{Hoffman:2011a}. We sampled the parameters in our model with
four chains of 5,000 steps each. The analysis can be done at any
radius, and so by repeating the analysis at increasing radii, we
produced a profile of the mean bias between the two mass measurement
methods.

In our analysis, the mass determination methods are independent and
use different definitions of the cluster centre (the x-ray centroid
for the hydrostatic mass, and the peak of the galaxy distribution on
the sky and the median of the line-of-sight velocity distribution of
the cluster members with the caustic method). This means that the
radial coordinates are not identical and a given radius does not
correspond to an identical aperture for both methods. For our main
analysis, we scale radial coordinates to the $\rf$ estimated from the
hydrostatic method. This is motivated by the expectation that the
hydrostatic mass should have lower intrinsic scatter with the true
mass. We show in \S \ref{sec:cluster-centring} that using $\rf$
computed independently for each method does not alter our conclusions.

\section{Results}
\label{sec.results}
In the following, we first present the comparisons between hydrostatic
and caustic masses for the full CHeCS sample. We then demonstrate that
the caustic masses for clusters with lower values of Ngal are likely
to be underestimated. Finally we present our main results, based on
the subset of clusters with higher values of Ngal.

\subsection{Initial results for the full sample} \label{section.results44}

\begin{center}
  \begin{table*}%
    \centering
    \scalebox{1.0}{
      \begin{tabular}{lccccccccccccccc}%
        \hline
        Cluster & z & Status & $R_{500}$ & $\MfX$ & $\MfC$    & Ngal  & Ngal & \textsc{nhtot} \\
                &   &             &   Mpc &  $10^{14}\text{M}_{\odot}$  &  $10^{14}\text{M}_{\odot}$ & sub-sample & & cm$^{-2}$ & \\
        \hline
        ZW1478  &     0.103 &    NRCC &   0.80 $\pm$ 0.03 &   1.57  $\pm$  0.27 &   0.81  $\pm$    0.04 &    low  &   82   &  3.86  \\
        A0655   &     0.127 &    NRCC &   1.08 $\pm$ 0.04 &   4.05  $\pm$  0.83 &   3.87  $\pm$    0.20 &    high &   315  &  4.39  \\
        A0697   &     0.281 &    NRCC &   1.47 $\pm$ 0.05 &   11.88 $\pm$  1.81 &   5.90  $\pm$    2.80 &    mid  &   185  &  3.28  \\
        MS0906  &     0.177 &    NRCC &   1.04 $\pm$ 0.02 &   3.74  $\pm$  0.33 &   1.78  $\pm$    0.24 &    low  &   101  &  3.61  \\
        A0773   &     0.217 &    NRCC &   1.32 $\pm$ 0.03 &   7.97  $\pm$  0.88 &   9.78  $\pm$    0.11 &    mid  &   173  &  1.33  \\
        A0795   &     0.137 &    NRCC &   1.09 $\pm$ 0.04 &   4.18  $\pm$  0.69 &   3.47  $\pm$    0.06 &    mid  &   179  &  3.62  \\
        ZW2701  &     0.216 &    RCC  &   1.07 $\pm$ 0.03 &   4.22  $\pm$  0.54 &   2.38  $\pm$    0.73 &    low  &   93   & 0.767  \\
        A0963   &     0.204 &    NRCC &   1.06 $\pm$ 0.02 &   4.03  $\pm$  0.31 &   4.10  $\pm$    0.04 &    high &   211  &  1.30  \\
        A0980   &     0.155 &    NRCC &   1.36 $\pm$ 0.07 &   8.33  $\pm$  2.20 &   5.66  $\pm$    1.80 &    high &   222  & 0.829  \\
        ZW3146  &     0.289 &    RCC  &   1.29 $\pm$ 0.03 &   8.02  $\pm$  0.95 &   3.99  $\pm$    1.78 &    low  &   106  &  2.69  \\
        A0990   &     0.142 &    NRCC &   1.29 $\pm$ 0.05 &   6.87  $\pm$  1.28 &   2.22  $\pm$    0.82 &    low  &   91   & 0.808  \\
        ZW3179  &     0.142 &    NRCC &   1.44 $\pm$ 0.13 &   10.19 $\pm$  4.57 &   1.52  $\pm$    0.11 &    low  &   69   &  4.08  \\
        A1033   &     0.122 &    NRCC &   1.11 $\pm$ 0.02 &   4.35  $\pm$  0.35 &   3.18  $\pm$    0.04 &    mid  &   191  &  1.73  \\
        A1068   &     0.139 &    RCC  &   1.15 $\pm$ 0.05 &   4.96  $\pm$  1.09 &   8.29  $\pm$    0.66 &    low  &   129  &  1.78  \\
        A1132   &     0.135 &    NRCC &   1.64 $\pm$ 0.08 &   14.09 $\pm$  2.87 &   5.12  $\pm$    0.21 &    mid  &   160  &  0.63  \\
        A1201   &     0.167 &    NRCC &   1.15 $\pm$ 0.02 &   5.03  $\pm$  0.47 &   3.05  $\pm$    0.06 &    mid  &   165  &  1.65  \\
        A1204   &     0.171 &    RCC  &   0.91 $\pm$ 0.02 &   2.52  $\pm$  0.28 &   1.42  $\pm$    0.18 &    low  &   92   &  1.36  \\
        A1235   &     0.103 &    NRCC &   1.05 $\pm$ 0.05 &   3.66  $\pm$  0.87 &   2.04  $\pm$    0.21 &    mid  &   131  &  1.78  \\
        A1246   &     0.192 &    NRCC &   1.12 $\pm$ 0.06 &   4.81  $\pm$  1.26 &   5.85  $\pm$    0.14 &    high &   226  &  1.65  \\
        A1302   &     0.115 &    NRCC &   1.05 $\pm$ 0.03 &   3.60  $\pm$  0.46 &   2.30  $\pm$    0.04 &    mid  &   162  & 0.924  \\
        A1413   &     0.141 &    NRCC &   1.33 $\pm$ 0.02 &   7.54  $\pm$  0.55 &   6.70  $\pm$    0.02 &    low  &   116  &  1.96  \\
        A1423   &     0.214 &    NRCC &   1.10 $\pm$ 0.03 &   4.61  $\pm$  0.50 &   4.14  $\pm$    0.07 &    high &   230  &  1.92  \\
        A1437   &     0.133 &    NRCC &   1.19 $\pm$ 0.02 &   5.31  $\pm$  0.38 &   9.37  $\pm$    1.19 &    mid  &   194  &  2.29  \\
        A1553   &     0.167 &    NRCC &   1.31 $\pm$ 0.06 &   7.42  $\pm$  1.64 &   5.72  $\pm$    0.03 &    mid  &   171  &  2.29  \\
        A1682   &     0.227 &    NRCC &   1.13 $\pm$ 0.03 &   5.05  $\pm$  0.65 &   6.75  $\pm$    0.04 &    mid  &   151  &  1.07  \\
        A1689   &     0.184 &    RCC  &   1.47 $\pm$ 0.02 &   10.56 $\pm$  0.70 &   10.42 $\pm$    3.26 &    high &   210  &  1.97  \\
        A1763   &     0.231 &    NRCC &   1.23 $\pm$ 0.03 &   6.47  $\pm$  0.75 &   10.90 $\pm$    1.24 &    high &   237  & 0.835  \\
        A1835   &     0.251 &    RCC  &   1.45 $\pm$ 0.02 &   10.73 $\pm$  0.67 &   9.57  $\pm$    0.66 &    high &   219  &  2.24  \\
        A1918   &     0.139 &    NRCC &   1.12 $\pm$ 0.05 &   4.61  $\pm$  1.14 &   2.55  $\pm$    0.06 &    low  &   80   &  1.42  \\
        A1914   &     0.166 &    NRCC &   1.44 $\pm$ 0.04 &   9.91  $\pm$  1.20 &   6.08  $\pm$    0.17 &    high &   255  &  1.10  \\
        A1930   &     0.131 &    NRCC &   1.06 $\pm$ 0.05 &   3.81  $\pm$  0.78 &   2.35  $\pm$    0.15 &    low  &   76   &  1.15  \\
        A1978   &     0.146 &    NRCC &   0.95 $\pm$ 0.02 &   2.79  $\pm$  0.32 &   1.19  $\pm$    0.33 &    low  &   63   &  1.80  \\
        A2009   &     0.152 &    NRCC &   1.33 $\pm$ 0.05 &   7.69  $\pm$  1.46 &   4.29  $\pm$    0.21 &    mid  &   195  &  3.89  \\
        RXJ1504 &     0.217 &    RCC  &   1.30 $\pm$ 0.01 &   7.65  $\pm$  0.36 &   3.09  $\pm$    2.17 &    low  &   120  &  8.38  \\
        A2034   &     0.113 &    NRCC &   1.31 $\pm$ 0.02 &   7.05  $\pm$  0.35 &   5.46  $\pm$    0.04 &    mid  &   182  &  1.62  \\
        A2050   &     0.119 &    NRCC &   1.04 $\pm$ 0.03 &   3.49  $\pm$  0.40 &   3.91  $\pm$    1.08 &    low  &   106  &  5.20  \\
        A2069   &     0.114 &    NRCC &   1.33 $\pm$ 0.04 &   7.46  $\pm$  1.14 &   6.10  $\pm$    0.06 &    high &   441  &  2.03  \\
        A2111   &     0.229 &    NRCC &   1.19 $\pm$ 0.03 &   5.85  $\pm$  0.66 &   3.71  $\pm$    0.42 &    mid  &   208  &  1.99  \\
        A2219   &     0.226 &    NRCC &   1.55 $\pm$ 0.02 &   12.86 $\pm$  0.58 &   10.07 $\pm$    2.61 &    high &   461  &  1.86  \\
        ZW8197  &     0.113 &    NRCC &   0.87 $\pm$ 0.03 &   2.08  $\pm$  0.36 &   1.86  $\pm$    0.03 &    low  &   76   &  2.39  \\
        A2259   &     0.161 &    NRCC &   1.16 $\pm$ 0.05 &   5.14  $\pm$  0.95 &   4.58  $\pm$    0.81 &    mid  &   165  &  3.81  \\
        RXJ1720 &     0.160 &    RCC  &   1.24 $\pm$ 0.03 &   6.30  $\pm$  0.68 &   5.06  $\pm$    0.32 &    high &   376  &  3.88  \\
        A2261   &     0.224 &    RCC  &   1.25 $\pm$ 0.03 &   6.83  $\pm$  0.70 &   3.34  $\pm$    1.20 &    high &   209  &  3.62  \\
        RXJ2129 &     0.234 &    RCC  &   1.31 $\pm$ 0.05 &   8.07  $\pm$  1.48 &   6.06  $\pm$    1.33 &    high &   325  &  4.21  \\
        \hline
      \end{tabular}
    }
    \caption{Summary of the caustic and hydrostatic masses
      as calculated within the hydrostatic $R_{500}$ for
      each cluster. Column 1 is the cluster name; column 2
      is the redshift of the cluster; column 3 is the
      dynamical state of the cluster; column 4 is the mean
      $R_{500}$; column 5 is the hydrostatic mass at
      $R_{500}$; column 6 is the caustic mass at the hydrostatic
      $R_{500}$; column 7 gives the Ngal sub-sample to
      which each cluster belongs (see \S
      \ref{subsection.ngal}); column 8 is the number of
      galaxies of each cluster within the caustics
      \citep{Serra:2013a}, as given in
      \citet{Rines:2013a}; column 9 gives the
      \textsc{nhtot} value \citep{Willingale:2013a} that
      we used for each cluster (see
      \S\ref{subsection.xray_analysis}). All errors are
      $1\sigma$ errors.}
    \label{table.massresults}
  \end{table*}
\end{center}

\begin{figure}
  \centering
  \includegraphics[width = 80mm]{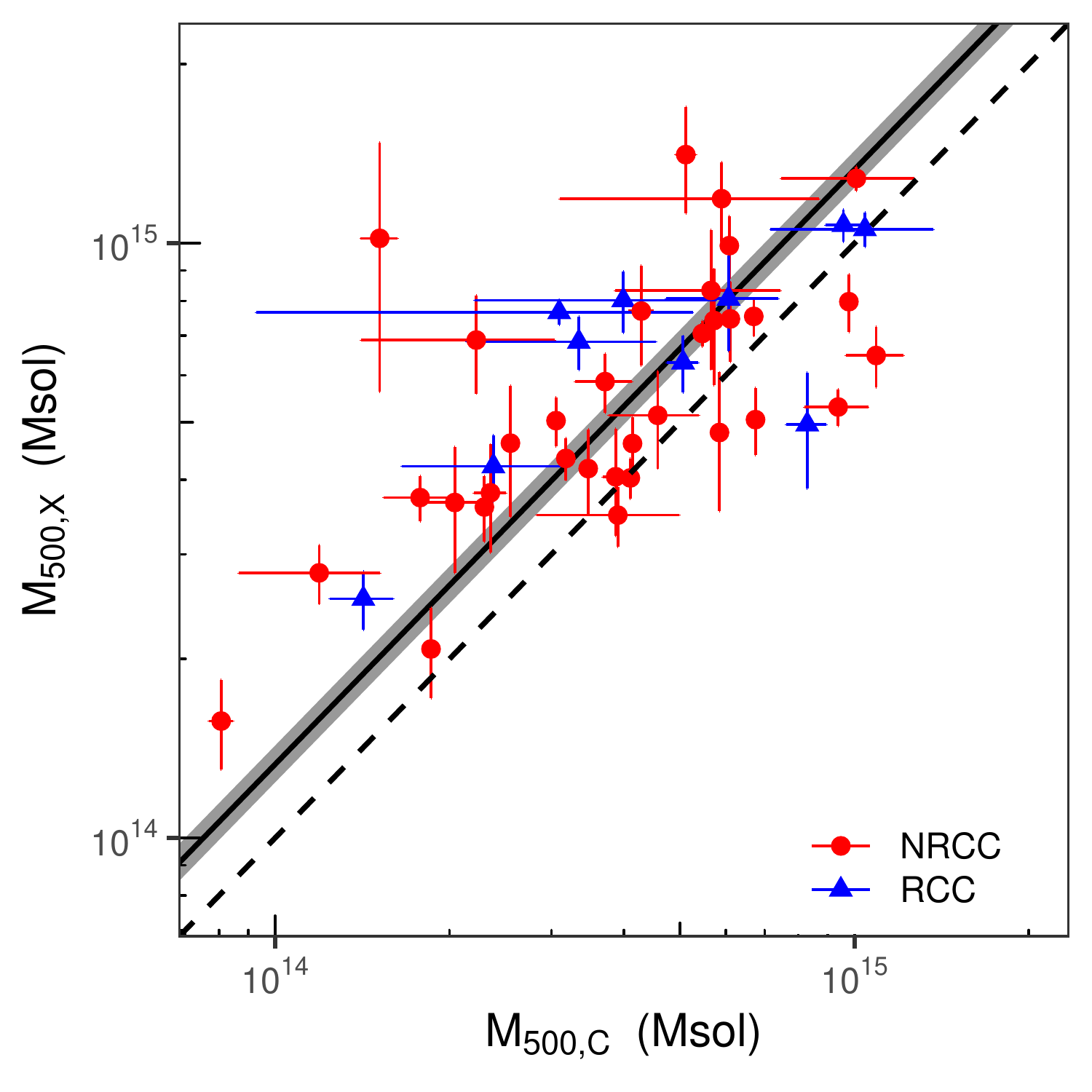}
  \caption{The hydrostatic versus the caustic masses are plotted
    for the full sample of clusters (with masses calculated at
    the hydrostatic $R_{500}$). NRCC clusters are red circles
    and RCC clusters are blue triangles. The solid black line
    shows the best fitting model to the full data, with the
    shaded region indicating the $1\sigma$ uncertainty. The 1:1
    line is also plotted as a dashed black line.}
  \label{figure.massscatterplot44RCC}
\end{figure}

The caustic and hydrostatic mass profiles for each cluster are shown
in Figure \ref{figure.caustichydrostaticmassprofiles}\footnote{For
  A1835, the hydrostatic mass profile decreases unphysically at around
  $R_{500}$. This is believed to be due to the hydrostatic assumption
  becoming worse at large radii and is discussed in detail in
  \citet{Bonamente:2013a}.}. These profiles were used to calculate the
$\MfX$ and $\MfC$, using the $R_{500}$ value
calculated from the hydrostatic mass profile for each cluster (unless
stated otherwise, $R_{500}$ always refers to that derived from the
hydrostatic mass profile)\footnote{We note that this is the maximum
  radius that can usefully be used for comparisons with X-ray data.
  The $R_{500}$ radius is not optimal for caustic mass profiles, as
  the caustic method performs better at larger radii, though the bias
  at this radius can be estimated from simulations \citep{Serra:2011a}
  which makes a meaningful comparison possible.}. The hydrostatic and
caustic masses are reported in Table \ref{table.massresults} and are
compared in Figure \ref{figure.massscatterplot44RCC}.

The temperature profiles required for the hydrostatic mass estimates
were measured out to radii close to, or beyond, $R_{500}$ for the
majority of clusters (see Figure
\ref{figure.caustichydrostaticmassprofiles}). For the mass profiles at
radii greater than the extent of the temperature profile, we
extrapolated the best fitting 3D temperature profile. As a
consequence, our hydrostatic mass profiles beyond the extent of the
measured temperature profiles are less robust. The median radius out
to which the temperature profiles were measured is 0.95 $R_{500}$, and
the range of radii is 0.51 - 1.62 $R_{500}$.

The use of the hydrostatic estimate of $R_{500}$ to determine both
sets of masses introduces a covariance between the hydrostatic and
caustic masses that is not included in our model. For this reason, we
verified that using a fixed aperture of radius 1 Mpc for the mass
measurements made a negligible difference to our results \citep[as
found in][]{Maughan:2016a}.

\begin{figure}
  \centering
  \includegraphics[width = 80mm]{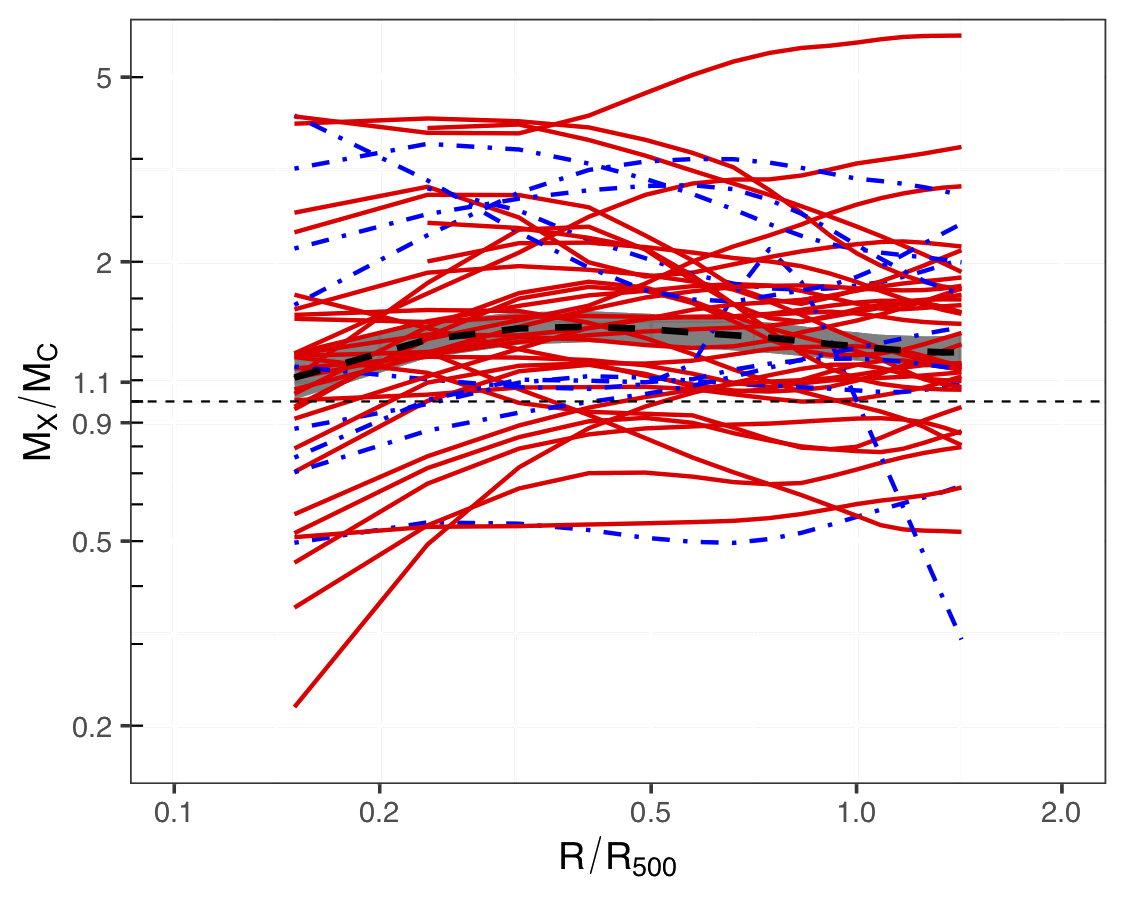}
  \caption{The $M_{X}/M_{C}$ profiles for all 44 clusters in our
    sample, scaled in radius by the hydrostatic $R_{500}$. NRCC
    clusters are solid red lines and RCC clusters are blue dot
    dashed lines. The 1:1 line is also plotted as a thin dashed
    black line. The thick dashed black line shows the average
    $M_{X}/M_{C}$ ratio, and the shaded region shows the
    1$\sigma$ uncertainty.}
  \label{figure.allprofs44RCC}
\end{figure}

The $M_{X}/M_{C}$ profile of each cluster (calculated as
$\muXh-\muCh$) is shown in Figure \ref{figure.allprofs44RCC}. We also
plot the mean $M_{X}/M_{C}$ profile (calculated as the mean bias
$\kappa$, see equation \ref{eqn.mx_mc_ratio}). The $M_{X}/M_{C}$
profile is consistent with the bias being constant with radius. At
smaller radii the $M_{X}/M_{C}$ value decreases, but this is likely
due to the caustic masses being overestimated at small radii (see e.g.
Figure 12 \citealp{Serra:2011a}).

\begin{figure}
  \centering
  \includegraphics[width = 80mm]{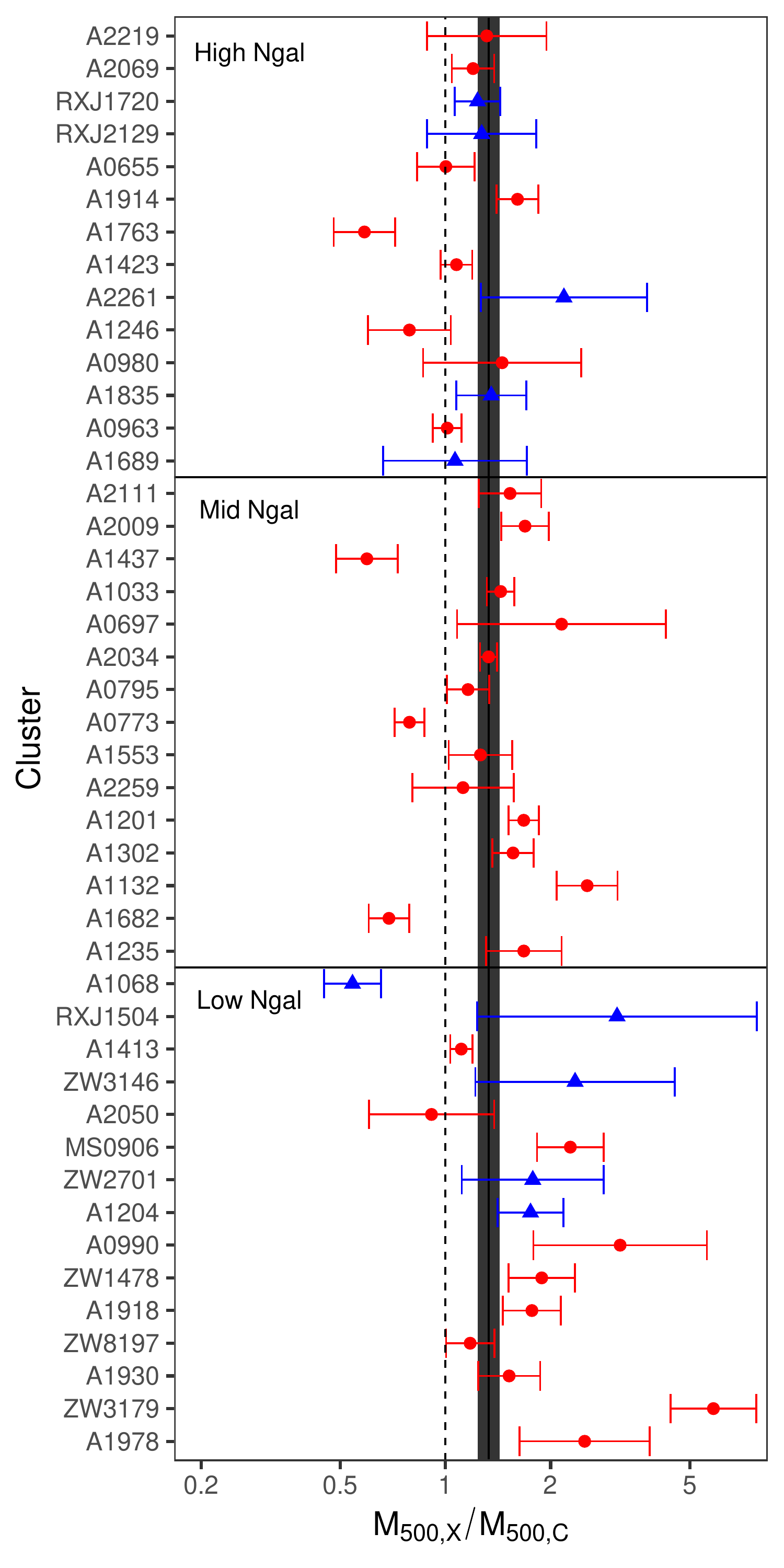}
  \caption{We show the $\MXMC$ ratio for all 44 clusters
    in our sample. NRCC clusters are red circles and RCC
    clusters are blue triangles. The solid black line and shaded
    region show the average $\MXMC$ ratio at the
    hydrostatic $R_{500}$. 1$\sigma$ errors are shown. The clusters are ranked by Ngal with the
    highest at the top, and the horizontal lines separate the
    clusters into the low, mid and high Ngal subsets (see \S \ref{subsection.ngal}).}
  \label{figure.allratios44RCC}
\end{figure}

The $\MXMC$ ratios at $R_{500}$ for the individual clusters is
shown in Figure \ref{figure.allratios44RCC}, and the average ratios
are summarised in Table \ref{table.resultssummary44RCC}. At $R_{500}$
the caustic and hydrostatic masses generally do not agree well. For
the full sample, the hydrostatic mass on average $\sim$ 30\% higher
than the caustic mass (significant at $\approx3\sigma$). Similar
results are found for both the NRCC and RCC clusters.

This $\MXMC$ value is consistent with the our earlier findings
\citep{Maughan:2016a}, but rules out $\MXMC=1$ at higher
significance, and remains counter-intuitive. The hydrostatic mass is
expected to be underestimated compared to the true mass (e.g.
\citealp{Nelson:2014a,Eckert:2019a}) while the caustic mass expected
to be overestimated compared to the true mass at $R_{500}$ (e.g.
\citealp{Serra:2011a}); thus we expect the $\MXMC$ ratio to be
less than 1.

\begin{table}
  \centering
  \def\arraystretch{1.4}%
  \scalebox{0.9}{
    \begin{tabular}{lccccccc}
      \hline
      Aperture &    Subset & $N_C$ &  $\kappa$             &     $\MXMC$                   &    $\delta$ (\%)\\
      \hline
      $R_{500}$ &    All   & 44          &   0.123$\pm$0.031 & $1.33^{+0.10}_{-0.09}$           &   $39^{+06}_{-06}$  \\
      $R_{500}$ &    RCC  & 10           &   0.126$\pm$0.102 & $1.32^{+0.33}_{-0.24}$           &   $51^{+25}_{-16}$  \\
      $R_{500}$ &    NRCC  & 34           &   0.125$\pm$0.035 & $1.33^{+0.11}_{-0.10}$           &   $41^{+07}_{-06}$  \\
      \hline
    \end{tabular}
  }
  \caption{Summary of the results from our mass comparison analysis.
    The first column gives the aperture within which the caustic and
    hydrostatic masses were calculated. For this column, $R_{500}$
    refers to the hydrostatic $R_{500}$. The second column gives the
    subset for which the summary statistics are given, and the third
    column $N_C$ gives the number of clusters in that subset.
    Columns 4, 5 and 6 give the summary statistics. The mean bias is
    given in column 4, the median ratio of masses is given in column
    5, and the intrinsic scatter between the caustic and hydrostatic
    mass is given in column 6. See \S\ref{subsection.modellingbias}
    for details on how these values were derived.}
  \label{table.resultssummary44RCC}
\end{table}

\subsection{Dependency on Ngal} \label{subsection.ngal} Figure
\ref{figure.massscatterplot44RCC} suggests that the high value of the
$\MXMC$ ratio for the full sample may be driven by clusters with lower
caustic masses (with hydrostatic masses being overestimated and/or
caustic masses being underestimated). However, simulations indicate
that dynamical mass estimators in general (including a different
implementation of the caustic method to that used here) tend to
underestimate masses when the number of galaxies used is low
\citep{Wojtak:2018a}. This is a confounding factor since the lower
mass clusters tend to have lower Ngal.

\begin{figure}
  \centering \includegraphics[width = 80mm]{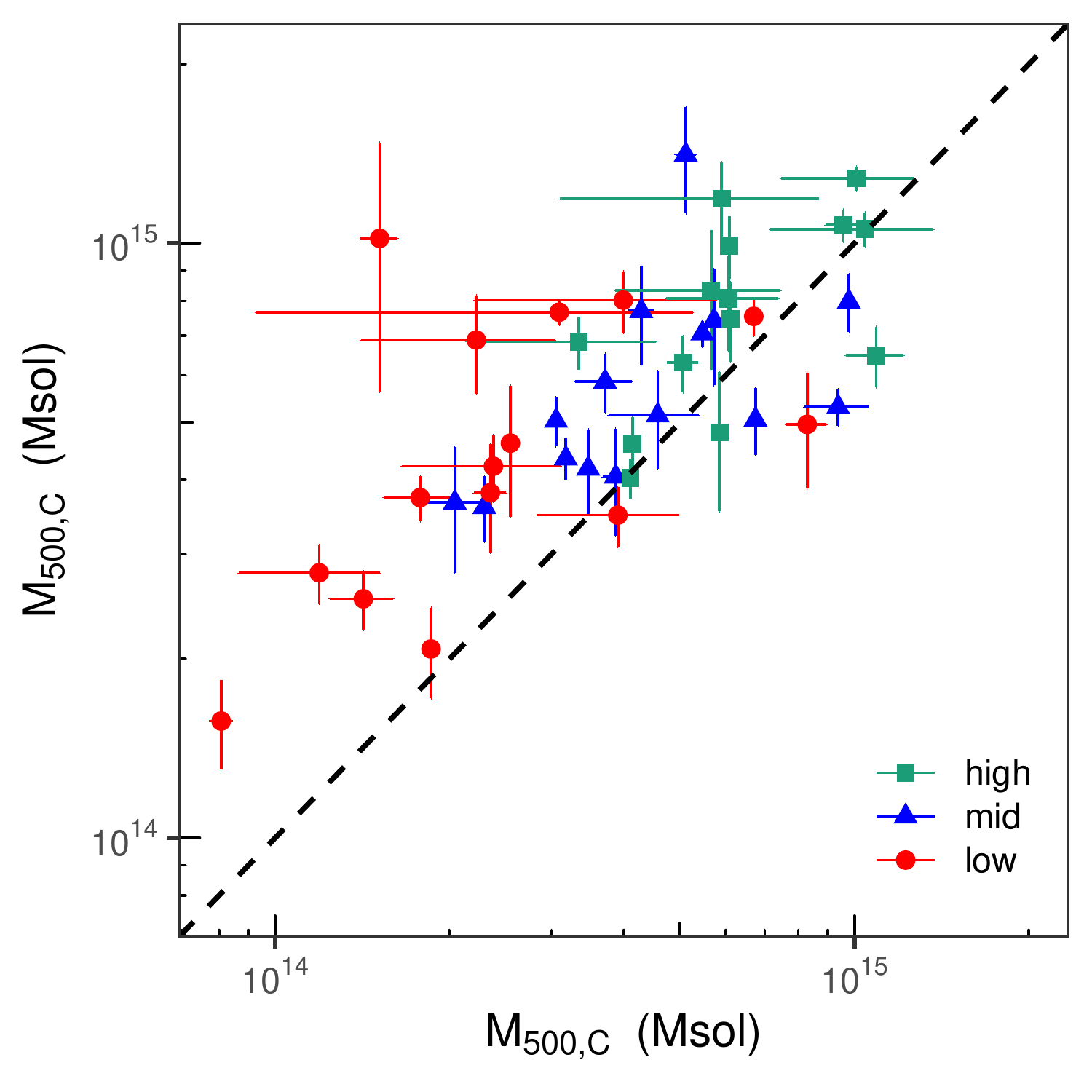}
  \caption{The hydrostatic and caustic masses measured within
    the hydrostatic $R_{500}$ are plotted for the full sample.
    This is the same data as plotted in Figure
    \ref{figure.massscatterplot44RCC}, but now showing the Ngal
    subsets. The high, mid and low Ngal bins are green squares,
    blue triangles and red circles respectively. The 1:1 line is
    also plotted as a dashed black line.}
  \label{figure.massscatterplot44Ngal}
\end{figure}

In order to investigate the role of Ngal in our results, the sample
was split on Ngal into three approximately equal subsets of 15, 15 and
14 clusters (see the Ngal sub-sample column in Table
\ref{table.massresults}). These low, mid and high Ngal sub-samples have
$63\le\Ngal\le129$ (mean 93), $131\le\Ngal\le208$ (mean 174), and
$210\le\Ngal\le461$ (mean 281) respectively. Figure
\ref{figure.massscatterplot44Ngal} shows the comparison between
hydrostatic and caustic masses with the different Ngal subsets
indicated. This illustrates the trend that lower mass clusters tend to
have lower Ngal. It is also clear that, at a given hydrostatic mass,
the clusters with the lowest Ngal are most discrepant in their caustic
mass.

\begin{figure*}
  \centering
  \includegraphics[width = 85mm]{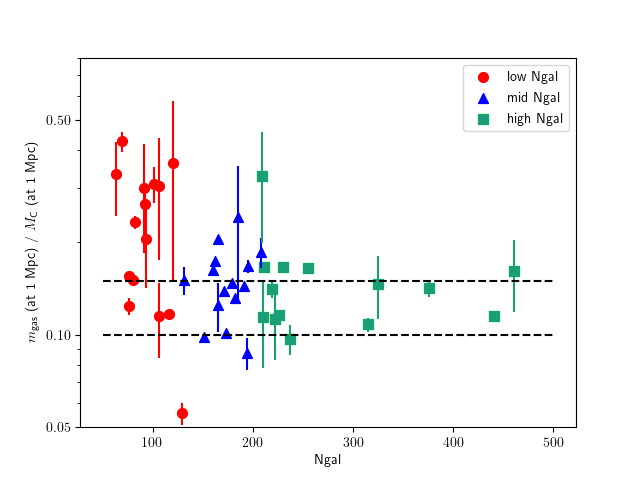}
  \includegraphics[width = 85mm]{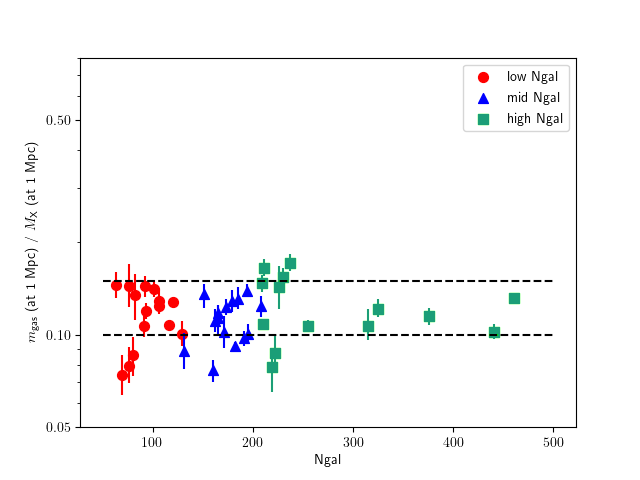}
  \caption{\textit{Left:} $\fgas$ measured at 1 Mpc using
    caustic mass in the denominator is plotted against Ngal.
    \textit{Right:} The same plot, but with $\fgas$ calculated
    with hydrostatic mass in the denominator. In both panels,
    the black dashed lines indicate $\fgas=0.10$ and
    $\fgas=0.15$ for reference, encompassing the approximate
    expected range of $\fgas$.
  }
  \label{figure.fgas_Ngal_1mpc}
\end{figure*}

To investigate this further, the gas fraction, $\fgas$ (defined as the
gas mass, $m_{\text{gas}}$, within some radius divided by the total
mass within the same radius), was found to be a useful discriminator
between possible systematics in the analysis. This is because
$m_{\text{gas}}$ can be measured accurately from X-ray observations
(e.g. \citealp{Nagai:2007a}), and there are robust theoretical
predictions for the range of $\fgas$ expected. We thus computed
$\fgas$ at a radius of 1 Mpc (chosen to avoid dependency of the radius
on any mass estimation method) with either the hydrostatic and caustic
masses in the denominator. The results are plotted in Figure
\ref{figure.fgas_Ngal_1mpc}. The plot also indicates the approximate
range of $\fgas$ ($\sim$ 0.1 - 0.15) expected at this radius (e.g.
\citealp{Eckert:2016a}).

When the caustic mass is used as the denominator in $\fgas$, there is
a strong indication of a trend with Ngal, with unphysically high
values of $\fgas$ for the clusters with lower values of Ngal. The mean
and standard error of $\fgas$ was 0.23$\pm$0.03, 0.15$\pm$0.01 and
0.15$\pm$0.01 for the low, mid and high Ngal sub-samples respectively.

In contrast, when the hydrostatic mass is used to compute $\fgas$, the
trend with Ngal disappears, and the average values of 0.12$\pm$0.01, 0.11$\pm$0.01,
0.12$\pm$0.01, for the low, mid and high Ngal sub-samples respectively
are consistent with the expected range.

These $\fgas$ measurements suggest that Ngal is responsible for the
high values of $\MXMC$ due to an underestimate of $M_C$. However,
due to the correlation between Ngal and mass, the possibility remains
that a trend with mass is more fundamental. During the completion of
the work presented here, additional caustic mass measurements became
available in the HeCS-Omnibus catalogue \citep{Sohn:2020a} that allow
the effects of mass and Ngal to be separated. The new catalogue
includes new caustic mass estimates for clusters in our sample, using
the same caustic method, but based (in most cases) on a larger number
of galaxies per cluster.

\begin{figure}
  \centering
  \includegraphics[width = 80mm]{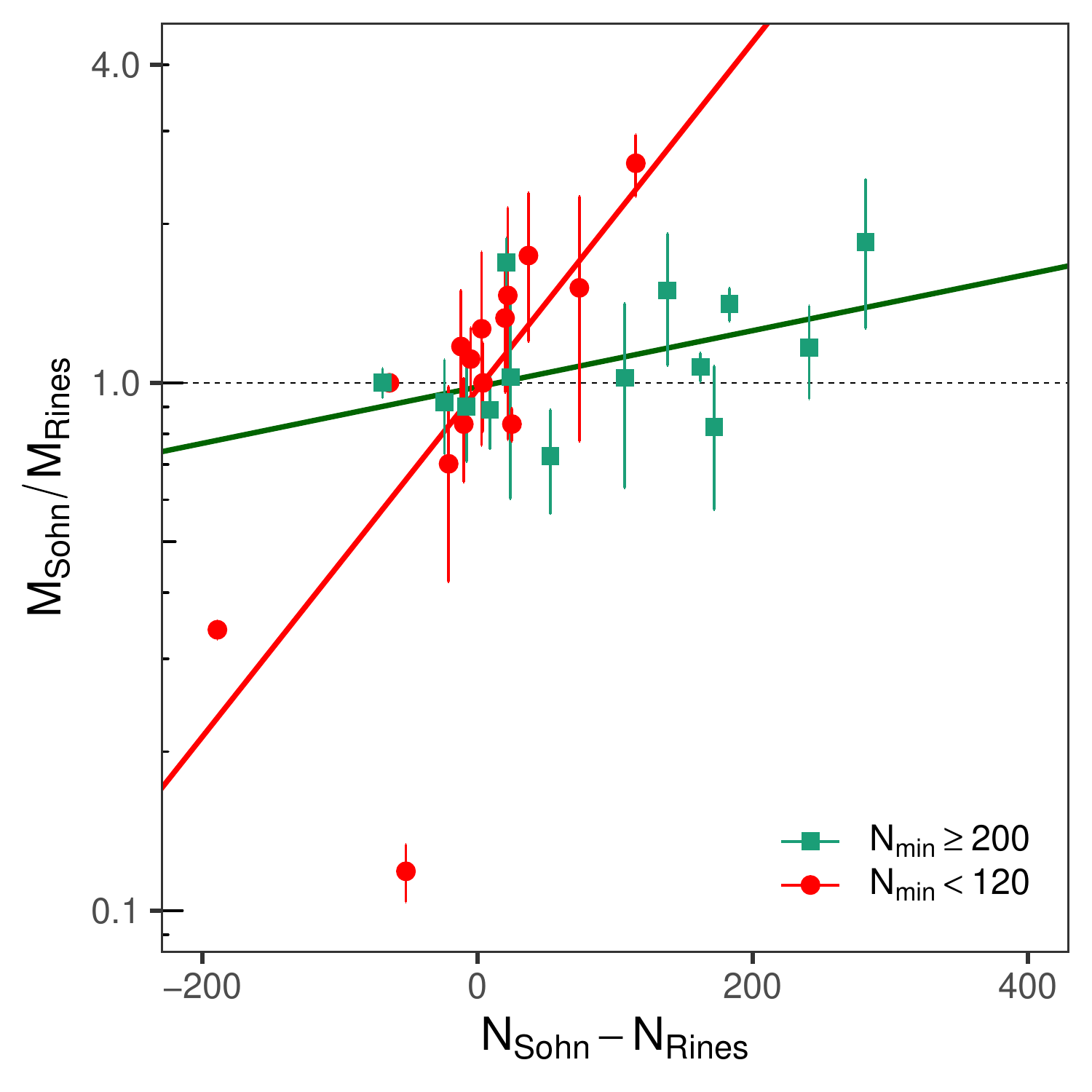}
  \caption{The ratio of the caustic masses (measured at the
    hydrostatic $R_{500}$) from \citet{Sohn:2020a} to those from
    \citet{Rines:2013a} are plotted as a function of the
    difference in Ngal between the analyses. The clusters are
    split into subsets containing those where the minimum Ngal
    used in the two analyses was at least 200 (green squares)
    and those where the minimum Ngal was less then 120 (red
    circles). The lines show the best fits to each subset.}
  \label{figure.sohn_rines_mprofs_and_ngal}
\end{figure}

When comparing the earlier and newer caustic mass measurements, it was
found that for well-sampled clusters, increasing the sampling had
little effect on the mass. On the contrary, for more poorly sampled
clusters, increasing Ngal systematically led to an increased mass
estimate. This is illustrated in Fig.
\ref{figure.sohn_rines_mprofs_and_ngal}, which shows the ratio of the
caustic masses (measured at the hydrostatic $R_{500}$) from
\citet{Sohn:2020a} to those from both \citet{Rines:2013a} (as used in
our work), plotted against the difference in Ngal between the two
analyses.

The clusters were divided into subsets based on the minimum Ngal used
in the two caustic calculations. For clusters where at least 200
galaxies were used in both analyses (green squares in Fig.
\ref{figure.sohn_rines_mprofs_and_ngal}), increasing the sampling has
a weak or null effect on the mass estimate. The green line is a
straight line fit to the data in this parameter space (i.e.
$\log(M_\text{Sohn}/M_\text{Rines})$ versus
$N_\text{Sohn} - N_\text{Rines}$) using BCES y on x regression
\citep{akr96}. This is a simplistic model and is not intended to fully
describe the systematic effects in the analyses, but is an adequate
description of the data presented here. The slope of the line is
$0.001\pm0.001$.

The impact of increasing Ngal for less well-sampled clusters is seen
in the blue triangles in Fig. \ref{figure.sohn_rines_mprofs_and_ngal}.
These are clusters for which Ngal is $<120$ in at least one of the two
analyses. It is clear in the figure that increasing the sampling leads
to systematically higher caustic mass estimates. The best fitting line
to this subset has a slope of $0.008\pm0.002$. The subsets were
defined to split the sample into three equal groups, and the
intermediate subset falls between the two outer cases and are omitted
from the figure for clarity.

We further test the role of Ngal as follows. We undersample the galaxy
catalogues of the 14 clusters in our high-Ngal sub-sample, by
progressively removing the faintest galaxies. When Ngal drops by
$\sim 50\%$, falling in the range $\sim 100-200$, the caustic mass
within $R_{500}$ decreases by $\sim 15\%$, on average. This average
mass underestimate is consistent with the analysis of Serra et al
(2011) who found that with $\text{Ngal}\sim 100$ the escape velocity
at $R_{500}$ is underestimated by $\sim 5-10\%$, whereas
$\text{Ngal}\gtrsim 200$ is required to return unbiased escape
velocity estimates \citep[see also][]{hal22}. The lower Ngal of the
undersampled clusters in our high-Ngal sample would move these
clusters in our mid-Ngal sample: the $\sim 15\%$ mass underestimate we
find in this test is thus also consistent with the larger ratio
$\MXMC$ we find for our mid-Ngal sample, as illustrated in
\textsection \ref{sec:final-results} below (Table
\ref{table.resultssummary44Ngal}).

Based on these results, we conclude that a systematic underestimate of
the caustic mass at $R_{500}$ for low values of Ngal is responsible
for the high apparent $\MXMC$ ratio found at low masses, and that
the high Ngal sub-sample ($\Ngal\ge210$) is not significantly affected
by this systematic. We thus repeat our analysis on the separate Ngal
subsets in the following section, and will draw our final conclusions
based on the results for the high Ngal subset alone.

\subsection{Final results}
\label{sec:final-results}
\begin{figure}
  \centering
  \includegraphics[width = 80mm]{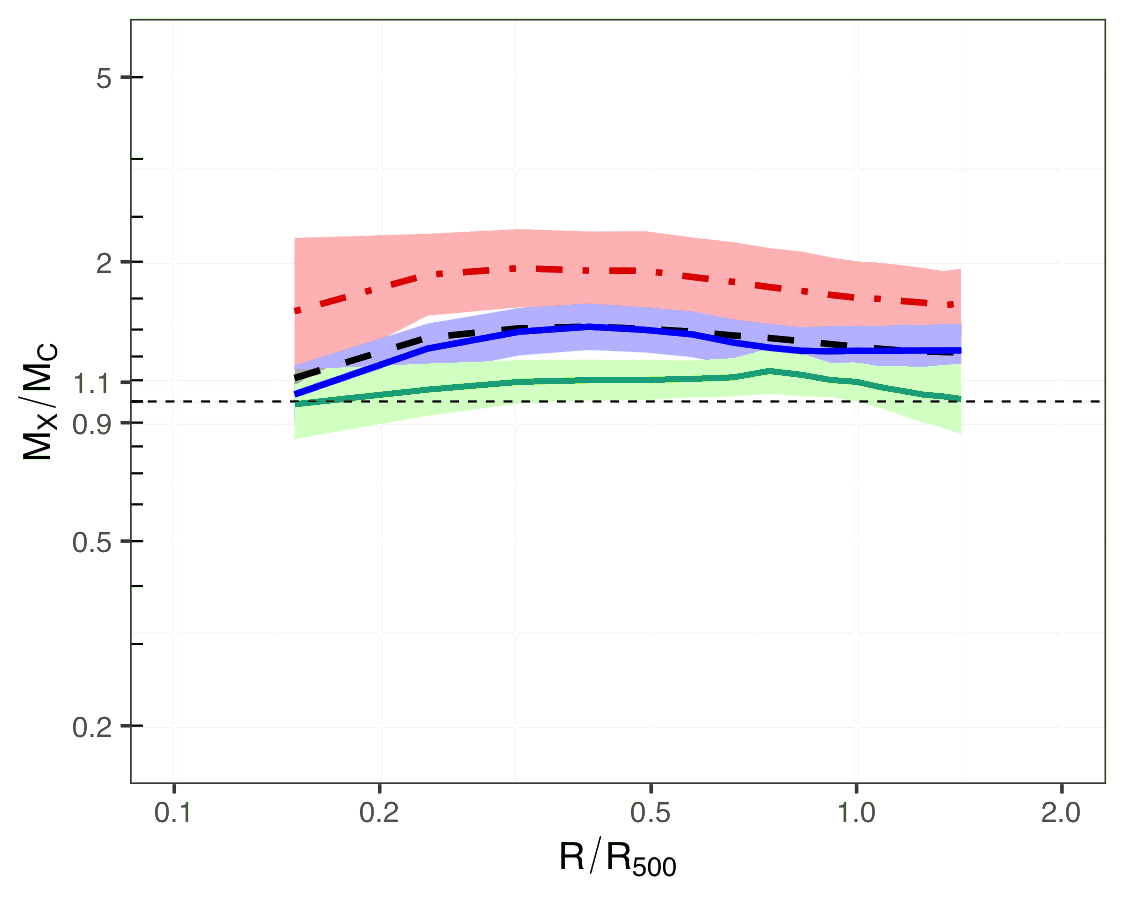}
  \caption{The average $M_{X}/M_{C}$ profiles for the high Ngal,
    mid Ngal and low Ngal clusters in green lines, blue lines,
    and red dot-dashed lines respectively. The 1:1 line is also
    plotted as a thin black dashed line. The thick dashed black
    line shows the average $M_{X}/M_{C}$ profile for all
    clusters. The profiles are scaled in radius by the
    hydrostatic $R_{500}$, and the shaded regions show the
    1$\sigma$ uncertainties.}
  \label{figure.allprofscondensed44Ngal}
\end{figure}

The analysis described in \S \ref{section.results44} was repeated for
each of the Ngal subsets separately, and the results are summarised in
Table \ref{table.resultssummary44Ngal}. Figure
\ref{figure.allprofscondensed44Ngal} shows the profiles of $M_X/M_C$
for the different Ngal subsets. The separation between the subsets is
clear, with no radial dependency.

\begin{table}
  \centering
  \def\arraystretch{1.5}%
  \scalebox{0.9}{
    \begin{tabular}{lccccccc}
      \hline
      Aperture &    Subset &  $N_C$ &  $\kappa$             &     $\MXMC$                   &    $\delta$ (\%)\\
      \hline
      $\mathbf{R_{500}}$ &    \textbf{High}    & \textbf{14}           &    $\mathbf{0.050\pm0.043}$ & $\mathbf{1.12^{+0.11}_{-0.10}}$           &   $\mathbf{25^{+12}_{-08}}$ \\
      $R_{500}$ &    Mid     & 15          &   0.109$\pm$0.053 & $1.28^{+0.16}_{-0.14}$           &   $41^{+12}_{-09}$  \\
      $R_{500}$ &    Low    & 15           &   0.230$\pm$0.082 & $1.69^{+0.33}_{-0.27}$           &   $59^{+18}_{-13}$  \\
      \hline
    \end{tabular}
  }
  \caption{Summary of the results from our mass comparison
    analysis when we define subsets of our clusters based on the
    Ngal value. The columns are the same as in Table
    \ref{table.resultssummary44RCC}. The line in bold represents
    the most reliable results, based on the high Ngal subset.}
  \label{table.resultssummary44Ngal}
\end{table}

As expected based on the preceding analysis, the results in Table
\ref{table.resultssummary44Ngal} show a trend towards lower
$M_X$/$M_C$ ratios with increasing Ngal. The most reliable results are
those based on the sample of 14 clusters with the highest values of
Ngal, and these are highlighted in boldface in the table. For this
subset, the average ratio of hydrostatic to caustic masses is
$\MXMC=1.12^{+0.11}_{-0.10}$ with an intrinsic scatter between the masses of
$\delta=(25^{+12}_{-08})\%$.

\begin{figure}
  \centering
  \includegraphics[width = 80mm]{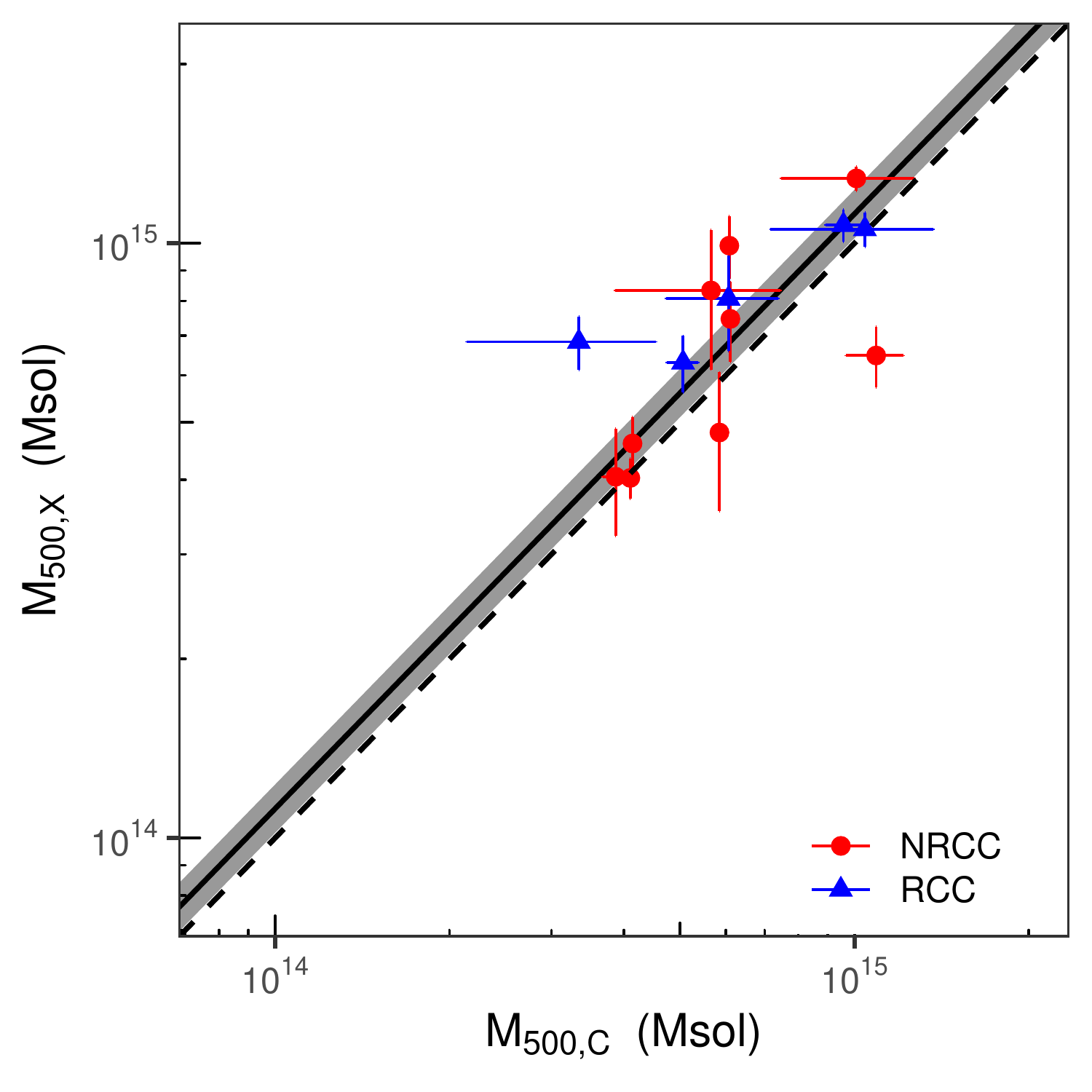}
  \caption{We show the hydrostatic masses versus the caustic
    masses for the 14 clusters in our sample with the highest
    Ngal values, both calculated at the hydrostatic $R_{500}$.
    We show RCC clusters as blue triangles and NRCC clusters as
    red circles. The solid black line
    shows the best fitting model, with the
    shaded region indicating the $1\sigma$ uncertainty. The 1:1
    line is also plotted as a dashed black line.}
  \label{figure.massscatterplot14Ngal}
\end{figure}

The comparison between the hydrostatic and caustic masses for the high
Ngal subset is shown in Figure \ref{figure.massscatterplot14Ngal},
with the RCC/NRCC classification of the clusters indicated. There does
not appear to be a significant trend with dynamical state but given the
reduced size of this high Ngal subset, a more detailed analysis would
not be useful.

Recall that 14 of the 16 clusters used in our earlier analysis in
\citet{Maughan:2016a} are in common with the current analysis. Of
these, six are in our high Ngal subset, seven are in the mid Ngal
sub-sample, and one is in the low Ngal sub-sample. This mix of mid-
and high-Ngal clusters naturally explains why
\citet{Maughan:2016a}
found an $M_X$/$M_C$ ratio that was slightly higher
($\MXMC=1.20\pm0.12$) than our final result for the high Ngal
subset.

\section{Discussion} \label{section.discuss_results_mxmc} In the
following section we consider the possible systematics that could have
impacted our results. We explore in more detail our conclusion that
the apparent increase in $\MXMC$ at low masses is driven by Ngal
rather than by mass (or some other factor). Finally, we compare our
results with other related analyses, and discuss the constraints that
can be derived on the level of hydrostatic bias in massive clusters.

\subsection{The impact of Ngal}
\label{sec:impact-ngal}

Our investigation of the comparison between hydrostatic and caustic
masses suggested that the caustic masses were underestimated when
fewer galaxies were used. While a general result of dynamical masses
being biased low when based on smaller numbers of galaxies was shown
in \citet{Wojtak:2018a}, the caustic method examined in that work
differed from the implementation used here.

The dependence on Ngal of the caustic mass was investigated for our
implementation of the caustic method by \citet{Serra:2011a}. They
found that the true escape velocity profile will be underestimated by
$\sim$5-10\% at $R_{500}$ (Figure 20, \citealp{Serra:2011a}) for
clusters with Ngal $\sim$ 100 (compared to clusters with a higher Ngal
for which the true escape velocity is not underestimated). This would
lead to the caustic mass being underestimated by $\sim$10-20\% at
$R_{500}$, were it not for the assumption of a constant
$\mathcal{F}_\beta$ (see \S\ref{subsection.causticmasses}), which
leads to a $\sim$10-15\% overestimate of the true mass at $R_{500}$
for clusters with a higher Ngal (Figure 12, \citealp{Serra:2011a}).
Thus, the simulations suggest that these two effects approximately
cancel out at our comparison radius of $R_{500}$, such that a large
underestimation of the cluster mass using the caustic method for low
Ngal clusters (like that implied by our results) is not expected.

As discussed in \S \ref{subsection.ngal}, we consider the caustic
masses for the high Ngal subset to be unaffected by this apparent
bias, and base our main conclusions on this subset. As a test of the
robustness of our results, we repeated the measurement of $\MXMC$
using the HeCS-Omnibus caustic data of \citet{Sohn:2020a} in place of
the original HeCS caustic data of \citet{Rines:2013a} that was used
for our main analysis. We selected the 14 clusters in our sample with
the highest Ngal in HeCS-Omnibus ($\text{Ngal}>274$ with a median of
$413$). For these best-sampled clusters, using the HeCS-Omnibus
caustic masses gives an average hydrostatic to caustic mass ratio of
$\MXMC=1.10$ at the hydrostatic $R_{500}$. This agrees very well with
the $\MXMC=1.12$ we found for the 14 clusters in our high Ngal subset
when using the \citet{Rines:2013a} data (for which $\text{Ngal}>210$
with a median of 234). This supports our conclusion that our high Ngal
subset is unaffected by systematics due to low numbers of galaxies
(i.e. $\text{Ngal}\gta200$ is sufficient to avoid biases in the
caustic mass at $R_{500}$). Indeed, when we apply the caustic
technique to the cluster catalogues in the high-Ngal sample where we
undersampled Ngal such that $\Ngal\lesssim 200$, the caustic mass
within $R_{500}$ is underestimated by $\sim 15\%$, on average,
suggesting that the increase of the ratio $\MXMC$ from $\sim 1.12$ of
the high-Ngal sample to $\sim 1.28$ of the mid-Ngal sample is mostly
driven by the lower number of the caustic members.

Poorer sampling of the caustics is likely to be responsible (at least
in part) for some of the outliers seen in the comparisons between the
hydrostatic and caustic masses (e.g in Figure
\ref{figure.allratios44RCC}). When restricted to the high Ngal subset,
there are no significant outliers in the comparison between
hydrostatic and caustic masses, given the $\sim30\%$ intrinsic scatter
between the mass estimators (e.g. Figure
\ref{figure.massscatterplot14Ngal}).

\subsection{Cluster centres and apertures}
\label{sec:cluster-centring}
In the caustic analysis, the cluster centre is located as part of the
process \citep{Rines:2013a}, while the centre is fixed at the X-ray
centroid for the hydrostatic analysis. This means that the profiles
are not generally centred at the same location. The separation between
the centres used for the X-ray and caustic analyses were generally
small, and showed no trend with Ngal or with $\MXMC$ (the median
separation was $\approx130\kpc$ in each of the Ngal subsets). We thus
conclude that the increasing disagreement between hydrostatic and
caustic masses at low Ngal is not being driven by miscentring.

\citet{Serra:2011a} investigated the impact of miscentering of the
caustic analyses on the mass determination in their simulations. At
$R_{500}$, the caustic mass estimate was increased by $\approx10\%$
when the profile was forced to use the true centre. This suggests that
our measurements of $\MXMC$ could be overestimated by $\approx 0.1$
due to miscentring of the profiles.

When the separation between the centres used for the caustic and
hydrostatic analyses is small, then the use of a common aperture size
(i.e. the hydrostatic $\rf$) allows the mass estimators to be compared
without adding scatter from different aperture definitions. However,
given that the centres are not identical, it is also helpful to
compare the masses when they are derived fully independently (i.e.
comparing the caustic mass within the caustic $\rf$ with the
hydrostatic mass within the hydrostatic $\rf$).

We repeated our analysis using these independent definitions of $\rf$.
For the full sample of 44 clusters, we found that mass ratio increased
from $\MXMC = 1.33^{+0.10}_{-0.09}$ (using the hydrostatic $\rf$) to
$\MXMC = 1.52^{+0.16}_{-0.14}$ (using independent $\rf$). In the case
of the high Ngal sub-sample, the original
$\MXMC = 1.12^{+0.11}_{-0.10}$ (hydrostatic $\rf$) increased slightly
to $\MXMC = 1.16^{+0.16}_{-0.13}$ (independent $\rf$).

These results make sense. When the masses agree well (as in the high
Ngal sub-sample) then the $\rf$ definitions also agree so the results
do not change. If the masses are more discrepant (as in the full
sample) then using the independent $\rf$ will tend to amplify the
difference.

As might be expected, the scatter between the mass estimators is
somewhat larger when the independent $\rf$ are used. The scatter
increases from $(39\pm6)\%$ to $(55^{+9}_{-8})\%$ for the full sample,
and from $(25^{+12}_{-08})\%$ to $(35^{+15}_{-11})\%$ for the high
Ngal sub-sample.

Based on this analysis, we conclude that the use of the hydrostatic
$\rf$ for both mass determinations does not have a strong effect on
our results.

\subsection{Impact of selection effects}
\label{sec:impact-select-effect}
In our work we do not account for the impact of selection effects in
our model. The sample is selected on the basis of X-ray flux, so
covariance between X-ray flux and $M_X$ could have an impact on the
inferred $\MXMC$ ratio. Specifically, if there were a strong positive
covariance between X-ray luminosity, $L_X$, and hydrostatic mass,
$M_X$, given the true mass, then X-ray selection would select clusters
that were both more luminous than average and had higher hydrostatic
masses than average, given their true mass. The size of this effect
would depend both on the scatter in the relevant relations and the
degree of covariance between $M_X$ and $L_X$.

We assessed the possible impact of this selection effect by simulating
cluster populations with different degrees of covariance between $L_X$
and $M_X$, imposing the sample selection function, and then applying
our model (described in \S \ref{subsection.modellingbias}) to the
simulated population to determine the impact of the sample selection
on the recovered $\MXMC$ ratio.

In more detail, we sampled a large number of points from a mass
function where each point consists of a true mass, $M_{\text{true}}$,
and a redshift. We use the mass - luminosity relation from
\citet{Mantz:2010a} to assign the true luminosity, $L_{\text{true}}$
to each cluster. We then assume a scatter between $L_{\text{true}}$
and $L_X$ of 30\%, a scatter between $M_{\text{true}}$ and $M_X$ of
10\% and a scatter between $M_{\text{true}}$ and $M_C$ of 30\%. We
also require a correlation coefficient, $\rho_{LM}$, between $L_X$ and
$M_X$ to create the covariance matrix for the multi-dimensional
Gaussian from which we sample $L_X$ and $M_X$ from $L_{\text{true}}$
and $M_{\text{true}}$. It is the strength of this correlation
coefficient that dictates how much a selection on X-ray flux might
bias an inference based on hydrostatic masses, and we investigate the
impact of different assumed values below. Finally, we apply our X-ray
selection to the fluxes derived from the $L_{\text{true}}$ and
redshift values, and use our model to measure the $\MXMC$ ratio for
the resulting simulated sample. We do not set any hydrostatic
or caustic bias in the simulation, so in the absence of selection
biases, we would measure $\MXMC=1$.

To our knowledge, covariances between $L_X$ and hydrostatic mass have
not been measured in the literature, but the covariances between other
ICM properties have been investigated in a few studies
\citep[e.g.][]{mau14,Mantz:2016a,Andreon:2017b,Farahi:2018a}. For
example, for the covariance between $L_X$ and $m_{\text{gas}}$,
\citet{Mantz:2016a} found a correlation coefficient of 0.43, while
\citet{Farahi:2018a} found 0.76. Meanwhile, for the covariance between
$L_X$ and $T$, \citet{Mantz:2016a} found a correlation coefficient of
0.53, while \citet{Farahi:2018a} found 0.49. In both cases, the values
obtained by \citet{Mantz:2016a} are the more relevant here since they
make use of a soft-band, core-included X-ray luminosity, which matches
the selection method used for our sample \citep[][made use of
bolometric luminosities with the core regions removed, but we include
their results here for comparison]{Farahi:2018a}. We therefore
conclude that a correlation coefficient of $\rho_{LM}=0.5$ is a
reasonable upper limit on the strength of the covariance between $L_X$
and $M_X$ and use this as a reference value for our simulations. We
also performed the simulations with with $\rho_{LM}=1$ as a maximally
pessimistic case.

We find that for $\rho_{LM}=0.5$, the recovered $\MXMC$ value is
1.03$\pm$0.01 and for $\rho_{LM}=1$ we measured $\MXMC=1.07\pm0.01$
(for a sample of 500 simulated clusters).

We also checked for a mass dependence of the selection bias by
splitting the simulated clusters into low (log($M_X$)<14.5), mid
(14.5<log($M_X$)<14.75), and high (log($M_X$)>14.75) mass sub-samples
(each containing 500 simulated clusters). The measured $\MXMC$ ratio
was largest for the low-mass sub-sample, due to a larger proportion of
the clusters being close to the flux limit. For the low-mass
sub-sample, assuming $\rho_{LM}=0.5$ we found $M_X/M_X=1.06\pm0.01$,
increasing to $1.15\pm0.01$ for $\rho_{LM}=1$. For the high mass
sub-sample (which is similar in mass to our high Ngal subset), the
measured $\MXMC$ ratios dropped to 1.02$\pm$0.01 and 1.03$\pm$0.01 for
$\rho_{LM}=0.5$ and $\rho_{LM}=1$ respectively.

We conclude that, while neglecting selection effects tends to bias the
$\MXMC$ ratio high, the impact is not significant for the mass range
of the high Ngal subset, and that selection biases alone cannot
explain the high values of $\MXMC$ that we find in the lower mass,
lower Ngal clusters.

\subsection{Impact of X-ray temperature calibration}
\label{sec:impact-x-ray}

It is established that there are systematic differences in cluster
temperatures measured by \emph{Chandra} and \emph{XMM-Newton},
temperatures measured with \emph{Chandra} being systematically hotter
\citep{Schellenberger:2015a}. This has been found to have a
corresponding systematic impact on the derived hydrostatic masses
\citealp{Mahdavi:2013a,Schellenberger:2015a}). We note, however, that
\citet{Martino:2014a} found no systematic differences in hydrostatic
masses for a sample of 50 clusters with both \emph{XMM-Newton} and
\emph{Chandra} data.

\citet{Schellenberger:2015a} found that the hydrostatic masses
measured from \textit{Chandra} data were $14\pm2\%$ higher than those
from \textit{XMM-Newton}. They found no significant trend with cluster
mass, suggesting that the apparent trend of $\MXMC$ with mass or Ngal
that we find is not driven by this systematic. If our hydrostatic
masses were systematically overestimated by $\approx15\%$, this would
have a direct effect of reducing our derived $\MXMC$ ratios by $0.15$.

\subsection{Summary of systematics}
\label{sec:summary-systematics}

We have identified a number of possible systematics that could impact
our comparison between caustic and hydrostatic masses. In the case of
the caustic masses, poorer sampling of the caustics was found to
result in an underestimate of the caustic masses for
$\text{Ngal}\la200$, producing an apparent trend of $\MXMC$ with mass.
For the better-sampled clusters, this systematic appears to be absent,
but the assumption of constant $\mathcal{F}_\beta$ is expected to lead
to an overestimate of the caustic mass by $10-15\%$ at $R_{500}$ (i.e.
biasing $\MXMC$ low by up to $0.15$). The miscentering of the caustic
profiles leads to an estimated underestimate of the caustic masses of
$\approx10\%$ at $R_{500}$ (biasing $\MXMC$ high by $\approx0.1$).

We find that the use of X-ray flux to select the sample combined with
a possible covariance between X-ray luminosity and hydrostatic mass
could bias the hydrostatic masses to be high. However, the effect is
estimated at a few percent for the more massive clusters comprising
our high Ngal subset, and the trend with mass is not sufficient to
explain the strength of the trend of $\MXMC$ with mass or Ngal.
Finally, systematics on the calibration of \textit{Chandra} could lead
to the hydrostatic masses being overestimated by $\approx15\%$
(biasing $\MXMC$ high by about $0.15$).

\subsection{Comparison with other work}
\label{sec:comp-with-other}

Direct comparisons between hydrostatic and caustic masses for samples
of clusters are scarce in the literature. \citet{Andreon:2017a}
compared the mean $f_{gas}$ for a sample of 34 clusters with caustic
masses, with the mean value from separate samples with hydrostatic
masses from \textit{Chandra} data. They inferred that caustic masses
were slightly higher than, but consistent with, hydrostatic masses at
$R_{500}$. This is consistent with the result we find for the high
Ngal subset. However, the clusters with caustic masses used in
\citet{Andreon:2017a} cover a range in mass and Ngal that is
comparable with our low and mid Ngal subsets, and are in tension (at
the $\approx2\sigma$ level) with those subsets for which we find the
caustic masses to be lower than hydrostatic masses (although
we consider our results for those subsets to be less reliable due to
the poorer sampling of the caustics).

\citet{Foex:2017a} compared hydrostatic and caustic masses for a
sample of 13 clusters with masses, redshifts and Ngal similar to our
high Ngal subset. They found a mean $M_C/M_X=1.32\pm0.18$ at $R_{500}$
(we note that this is the inverse of our mass ratio), and hence
caustic masses were higher than hydrostatic masses by $\sim30\%$
(although the ratio dropped to unity when some substructures were
removed in the dynamical analysis). While this contrasts with our
results, the difference is not strongly significant
($\approx2\sigma$). Furthermore, the \citet{Foex:2017a} analysis was
based on hydrostatic masses from \textit{XMM-Newton}; adjusting these
masses to match the \textit{Chandra} calibration would reduce the
tension with our result to less than $1\sigma$.

\citet{Ettori:2019a} compared hydrostatic masses (from
\textit{XMM-Newton}) and caustic masses at $R_{200}$ for 6 clusters,
finding that the caustic masses were lower than the hydrostatic masses
by about $30\%$ on average. This is in the same sense as the
difference we find for our sample, but the clusters include some with
lower Ngal than our high Ngal subset, so a direct comparison is not
possible.

\subsection{Implications for the hydrostatic bias}
\label{sec:impl-hydr-bias}

Conventionally, the hydrostatic bias is defined as $b=1-M_X/M$, where
$M_X$ is the hydrostatic mass and $M$ is the true mass. Our best
estimate of the ratio of hydrostatic to caustic masses, with
statistical error is $\MXMC=1.12^{+0.11}_{-0.10}$, for the high Ngal
subset. Under the strong assumption that the caustic masses are
unbiased, we can write $b=1-\MXMC$, in which case our results exclude
a hydrostatic bias of more than $20\%$ at the $3\sigma$ level based on
statistical uncertainty alone. However, taking the possible systematic
uncertainties on both the caustic and hydrostatic masses into account,
we cannot rule out larger values of the hydrostatic bias.

Constraints on the hydrostatic bias from direct comparisons between
hydrostatic and WL masses currently encompass a range from
approximately zero
\citep[e.g.][]{Gruen:2014a,Israel:2014a,Applegate:2016a,Smith:2016a}
up to around $30\%$
\citep[e.g][]{von-der-Linden:2014a,Donahue:2014a,Hoekstra:2015a}.
Meanwhile, constraints on the level of bias have also been estimated
indirectly by comparisons of WL masses with masses derived from
scaling relations of other observables that were calibrated to
hydrostatic masses (e.g. from X-ray luminosity \citet{ser17} or the
Sunyaev-Zel'dovich effect signals \citet{med18,miy19}). These also
tend to favour biases of $20-30\%$.

The question of hydrostatic bias has also been addressed with
simulations, which typically find biases of $10-30\%$, but with radial
and mass dependences, and sensitivity to the method used to
reconstruct the hydrostatic masses from the simulation data
\citep[e.g.][]{Nagai:2007a,Rasia:2012a,pea20,bar21}.

Simulations suggest a significant mass dependence of the hydrostatic
bias, with the bias increasing for higher mass clusters
\citep[e.g][]{bar21}. However, due to the relatively poorer sampling
of the caustics in our lower mass clusters, we are not able to test
these results observationally.

Overall, there is not a clear consensus on the magnitude of the
hydrostatic bias. Our results are more consistent with those estimates
from the low end of the range and add a useful new datapoint to this
still-open question, since the caustic mass determination is subject
to different systematics and assumptions to those affecting the WL
masses which are primarily used to address this question
observationally.

\section{Summary and conclusions}
\label{section.summary}

We compared masses of 44 galaxy clusters determined independently from
X-ray data under the assumption of hydrostatic equilibrium, and galaxy
velocities using the caustic method. Our initial results showed
significant discrepancies between the masses which appeared to
increase for lower mass clusters. We found that this was driven by the
poorer sampling of the caustics for the lower mass clusters leading to
an underestimate of the mass using the caustic method at $R_{500}$.
The caustic method is not optimised for mass determinations at these
radii, but this finding suggests that simulations have underestimated
the impact of poorer sampling at smaller radii.

Restricting our analysis to the 14 clusters with best-sampled caustics
(the high Ngal subset), we find an average ratio of the hydrostatic to
caustic masses of $\MXMC=1.12^{+0.11}_{-0.10}$. We find no evidence
for radial dependence of this ratio or dependence on dynamical state
of the cluster, but these conclusions are limited by the reduced
sample size.

We investigated possible systematics that could affect the mass
estimates for this high Ngal subset and find that the most significant
are as follows. The assumption of constant $\mathcal{F}_\beta$ and
possible miscentering of the caustic profiles could bias the caustic
masses high and low respectively by $10-15\%$. The uncertainties on
the absolute calibration of the X-ray temperature measurements could
bias the hydrostatic masses high by about $15\%$ compared to those
measured with \textit{XMM-Newton}.

We interpret our results as favouring a hydrostatic bias that is close
to zero. However, the systematics we identified would allow for
significantly larger levels of hydrostatic bias.

\begin{acknowledgements}
  CL, BJM and RTD acknowledge support from the UK Science and
  Technology Facilities Council (STFC) grant ST/R00700/1. BJM
  acknowledges further support from STFC grant ST/V000454/1. CL also
  acknowledges support from an ESA Research Fellowship. AD
  acknowledges partial support from the Italian Ministry of Education,
  University and Research (MIUR) under the Departments of Excellence
  grant L.232/2016, and from the INFN grant InDark. This research has
  made use of NASA’s Astrophysics Data System Bibliographic Services.
\end{acknowledgements}

\bibliographystyle{./aa}
\bibliography{./Snapshots}

\begin{appendix} %

  \section{Notes on individual clusters} \label{subsection.individualnotes} In this section we detail where we differed from the analysis methods presented in \S\ref{section.analysis}, or any particular points of interest for each cluster in our sample.

  When extended sources other than the cluster were detected in the images, we attempted to determine if they were physically associated with the cluster or were unrelated projected sources. We did this by querying the NASA Extragalactic Database\footnote{\url{https://ned.ipac.caltech.edu/}} to see if the source has a known counterpart.

  \begin{itemize}
  \item A773 - For ObsID 5006, due to flaring we only use the first 15 ks of the observation.
  \item MS0906 - At 9:08:58.02, 11:01:58.36 (RA, DEC = 137.24, 11.03),
    there is a large region of diffuse extended emission (which
    corresponds to cluster A750 \citep{Abell:1989a}) which we mask with
    a 230$''$ radius circle.
  \item ZW2701 - At 9:53:05.85, 51:49:16.39 (148.27, 51.82) there is a
    region of extended emission that we mask with a 150$''$ radius
    circle.
  \item A963 - For ObsID 903, a temperature of 0.28 keV was used when
    fitting the APEC model to the soft background residuals. We note
    that the standard temperature used for this is 0.18 keV, as it is
    found that this temperature generally models these residuals well
    \citep{Giles:2015a}. The higher temperature implies that this
    observation includes a hotter region of the galactic foreground
    emission.
  \item ZW3146 - For ObsID 9371, due to flaring we only use the first 34
    ks of the observation.
  \item A1423 - For ObsID 538, due to flaring we only use the first 11
    ks of the observation.
  \item A1553 - For ObsID 12254, due to flaring we only use the first 10
    ks of the observation.
  \item A1682 - At 13:06:59.904, 46:31:40.65 (196.75, 46.53) there is a
    cluster galaxy \citep{Morrison:2003a} and at 13:07:13.47,
    46:29:02.31 (196.81, 46.48) there is another small region of
    extended emission. Both sources are masked by radius 40$''$ circles.
  \item A1763 - At 13:34:52.8, 40:57:21.6 (203.72, 40.96) we use a
    90$''$ radius circle to mask a region of extended emission, which is
    likely to be associated with a previously known X-ray source
    \citep{Evans:2010a}.
  \item A1930 - At 14:32:42.72, 31:33:50.4 (218.18, 31.56) there is a
    region of extended emission that we mask with a circle of radius
    160$''$. The extended emission is the cluster RM J143242.6+313407.1
    \citep{Rozo:2015a} at redshift $z = 0.137$, and is likely part of
    the same dark matter halo as A1930 (z = 0.1308).
  \item A2009 - For ObsID 10438, a temperature of 0.27 keV was used when
    fitting the APEC model to the soft background residuals.
  \item A2069 - There is a separate cluster at 15:24:25.846,
    +30:00:16.039 (231.11, 30.00) at redshift $z = 0.119$ (MaxBCG
    J231.10029+30.00604, \citealp{Koester:2007a}); it is likely part of
    the same dark matter halo as A2069. We mask it with a rectangle of
    23$'$ by 14$'$ at an inclination angle of 35$\degree$. There is
    either a group of point sources or a filament at 15:23:38.87,
    +29:58:30.226 (230.91, 29.98) that we mask using an ellipse. In
    addition, for ObsID 4965, due to flaring we only use the first 40 ks
    of the observation.
  \item A2261 - At 17:22:12.78, 32:06:36.95 (260.55, 32.11) there is a
    region of diffuse extended emission that we mask with a 80$''$
    radius circle. This source is associated with a galaxy cluster
    (GMBCG J260.55436+32.11438) at $z = 0.304$ \citep{Hao:2010a}, so is
    not associated with A2261 which is at redshift $z = 0.2242$.
  \item RXJ1720 - For ObsID 4361, due to flaring we only use the first
    15 ks of the observation.

  \end{itemize}

  \section{Temperature and density profiles} \label{app.section.vikh_3D_models}
  In this Appendix we expand on the method used to model
  the temperature and density profiles used to determine the
  hydrostatic masses in \S\ref{section.analysis}.

  \subsection{Temperature Profiles}
  \label{app.subsection.3D_temp_prof} The model used
  for the 3D distribution of the temperature
  \citep{Vikhlinin:2006a} is sufficient to describe a temperature
  decline (if present) in the central core region of the cluster
  in addition to the rest of the cluster temperature profile
  outside the central cooling region of a cluster. The part
  outside the central cooling region is modelled by
  \begin{equation} \label{equation.temp_1}
    t(r)
    = - \frac{(r/r_t)^{-a}}{\left[ 1 + (r/r_t)^b \right]^{c/b}} \quad ,
  \end{equation}
  where $r$ is the radius of the cluster and all other symbols
  are free parameters: $a$ models the slope at small radii, $c$
  models the slope at large radii, and $b$ models the width of
  the transition between these slopes. This transition occurs at
  a radius $r_t$. The decrease in temperature found in the
  central cooling region of many clusters can be modelled as
  \begin{equation} \label{equation.temp_2}
    t_{cool}(r)
    = - \frac{(x + T_{min}/T_0)}{x + 1} \quad ,         \qquad
    x = \left( \frac{r}{r_{cool}} \right)^{a_{cool}},
  \end{equation}
  where $r$ is cluster radius and all other symbols are free
  parameters: $T_0$ is the normalisation, $T_{min}$ is the
  minimum temperature, $r_{cool}$ models radius of the central
  cooling region and $a_{cool}$ models the slope of the cool
  region which extends to a radius $r_{cool}$.

  The 3D temperature profile model is a product of equations
  \ref{equation.temp_1} and \ref{equation.temp_2}
  \begin{equation} \label{equation.temp_final}
    T_{3D}(r)
    = T_0    \,  t_{cool}(r)t(r) \quad .
  \end{equation}

  \subsection{ICM density profiles} \label{app.subsection.3D_gas_prof}

  The model used to describe the 3D distribution of the density \citep{Vikhlinin:2006a} is given by:
  \begin{equation} \label{equation.gasdensity}
    n_{\text{e}} n_{\text{p}}
    = n_{0}^2  \frac{(r/r_c)^{-\alpha}}{ (1 + r^2/r_c^2)^{3\beta - \alpha /2}}
    \frac{1}{(1 + r^{\gamma}/r_s)^{\epsilon / \gamma}}
    + \frac{n_{02}^2}{(1+r^2/r_{c2}^2)^{3 \beta_2}} ,
  \end{equation}
  with $r$ being the cluster radius and the nine free parameters
  able to model both the central and outer parts of the density
  profile sufficiently. We fix $\gamma = 3$ as in
  \citet{Vikhlinin:2006a}, and all the other parameters are
  free. $n_{\text{e}}$ and $n_{\text{p}}$ are the electron and
  proton density respectively, and $r$ is the distance from the
  centre of the cluster. The first term in the equation is a
  modification of the $\beta$-model \citep{Cavaliere:1978a} that
  allows for a power-law type cusp, with slope $\alpha$ at small
  radii, which is expected in the centres of clusters that are
  dynamically relaxed. The second term models the change of the
  slope by $\epsilon$ near the radius $r_s$, and the width of
  this transition region is controlled by $\gamma$. The final
  term is another $\beta$-model component, which increases the
  freedom of the overall model near the cluster centre.

  \subsection{Projection of 3D profiles}
  \label{sec:proj-3d-prof}

  The 3D temperature and density profiles must be
  projected before they can be compared with the observed
  profiles. For this projection we assume spherical symmetry
  \citep[see e.g.][for the geometry of this
  projection]{Ettori:2013a}. We note that whenever we use an
  abundance profile in our analysis, we use the measured
  projected abundance profile (we lack the very high quality
  data needed to deproject abundance profiles), and at large
  radii set the abundance to 0.3 $Z_{\odot}$, in line with
  observed flattening of abundance profiles at large radii
  \citep{Leccardi:2008a}.

  The projected temperature profile was computed using
  the method described in \citet{Vikhlinin:2006a}. Briefly,
  the temperature that would be measured when fitting a single
  temperature thermal plasma model to the projected emission
  is computed by weighting the different temperature
  components of the 3D model along the line of sight by their
  emissivity. The emissivity is derived from the projected
  abundance and 3D density model, and the weighting algorithm
  of \citet{Vikhlinin:2006b} was used.

  The 3D density profile is projected to produce a model
  surface brightness profile. The count rate from each volume
  element along the line of sight is computed assuming an APEC
  plasma emission model with the normalisation set by the
  density in the volume element, and the temperature given by
  the 3D temperature profile model. This conversion from 3D
  density to projected surface brightness uses the appropriate
  instrument response for the detector location.

  In the fitting process, the projection of the
  temperature and density profiles are performed
  simultaneously for a given set of model parameters to
  produce projected temperature and surface brightness
  profiles that are compared with the data. This is an
  improvement over the method used in \citet{Maughan:2016a},
  where a normalised average temperature profile was assumed
  when converting from density to emissivity.

  \section{Mass profile plots}

  Figure \ref{figure.caustichydrostaticmassprofiles} shows the X-ray hydrostatic and caustic mass profiles for the 44 clusters in our sample.

  \begin{figure*}
    \centering
    \subfloat{\includegraphics[angle=0,width=200px]{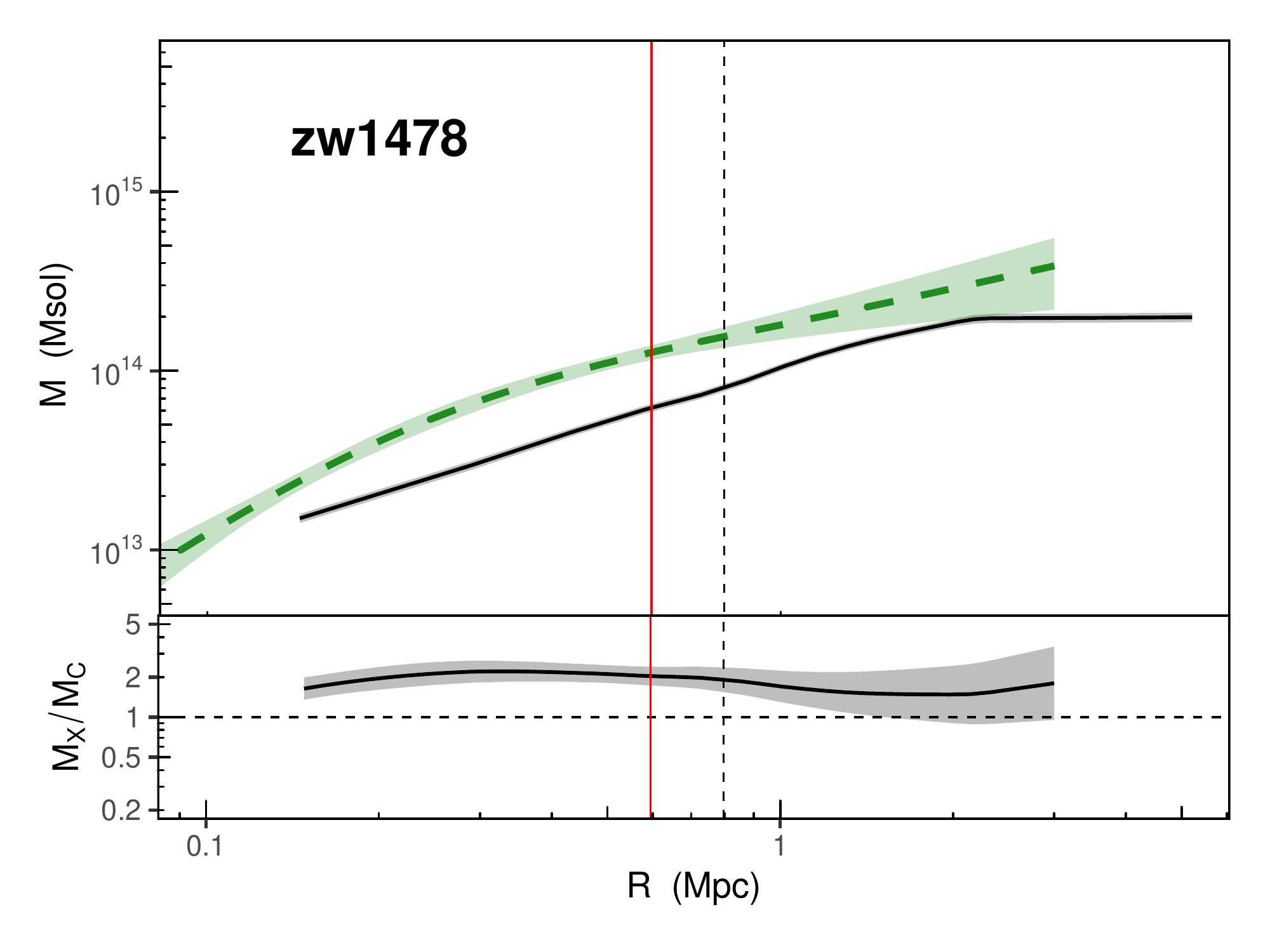}} \qquad
    \subfloat{\includegraphics[angle=0,width=200px]{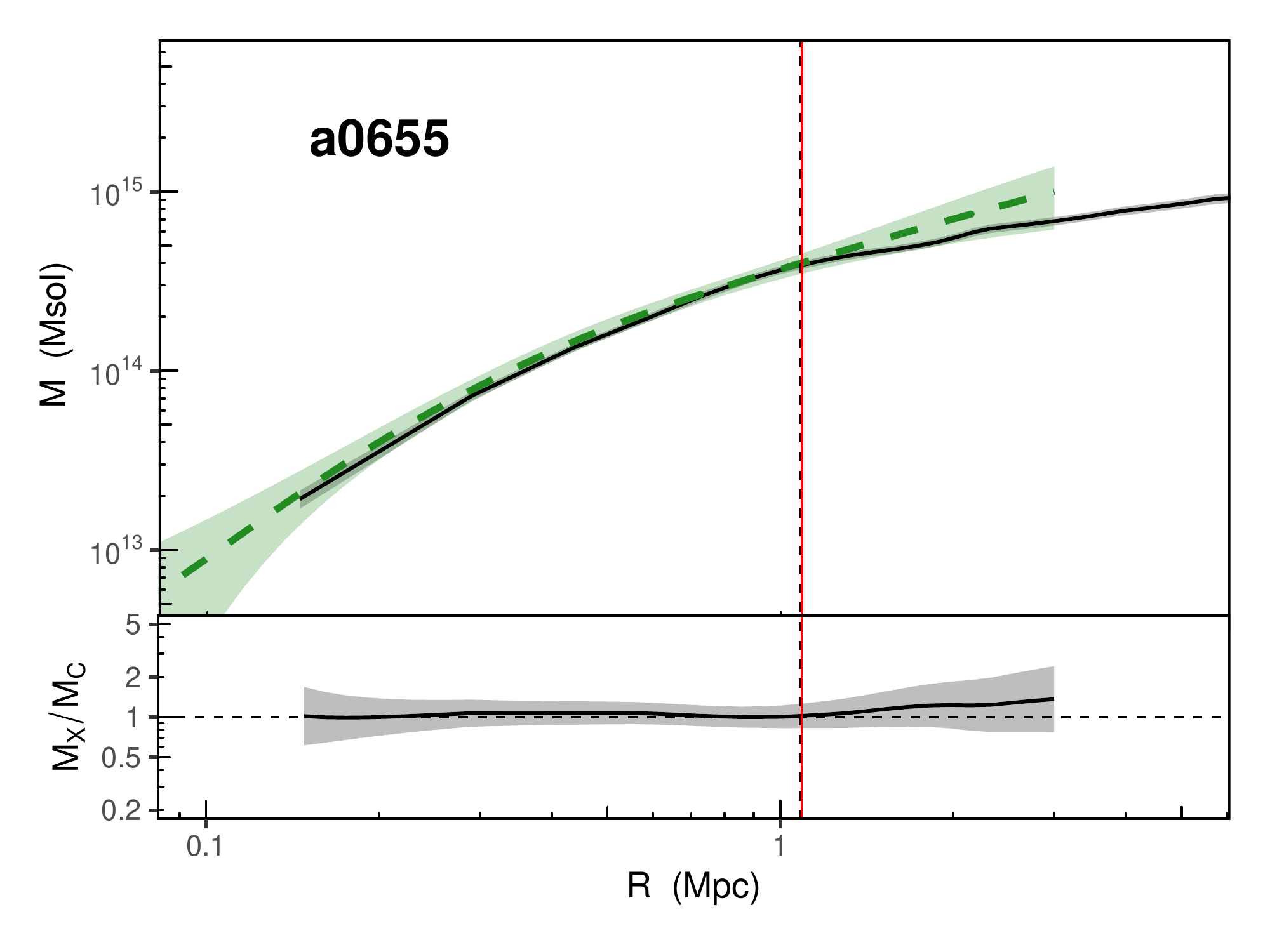}} \\
    \subfloat{\includegraphics[angle=0,width=200px]{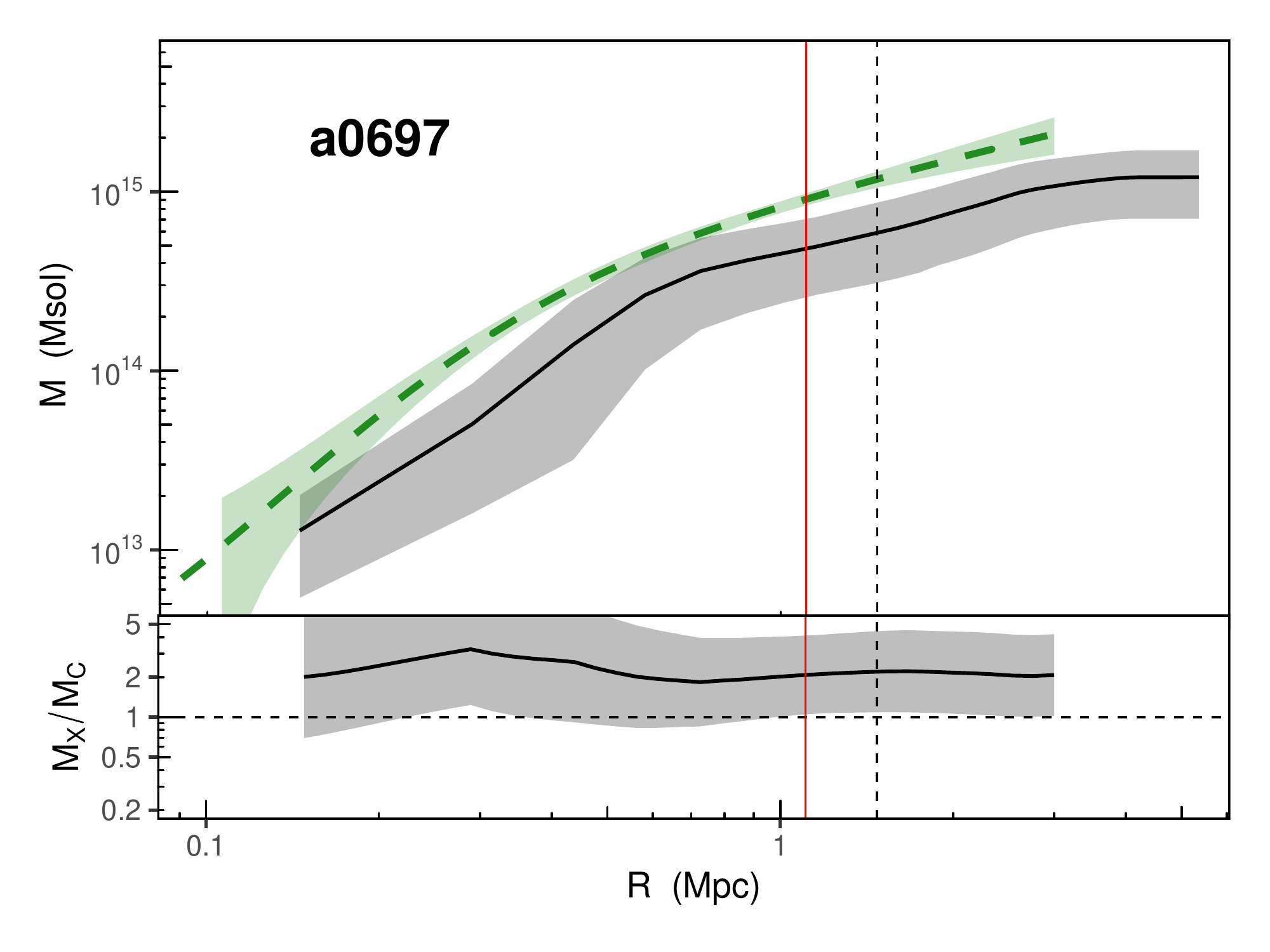}} \qquad
    \subfloat{\includegraphics[angle=0,width=200px]{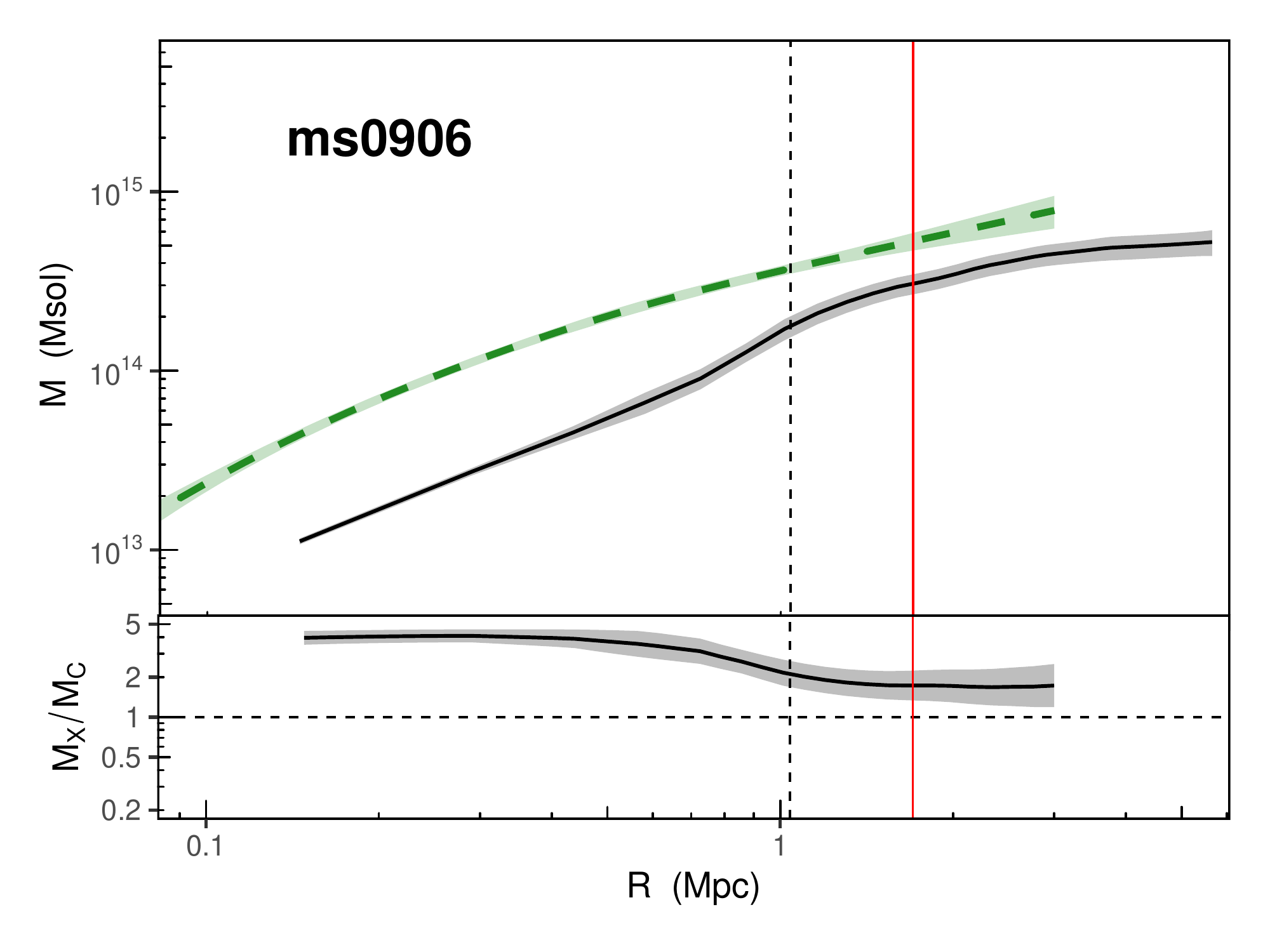}} \\
    \subfloat{\includegraphics[angle=0,width=200px]{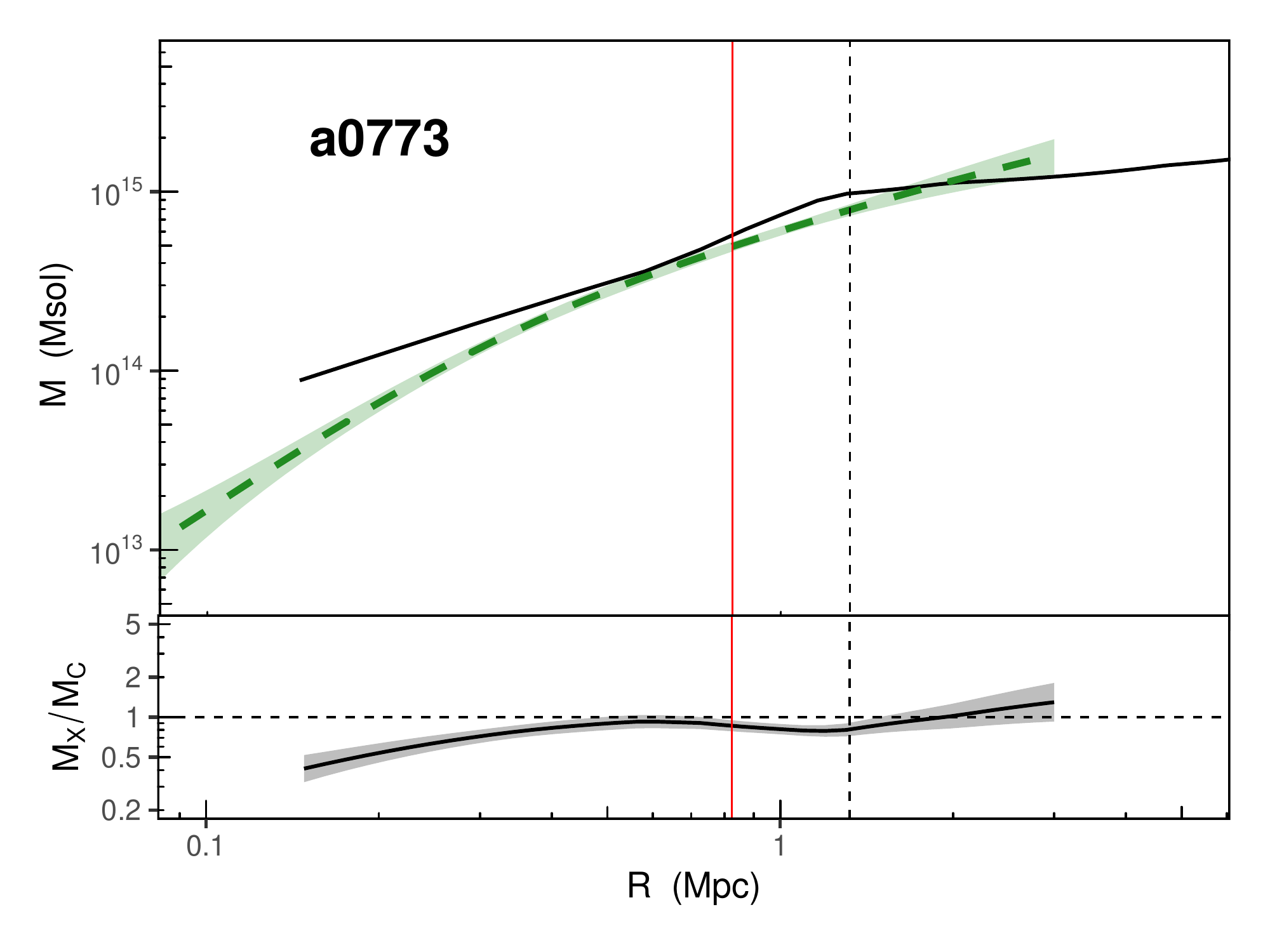}} \qquad
    \subfloat{\includegraphics[angle=0,width=200px]{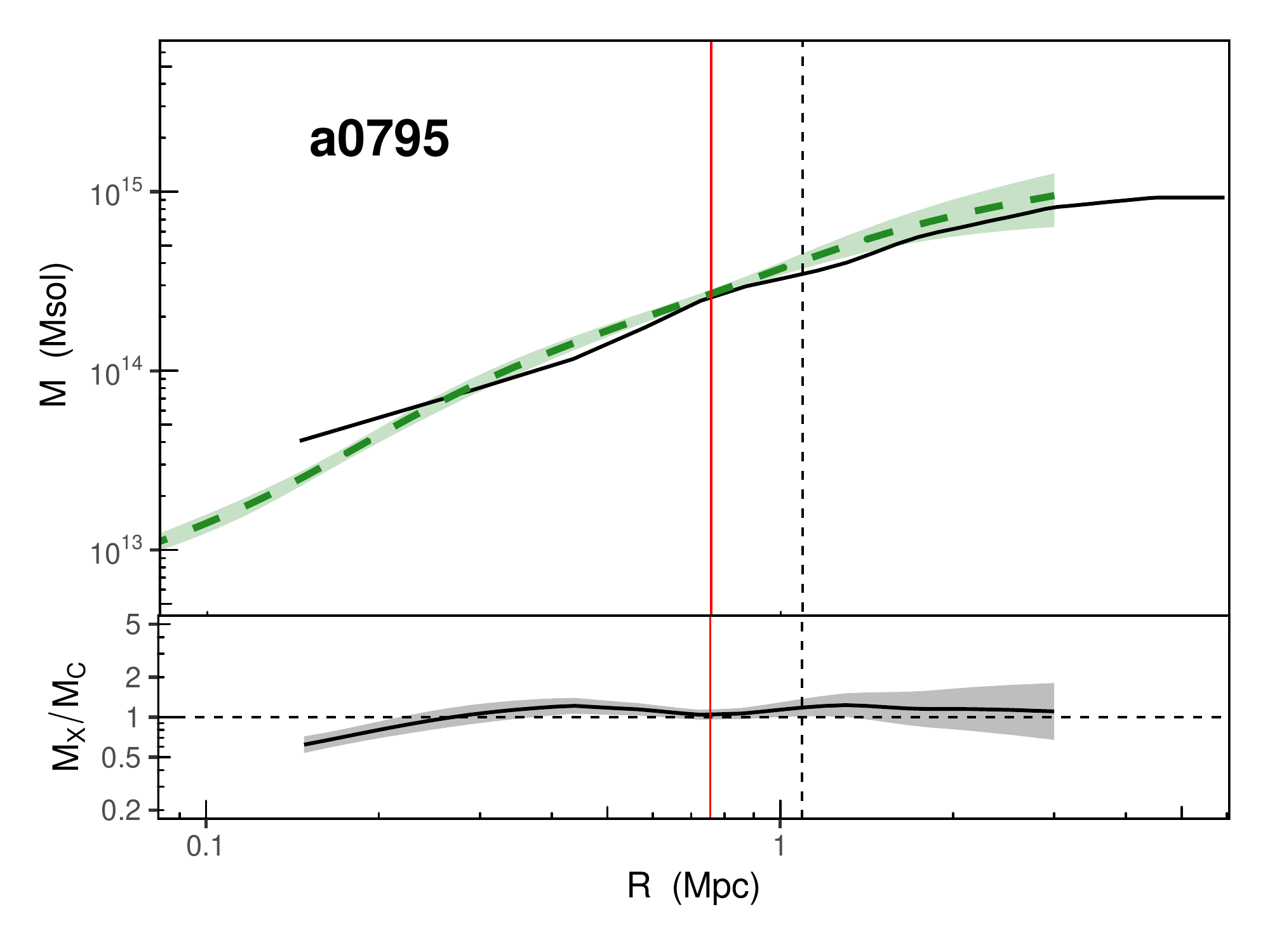}} \\
    \subfloat{\includegraphics[angle=0,width=200px]{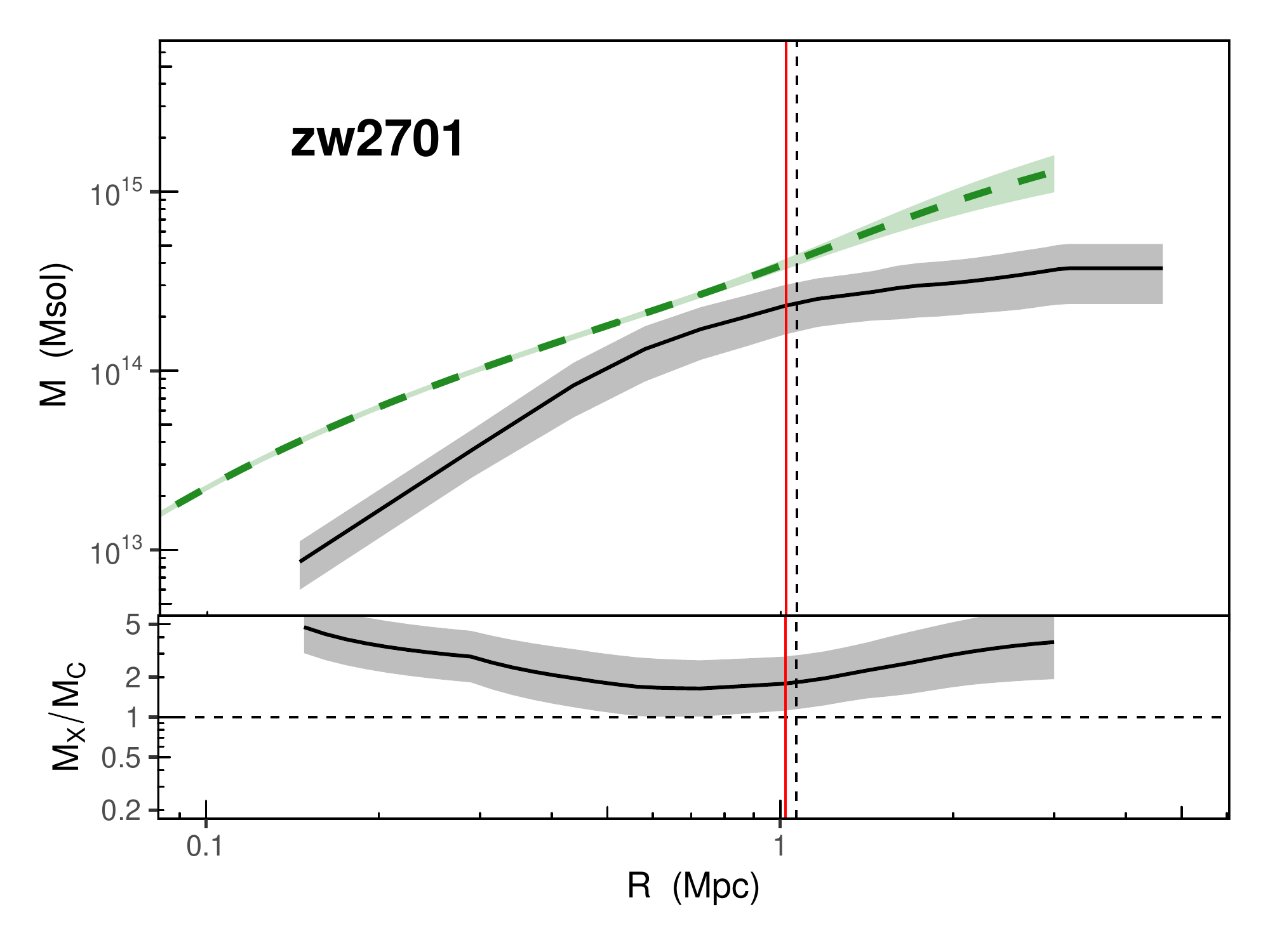}} \qquad
    \subfloat{\includegraphics[angle=0,width=200px]{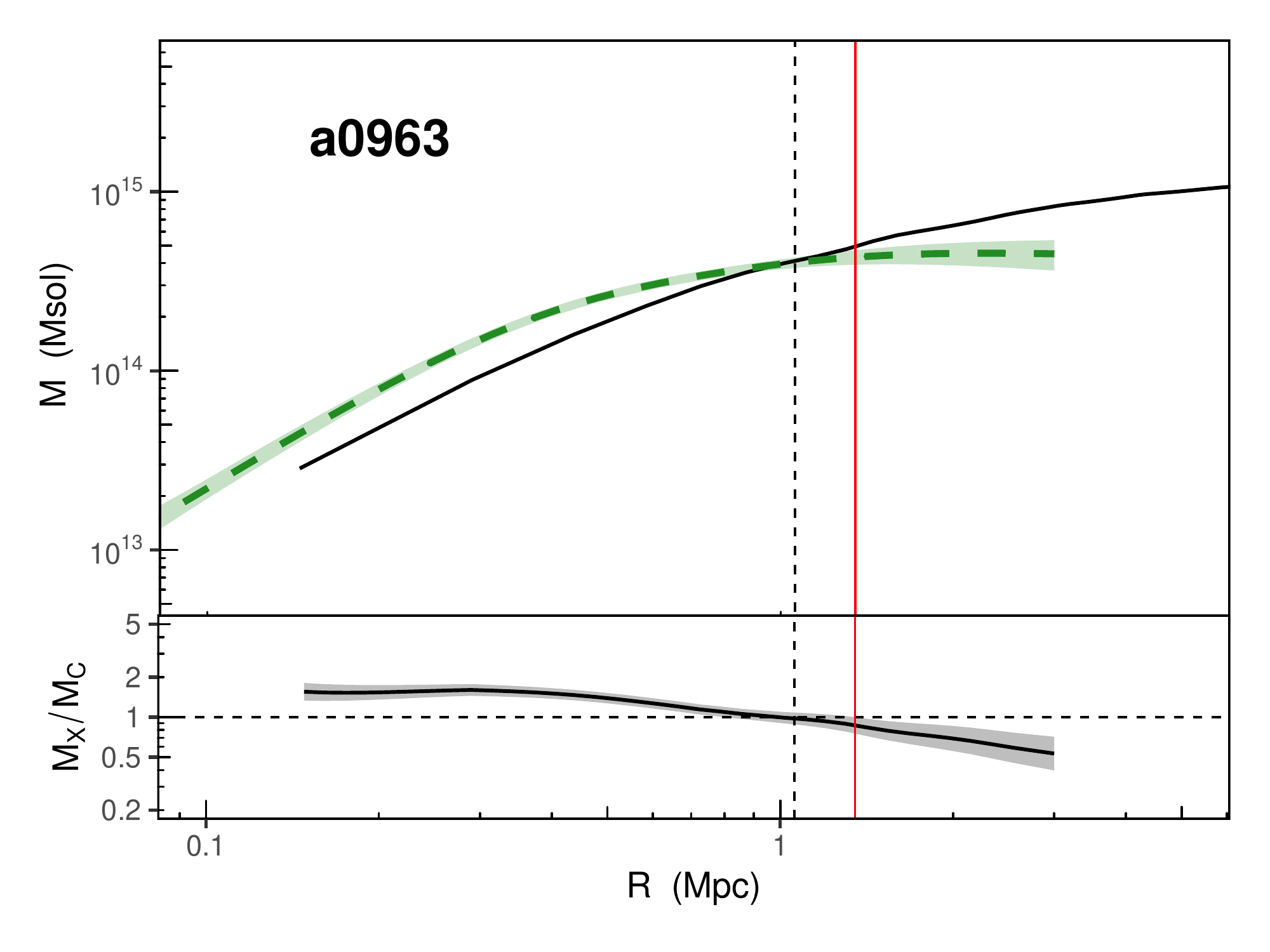}}
    \caption{The caustic and hydrostatic mass profiles for each cluster as black solid and green dashed lines respectively in the upper panels of each plot. The ratio of the hydrostatic to caustic mass, $M_{X}/M_{C}$, are shown in the lower panels of each plot. 1$\sigma$ uncertainties are shown by the shaded regions. The vertical black line is at the value of $R_{500}$ as calculated from the hydrostatic mass profile; the solid red vertical line is at the outer radius of the measured temperature profile (note that hydrostatic masses beyond this radius are based on extrapolation).}
    \label{figure.caustichydrostaticmassprofiles}
  \end{figure*}

  \begin{figure*}
    \ContinuedFloat
    \centering
    \subfloat{\includegraphics[angle=0,width=200px]{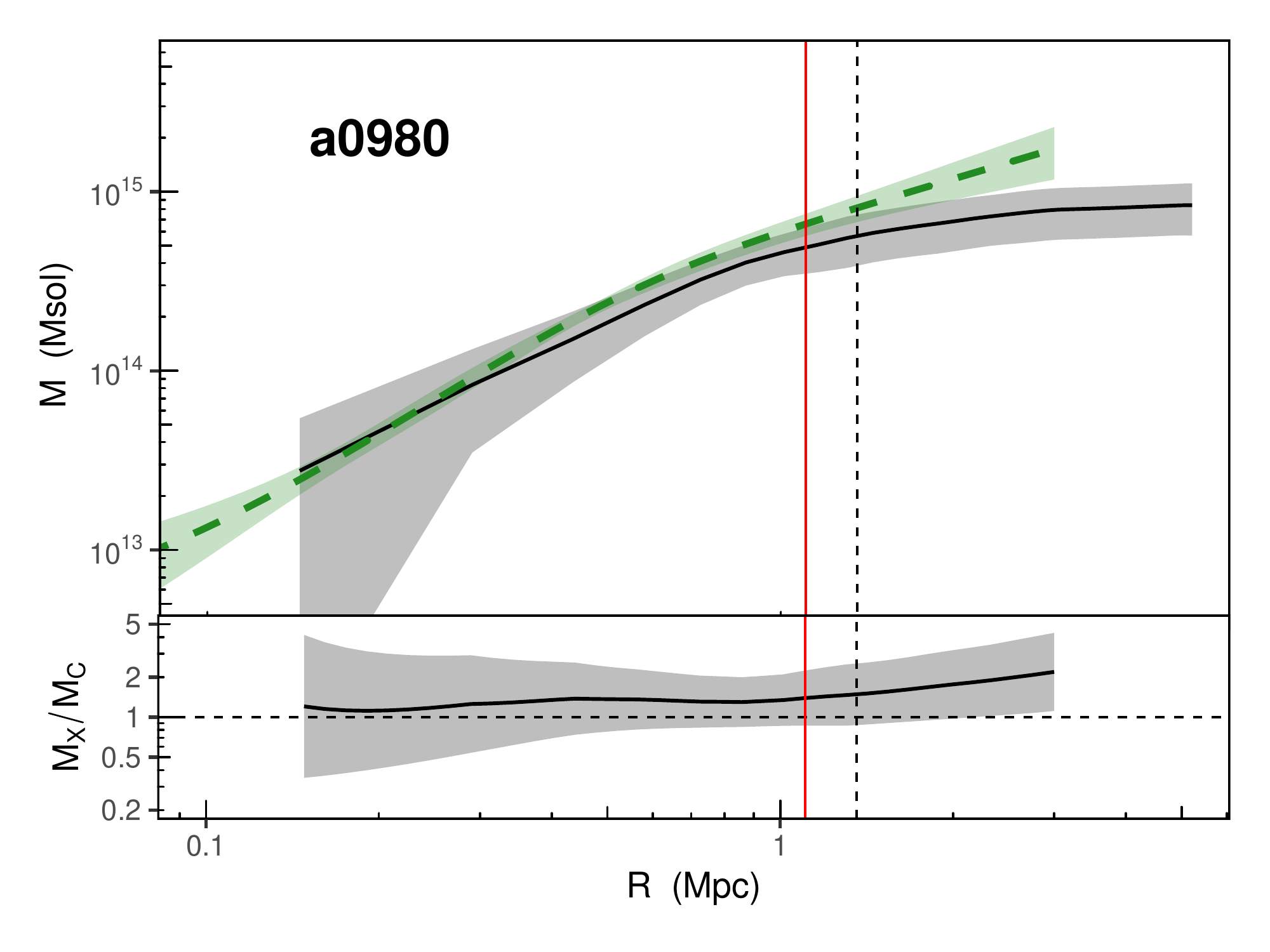}} \qquad
    \subfloat{\includegraphics[angle=0,width=200px]{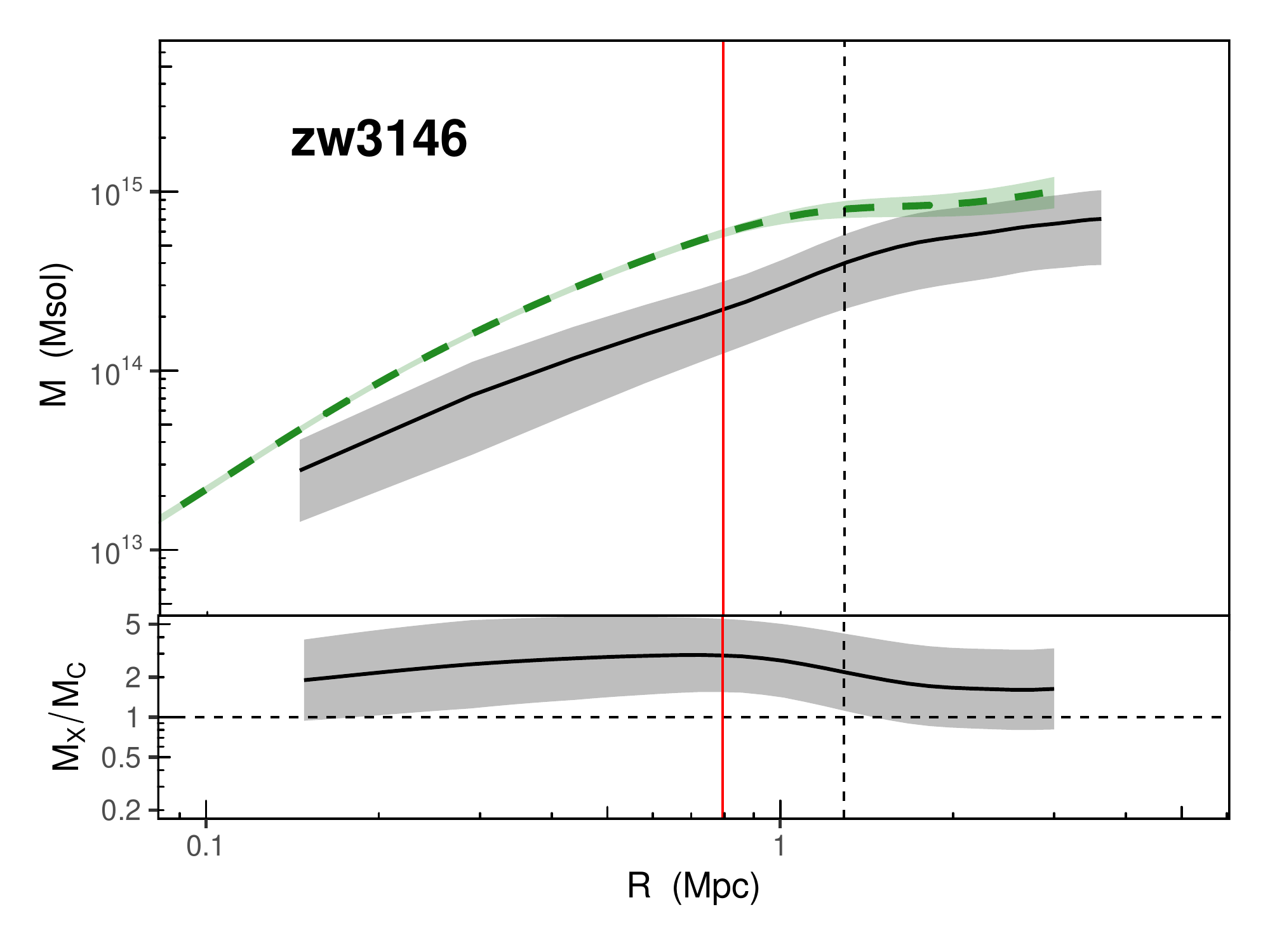}} \\
    \subfloat{\includegraphics[angle=0,width=200px]{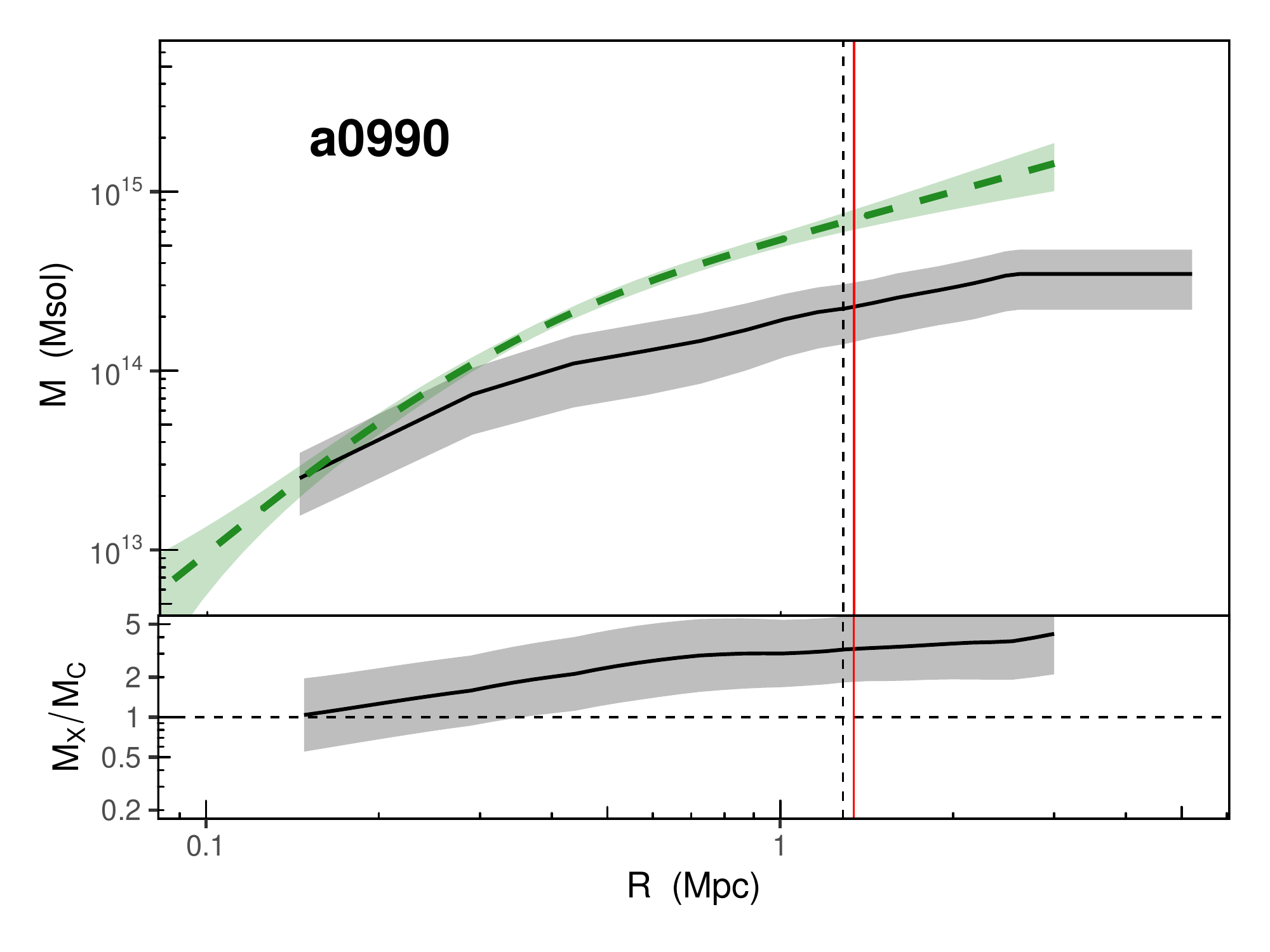}} \qquad
    \subfloat{\includegraphics[angle=0,width=200px]{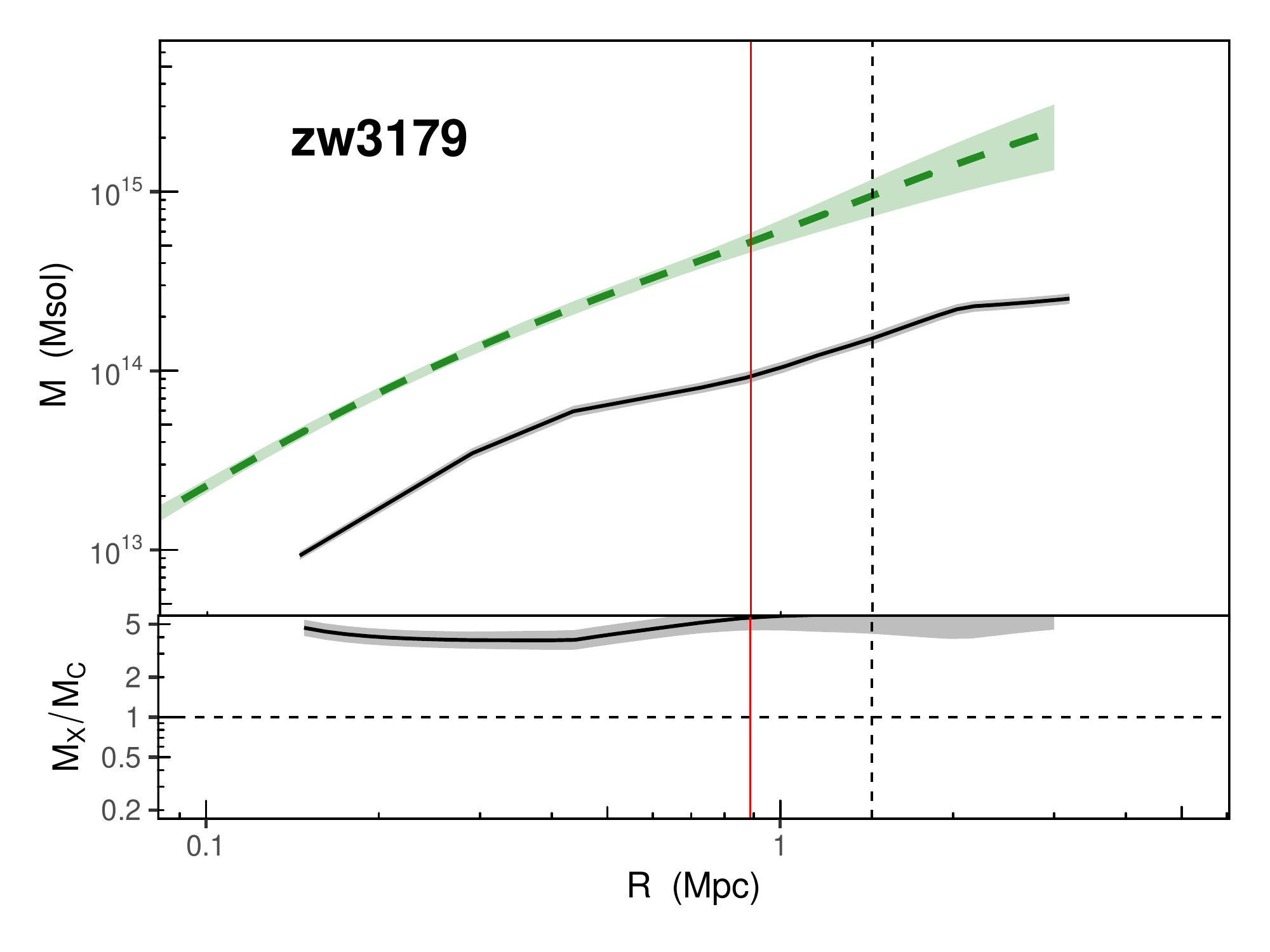}} \\
    \subfloat{\includegraphics[angle=0,width=200px]{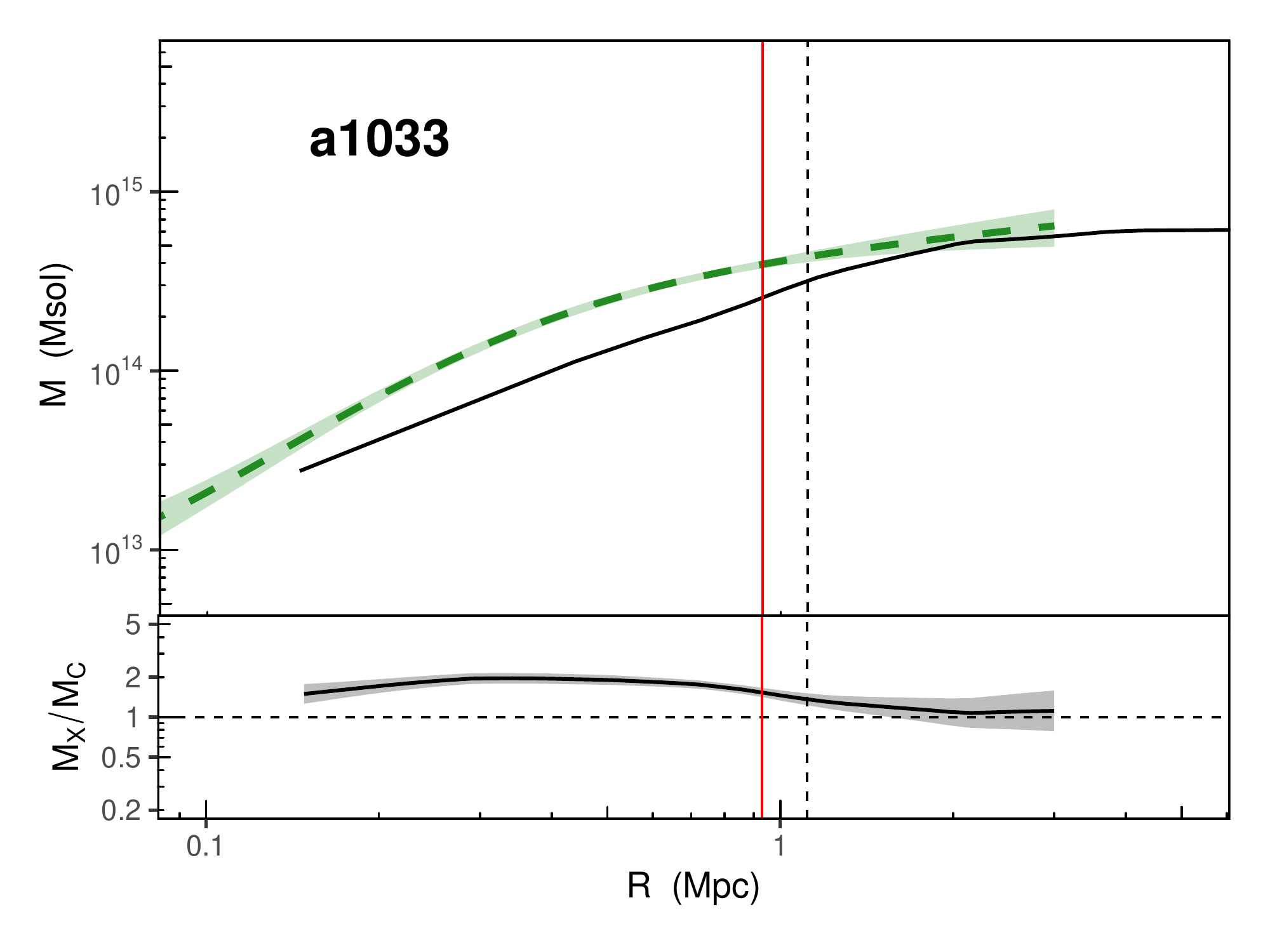}} \qquad
    \subfloat{\includegraphics[angle=0,width=200px]{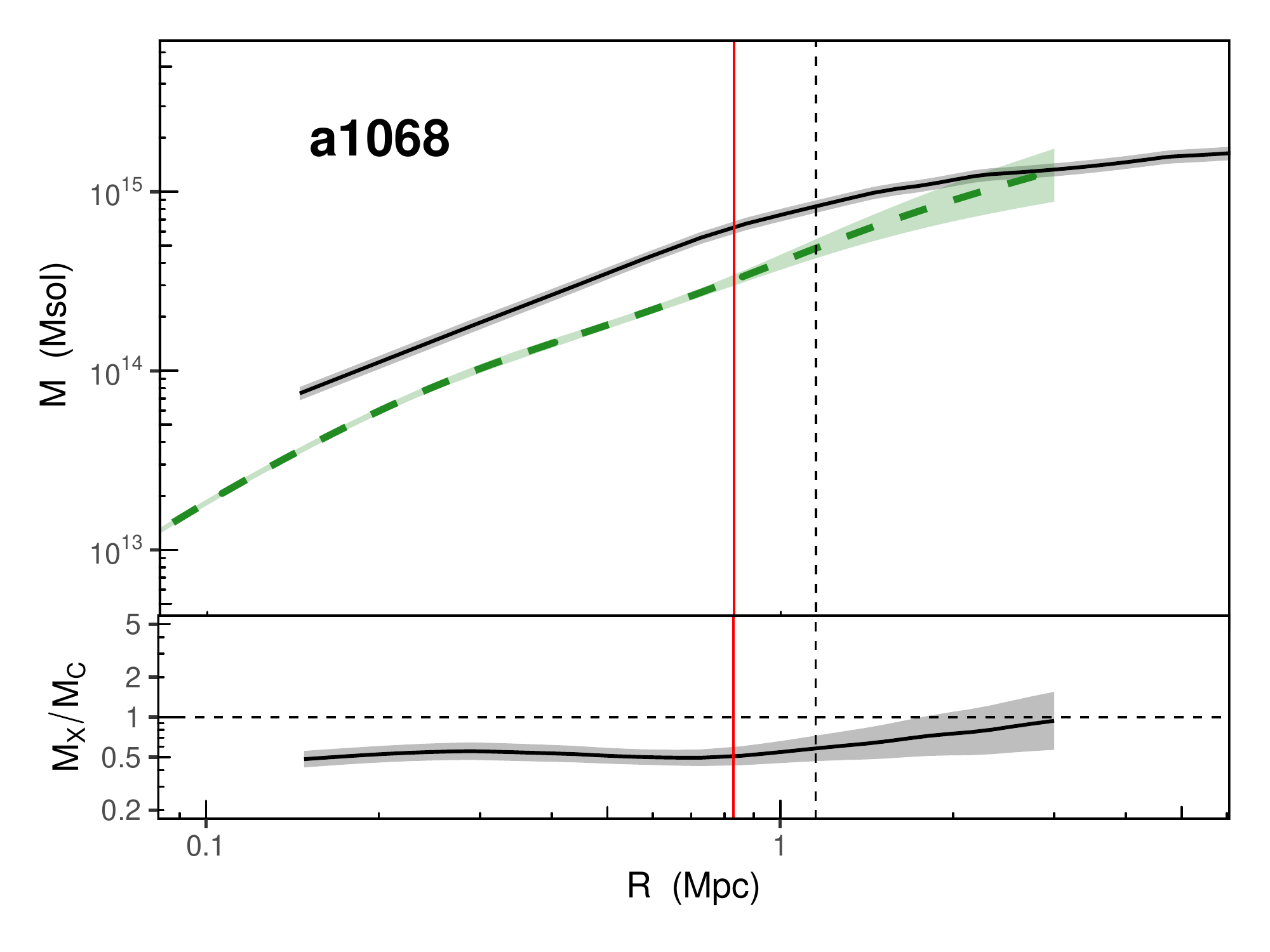}} \\
    \subfloat{\includegraphics[angle=0,width=200px]{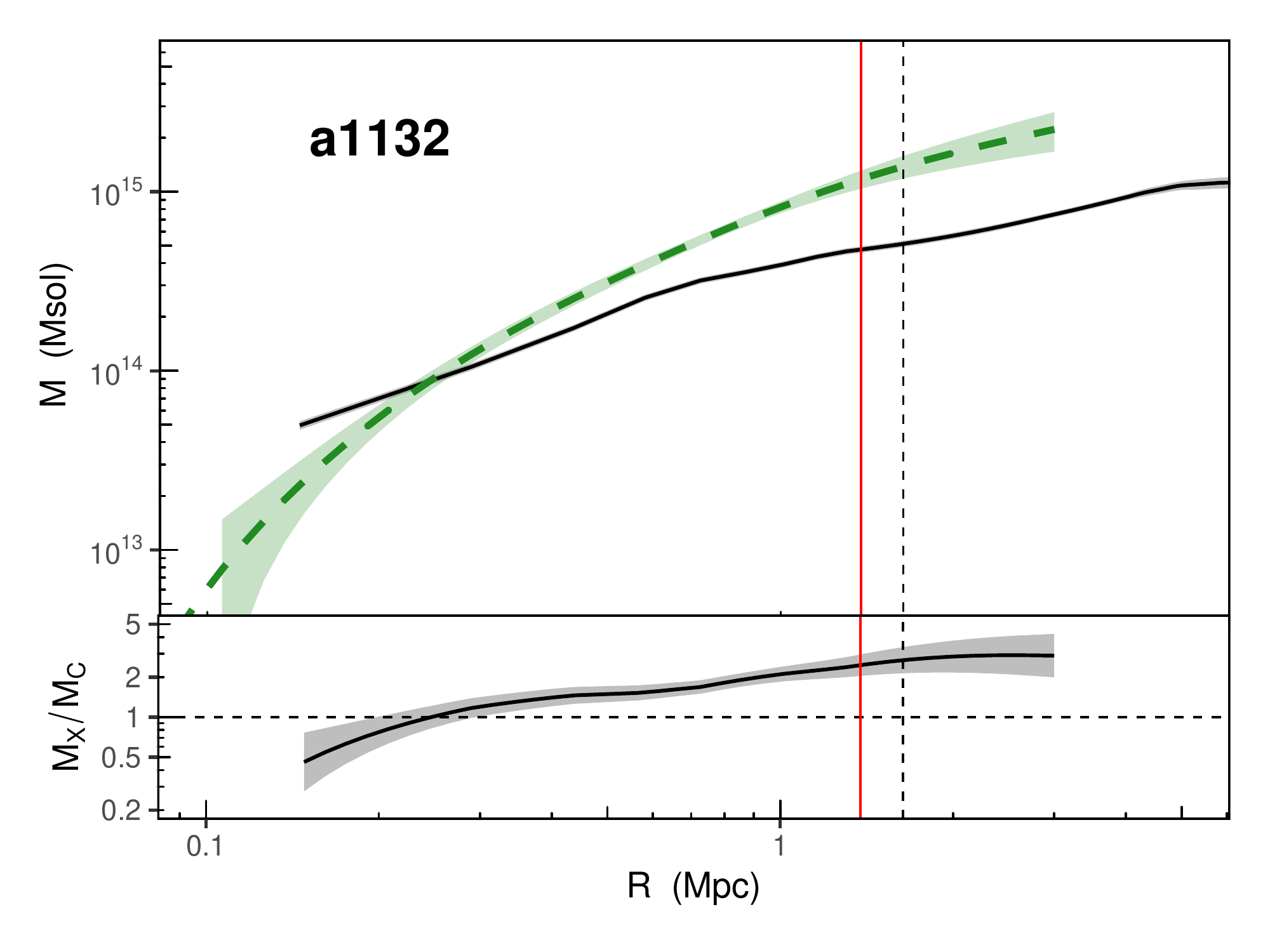}} \qquad
    \subfloat{\includegraphics[angle=0,width=200px]{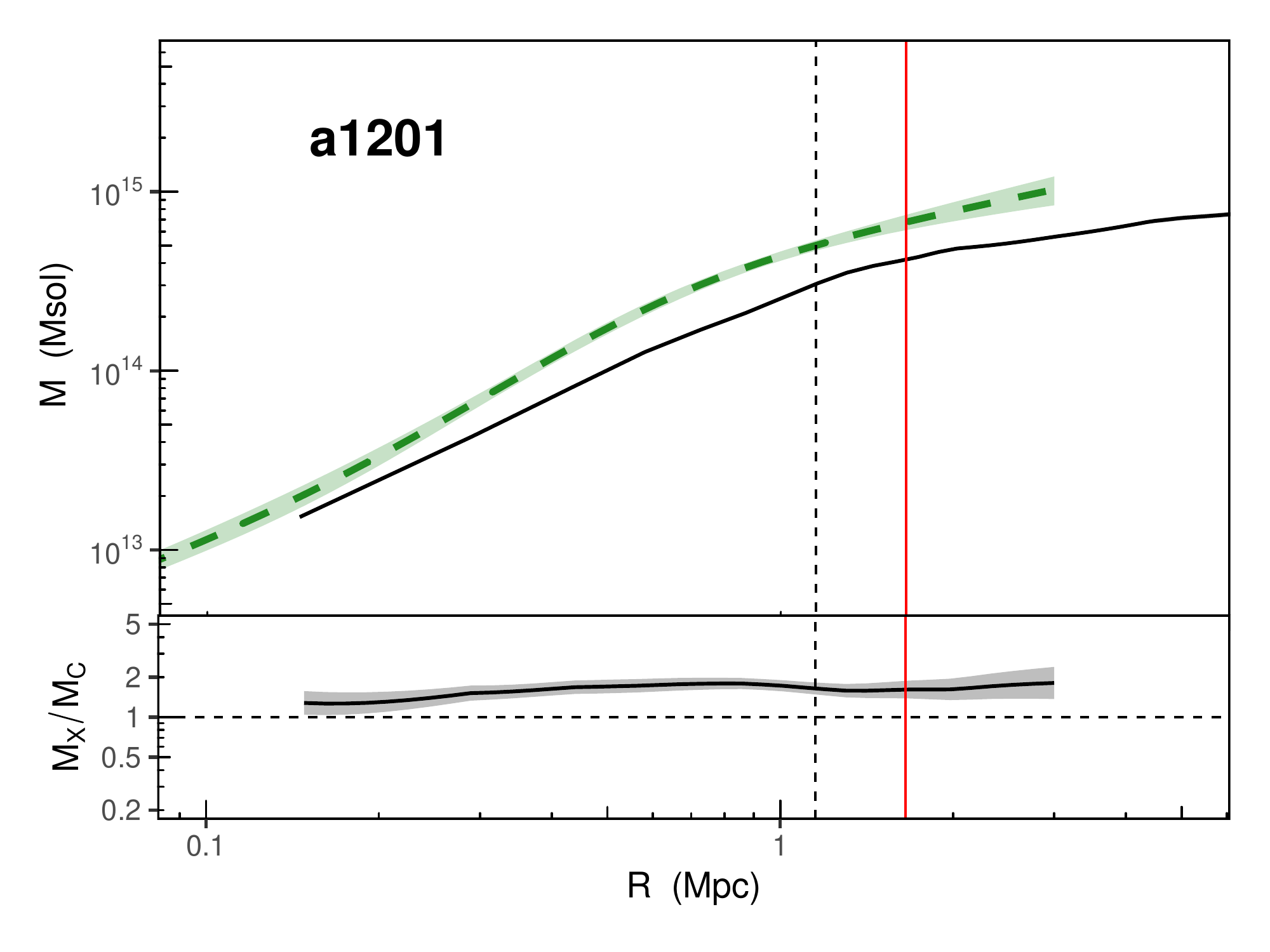}}
    \caption{\emph{- continued}}
  \end{figure*}

  \begin{figure*}
    \ContinuedFloat
    \centering
    \subfloat{\includegraphics[angle=0,width=200px]{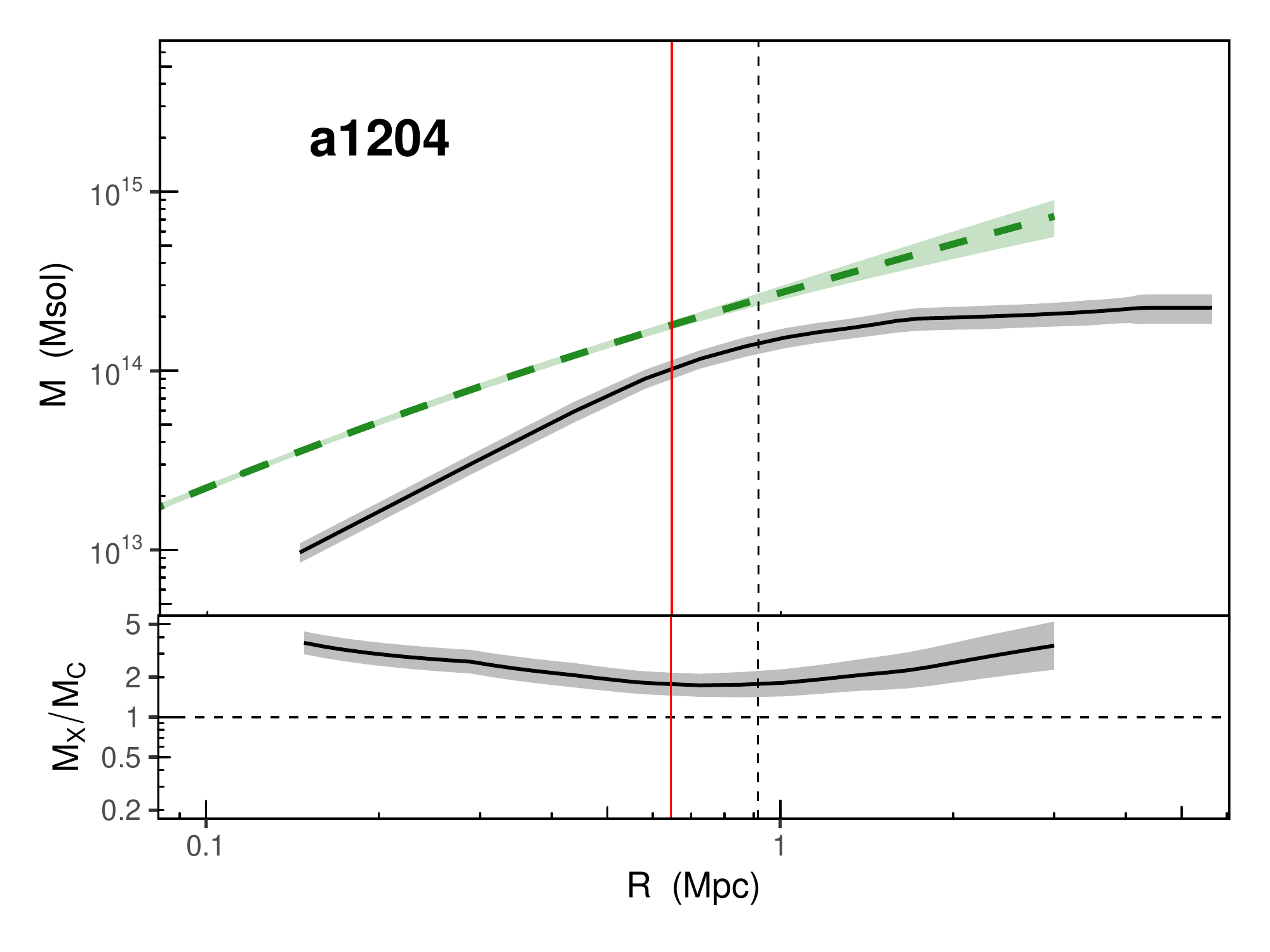}} \qquad
    \subfloat{\includegraphics[angle=0,width=200px]{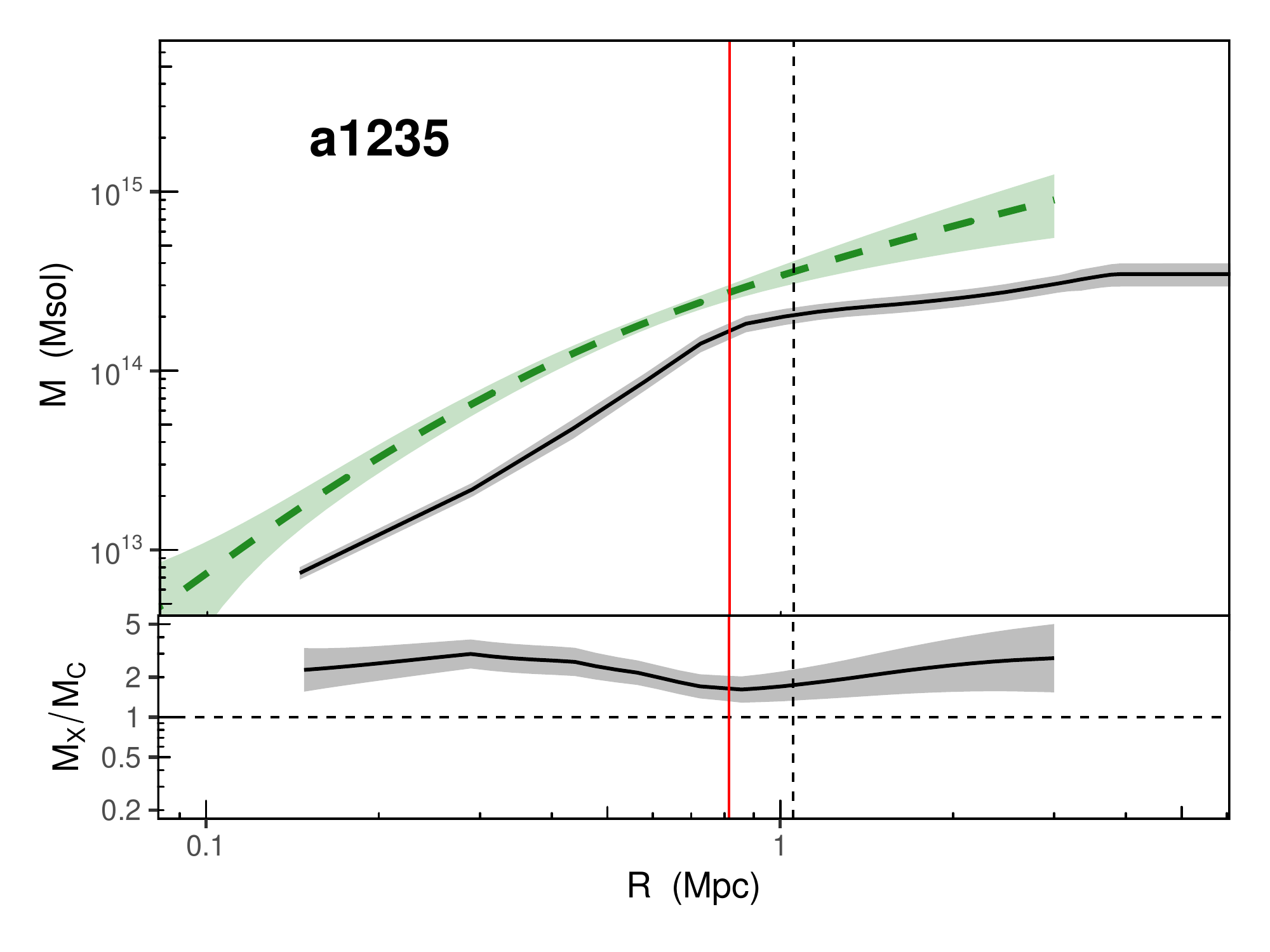}} \\
    \subfloat{\includegraphics[angle=0,width=200px]{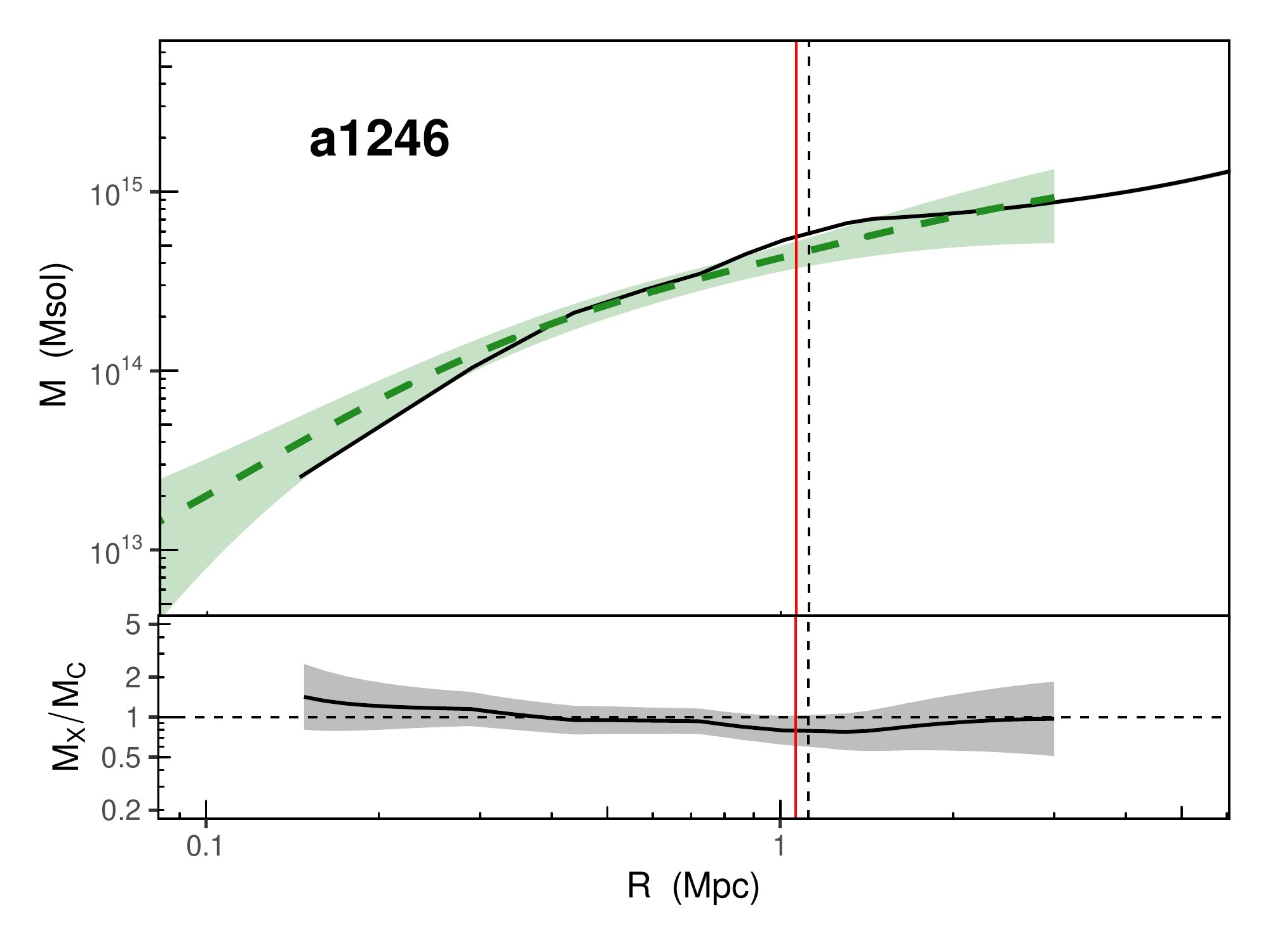}} \qquad
    \subfloat{\includegraphics[angle=0,width=200px]{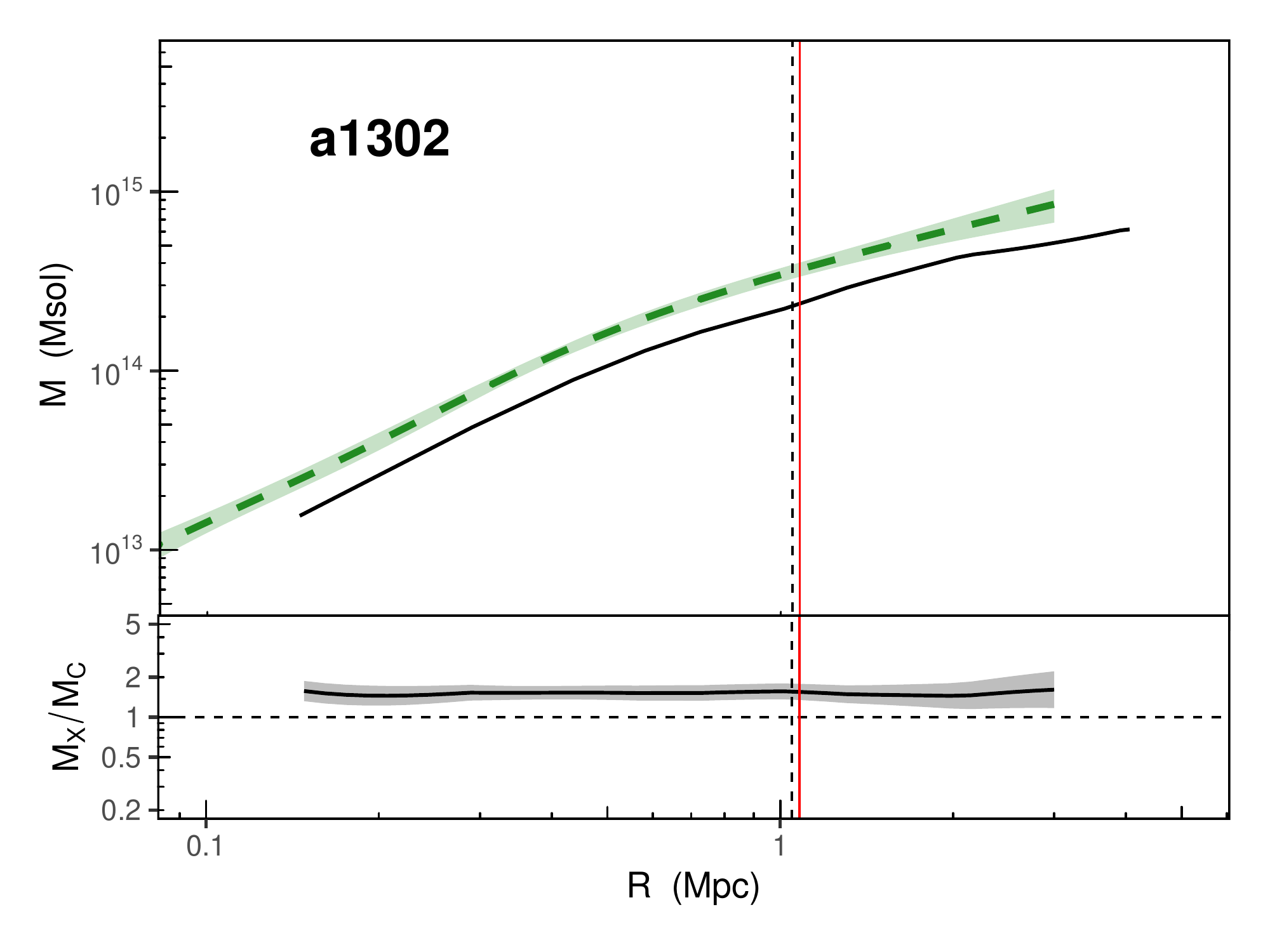}} \\
    \subfloat{\includegraphics[angle=0,width=200px]{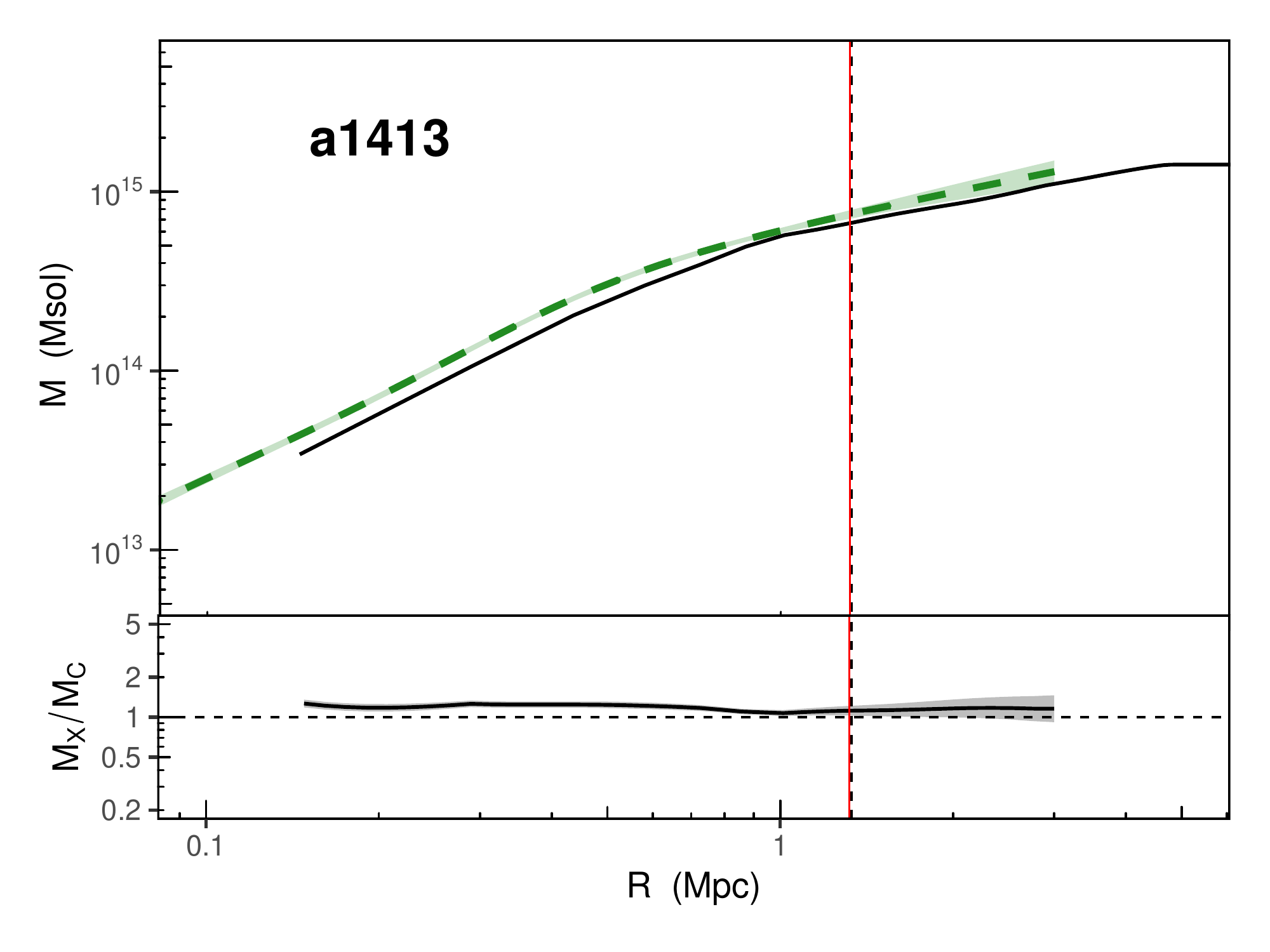}} \qquad
    \subfloat{\includegraphics[angle=0,width=200px]{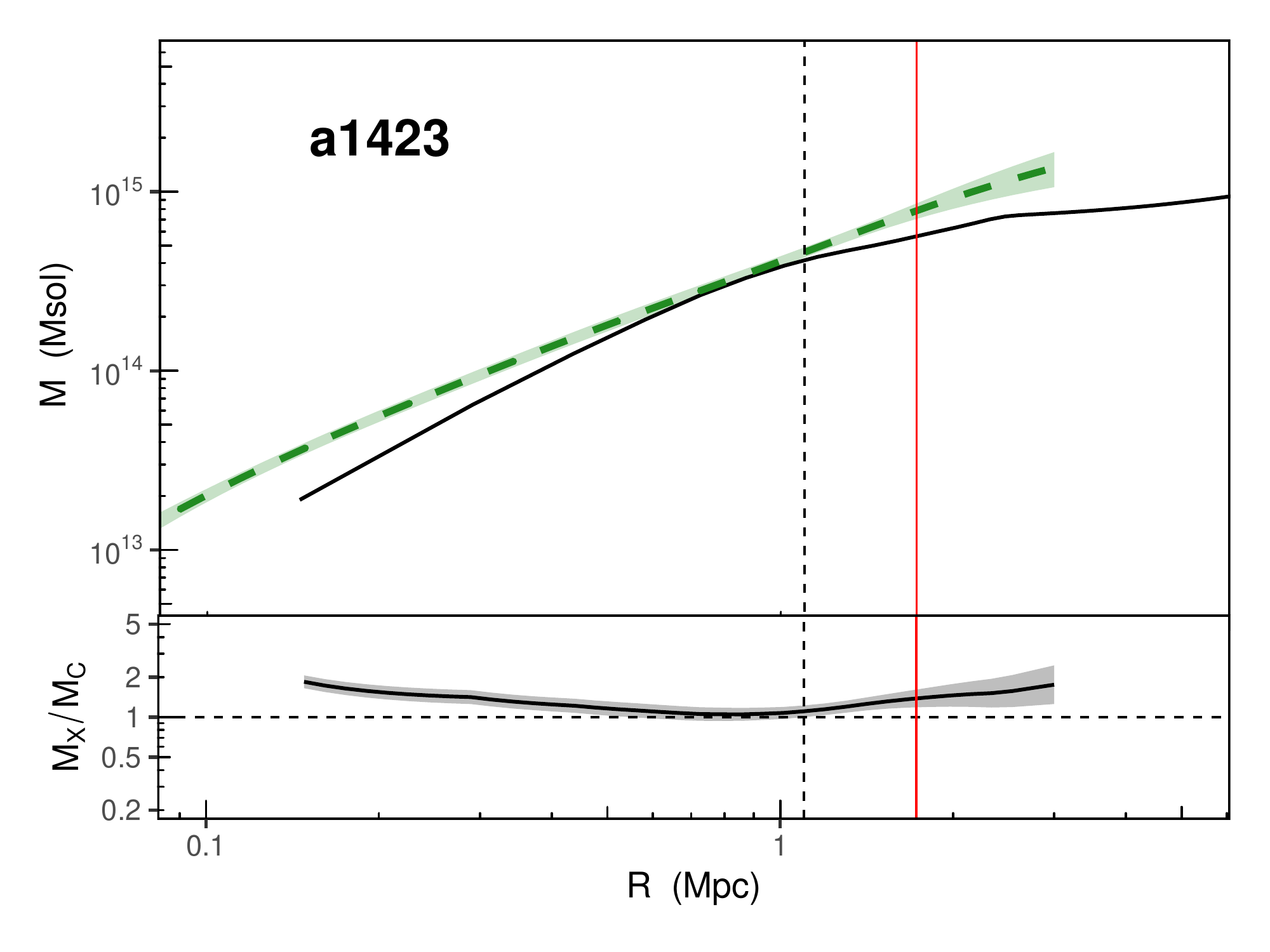}} \\
    \subfloat{\includegraphics[angle=0,width=200px]{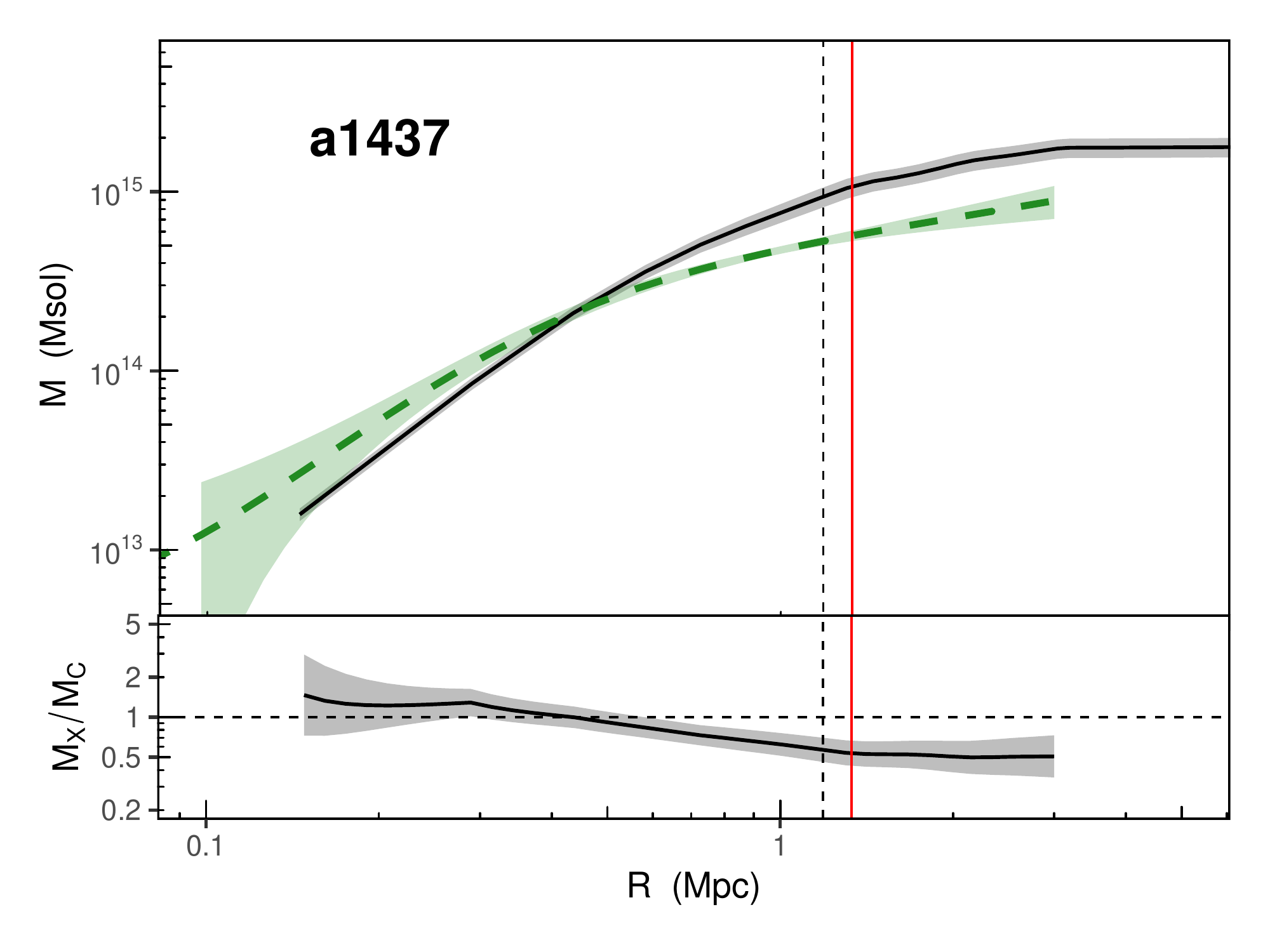}} \qquad
    \subfloat{\includegraphics[angle=0,width=200px]{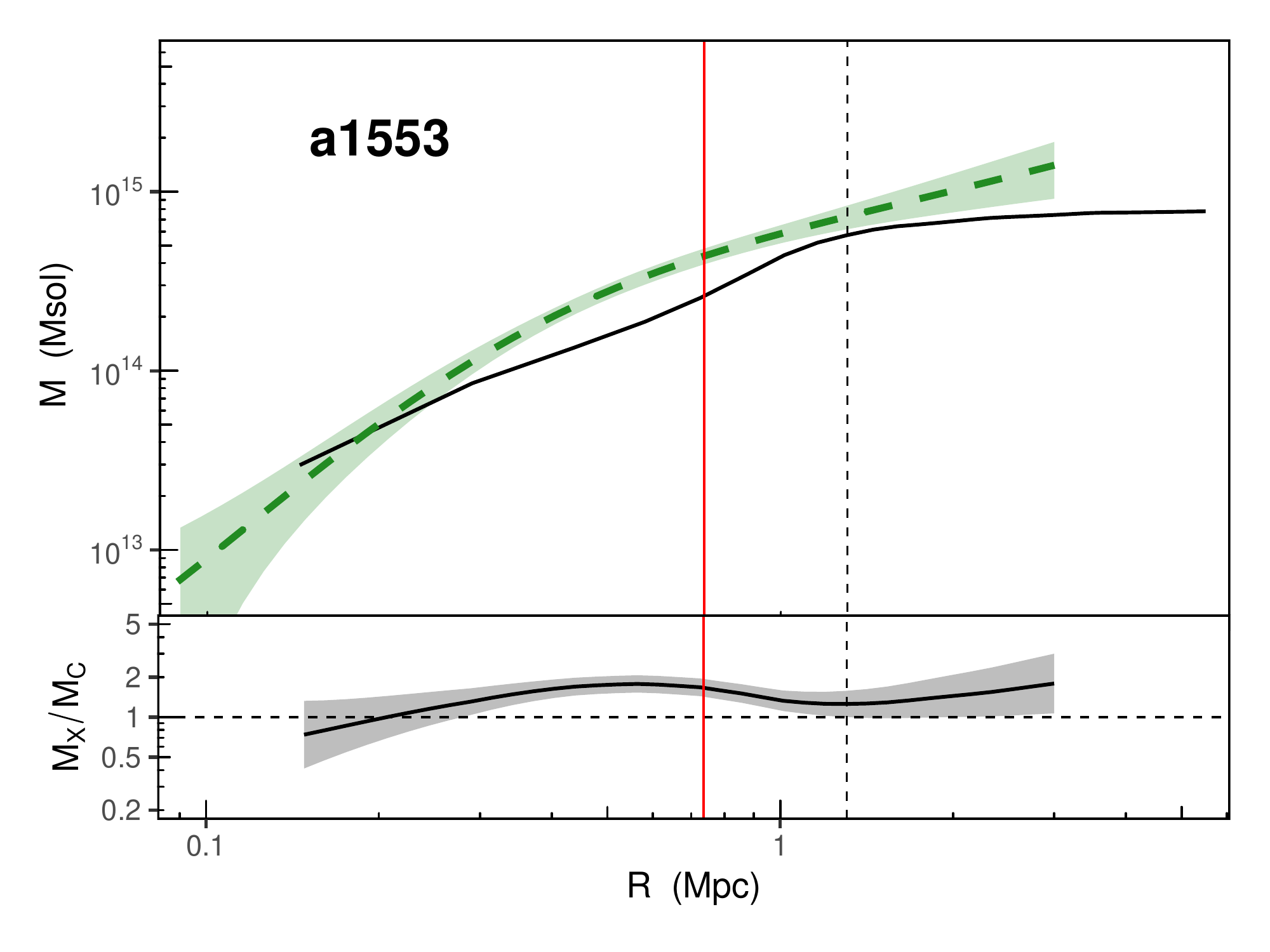}}
    \caption{\emph{- continued}}
  \end{figure*}

  \begin{figure*}
    \ContinuedFloat
    \centering
    \subfloat{\includegraphics[angle=0,width=200px]{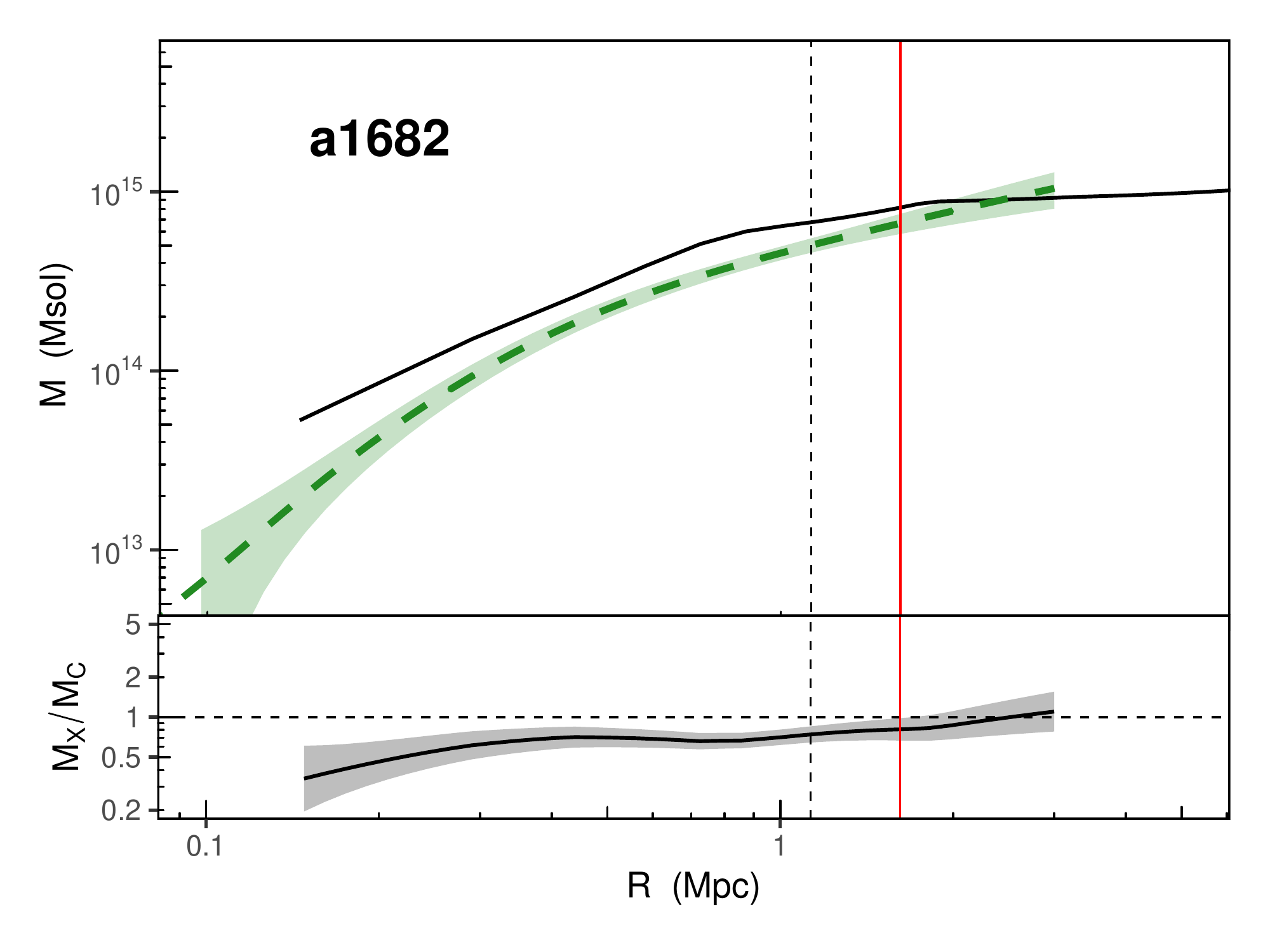}} \qquad
    \subfloat{\includegraphics[angle=0,width=200px]{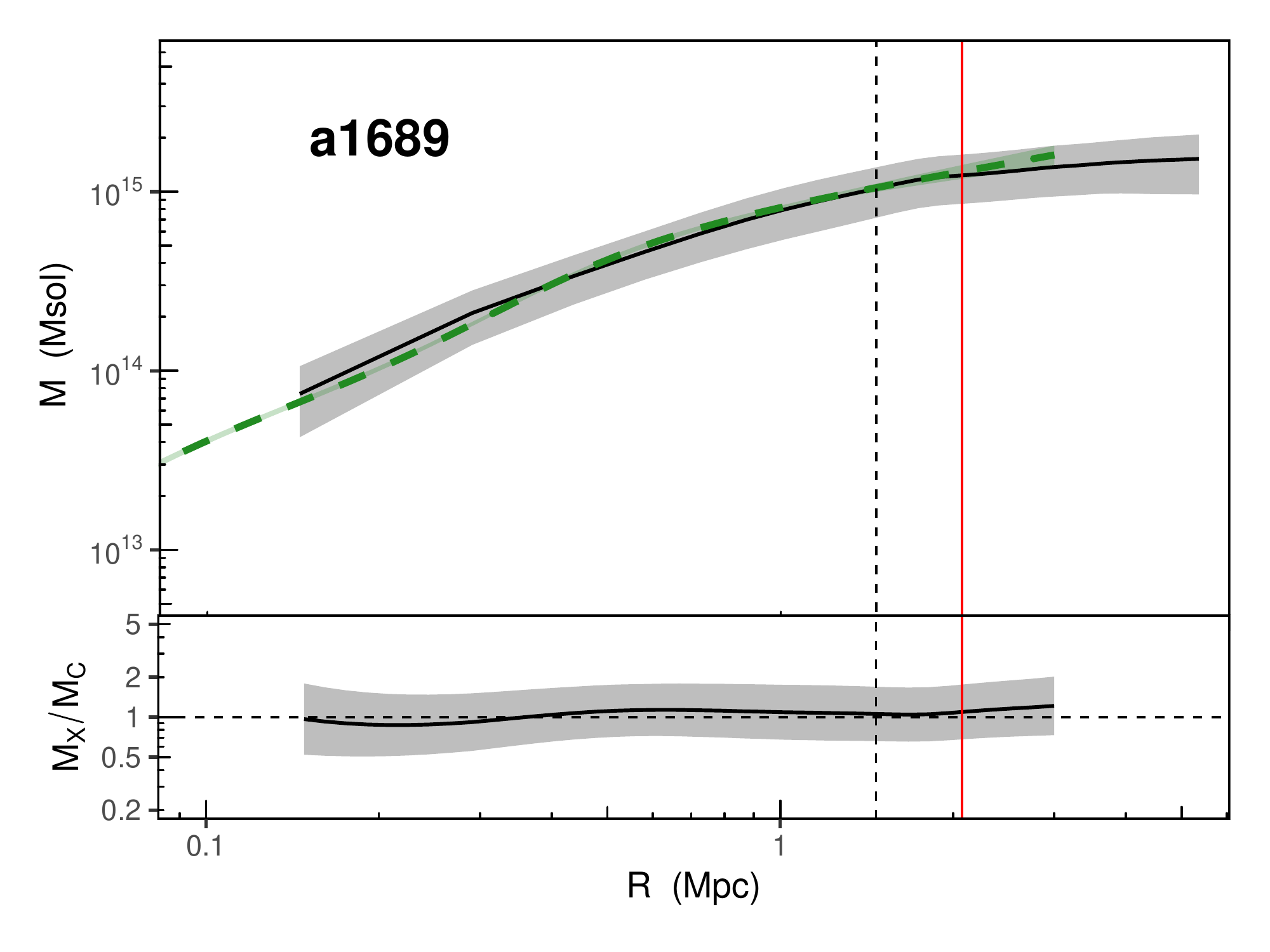}} \\
    \subfloat{\includegraphics[angle=0,width=200px]{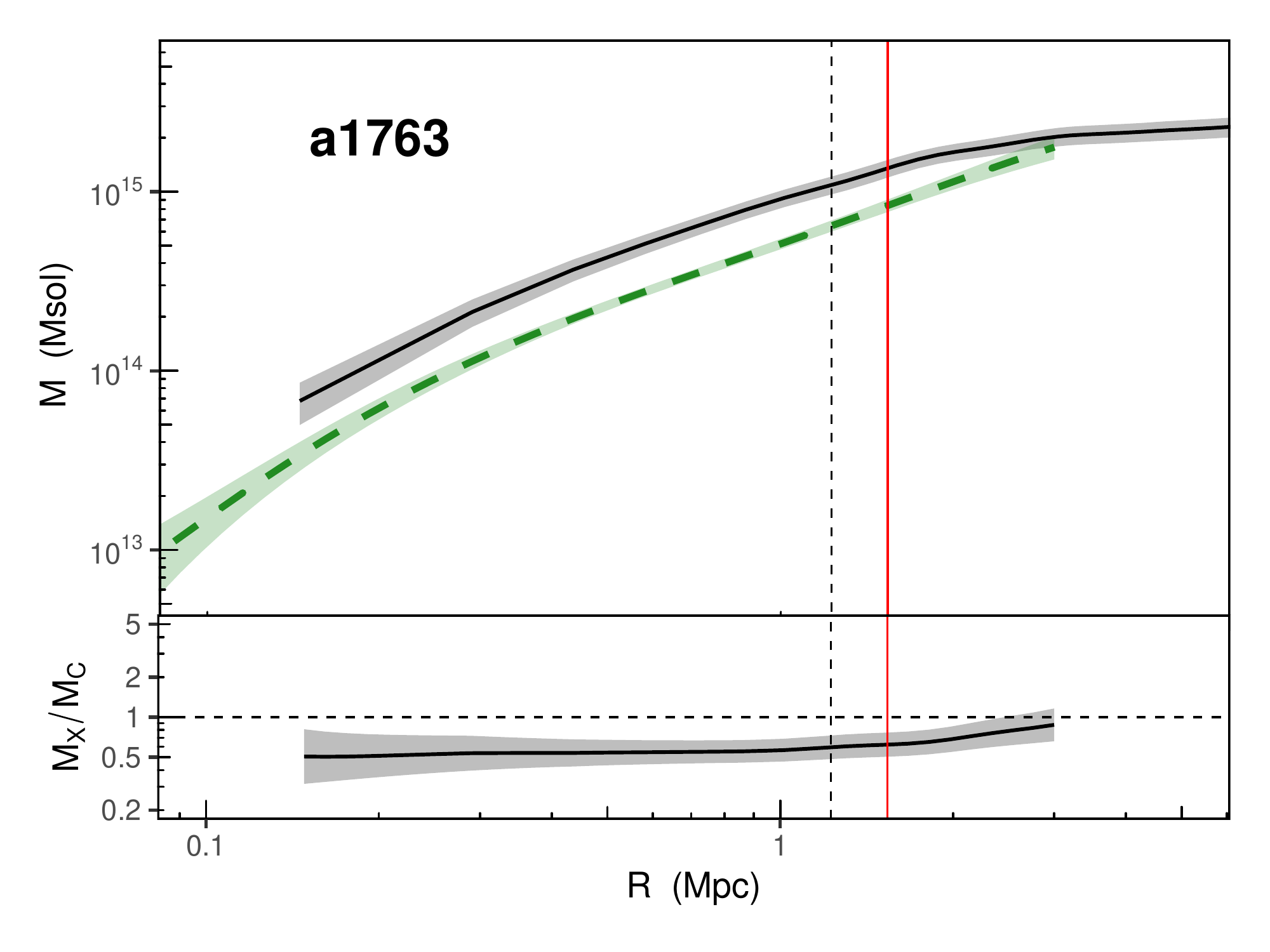}} \qquad
    \subfloat{\includegraphics[angle=0,width=200px]{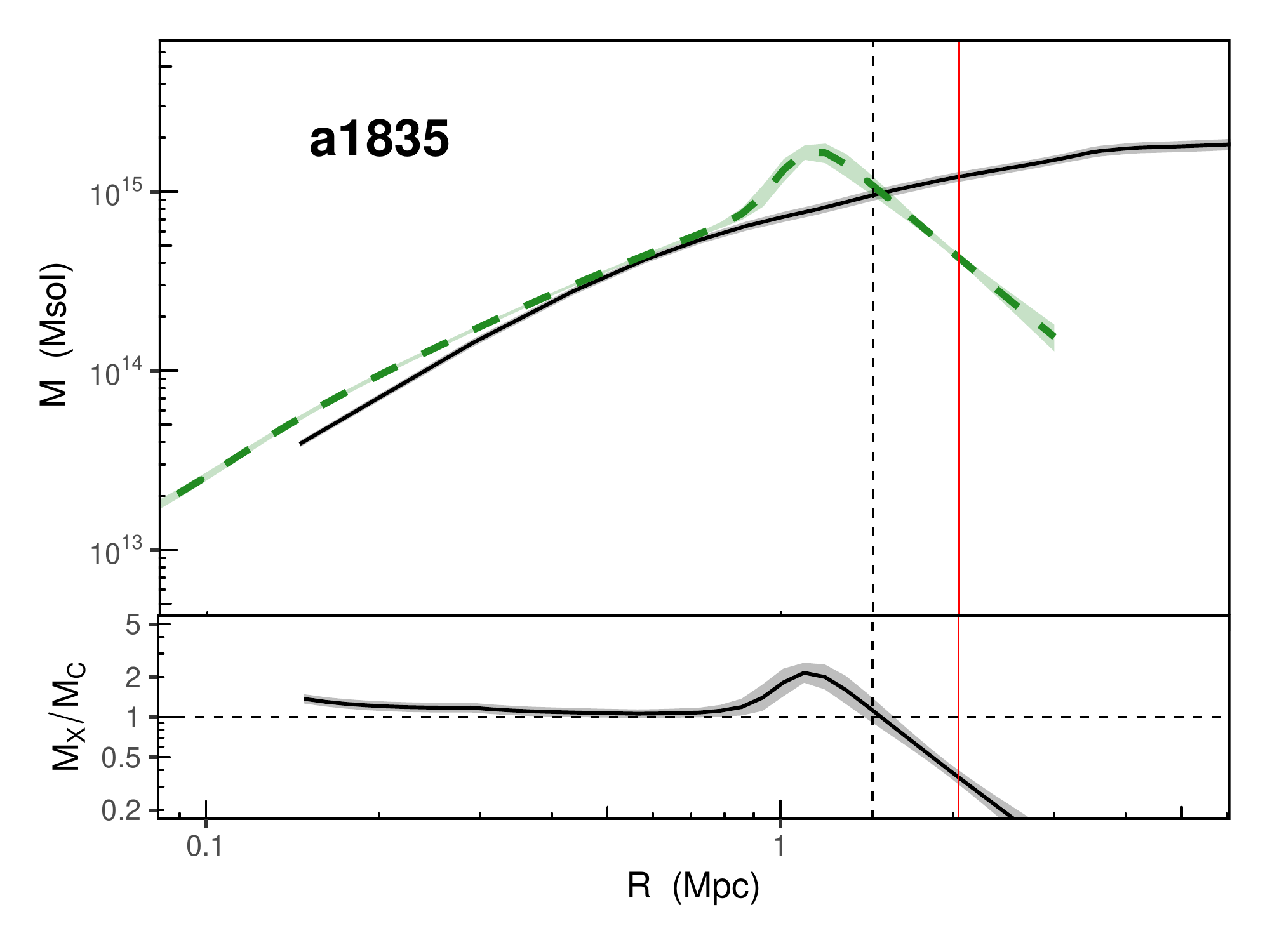}} \\
    \subfloat{\includegraphics[angle=0,width=200px]{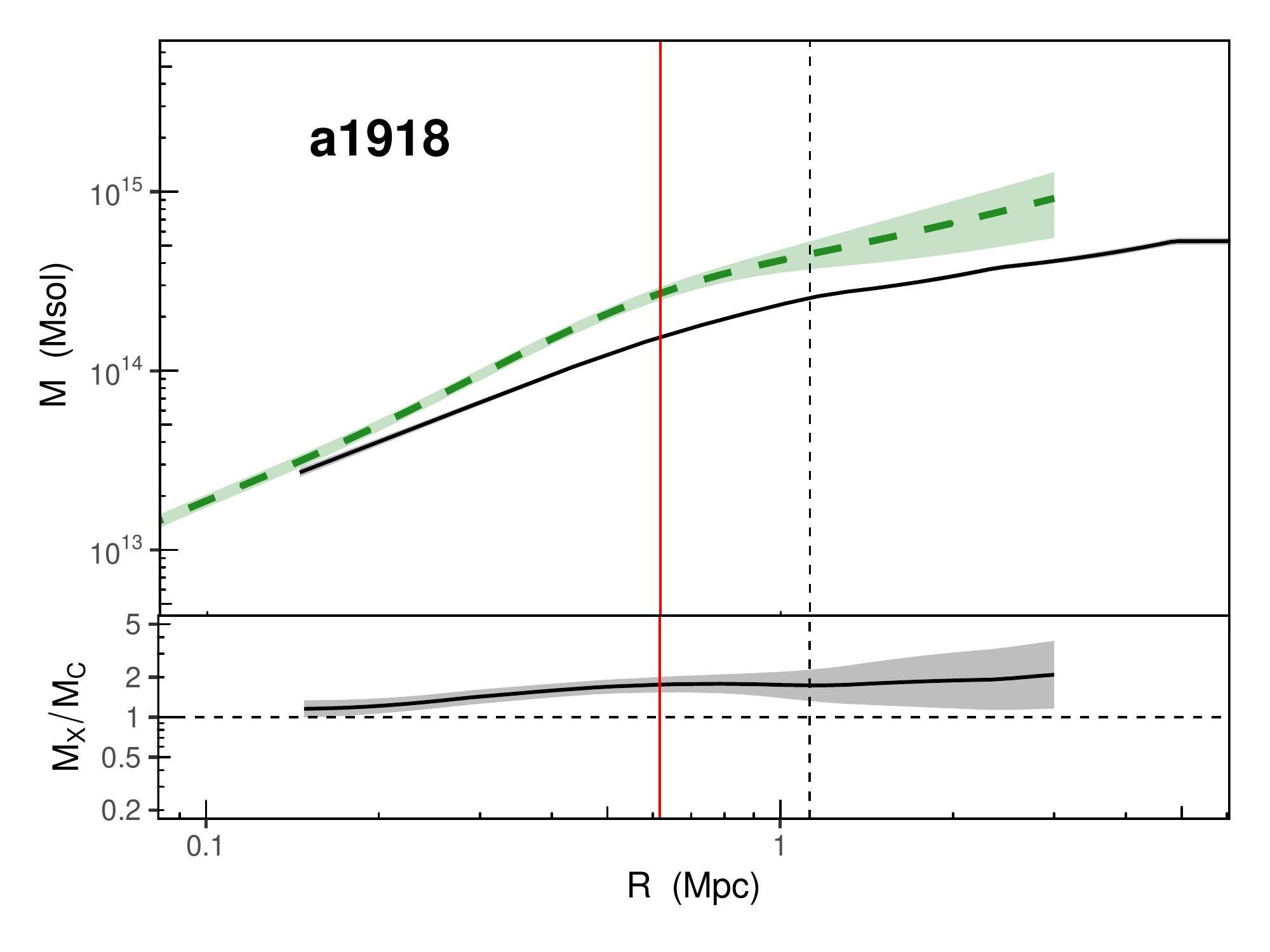}} \qquad
    \subfloat{\includegraphics[angle=0,width=200px]{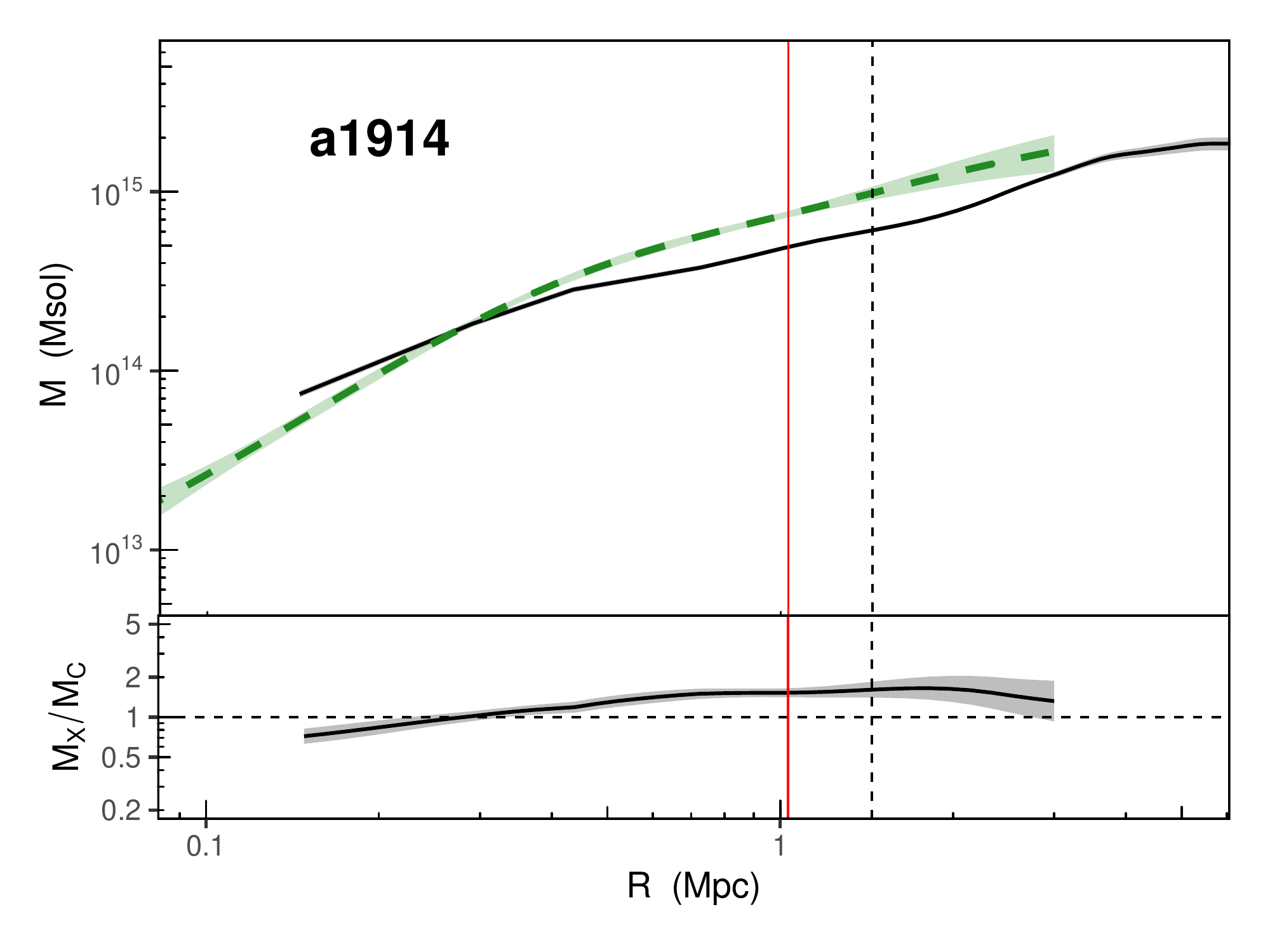}} \\
    \subfloat{\includegraphics[angle=0,width=200px]{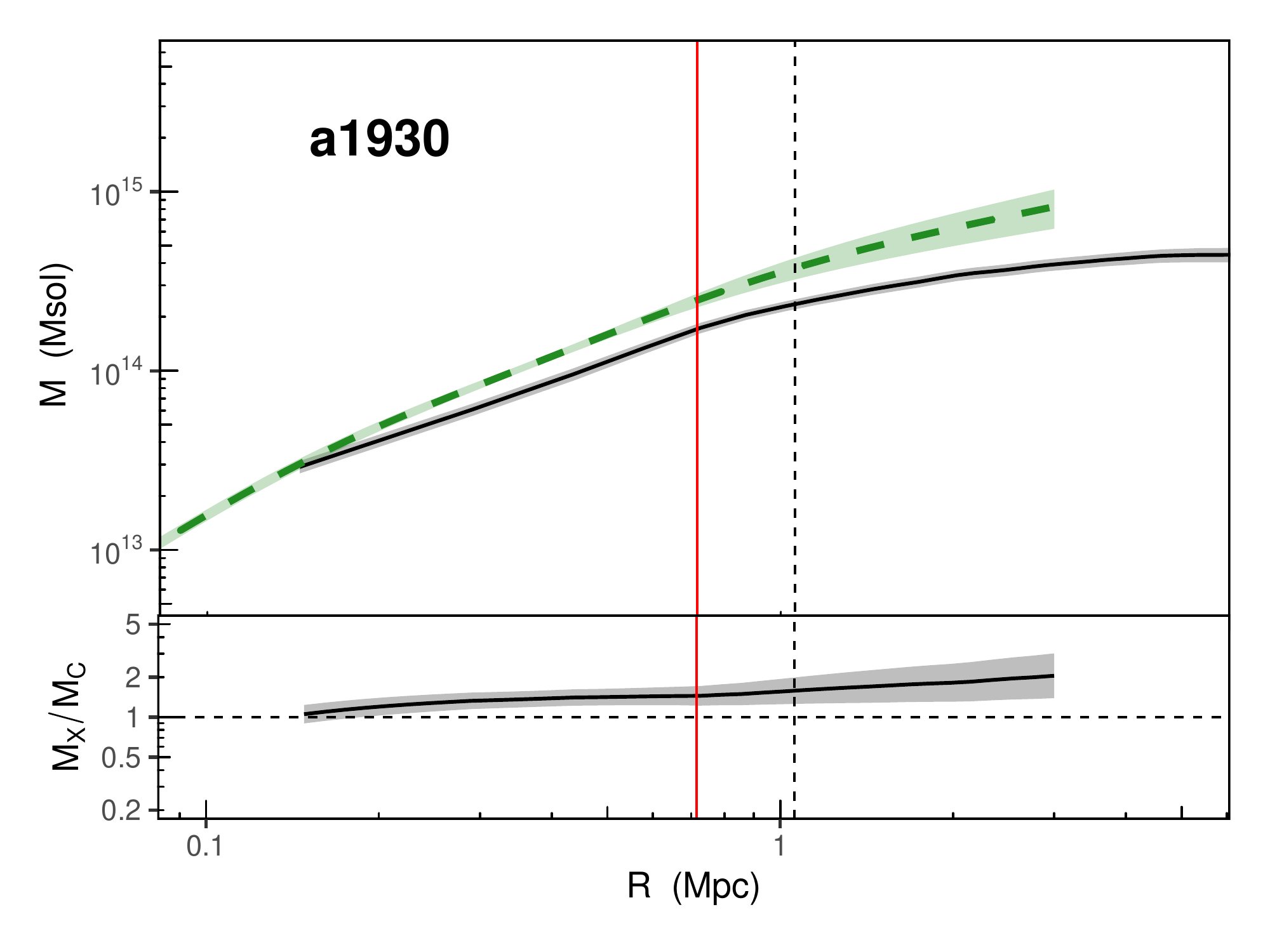}} \qquad
    \subfloat{\includegraphics[angle=0,width=200px]{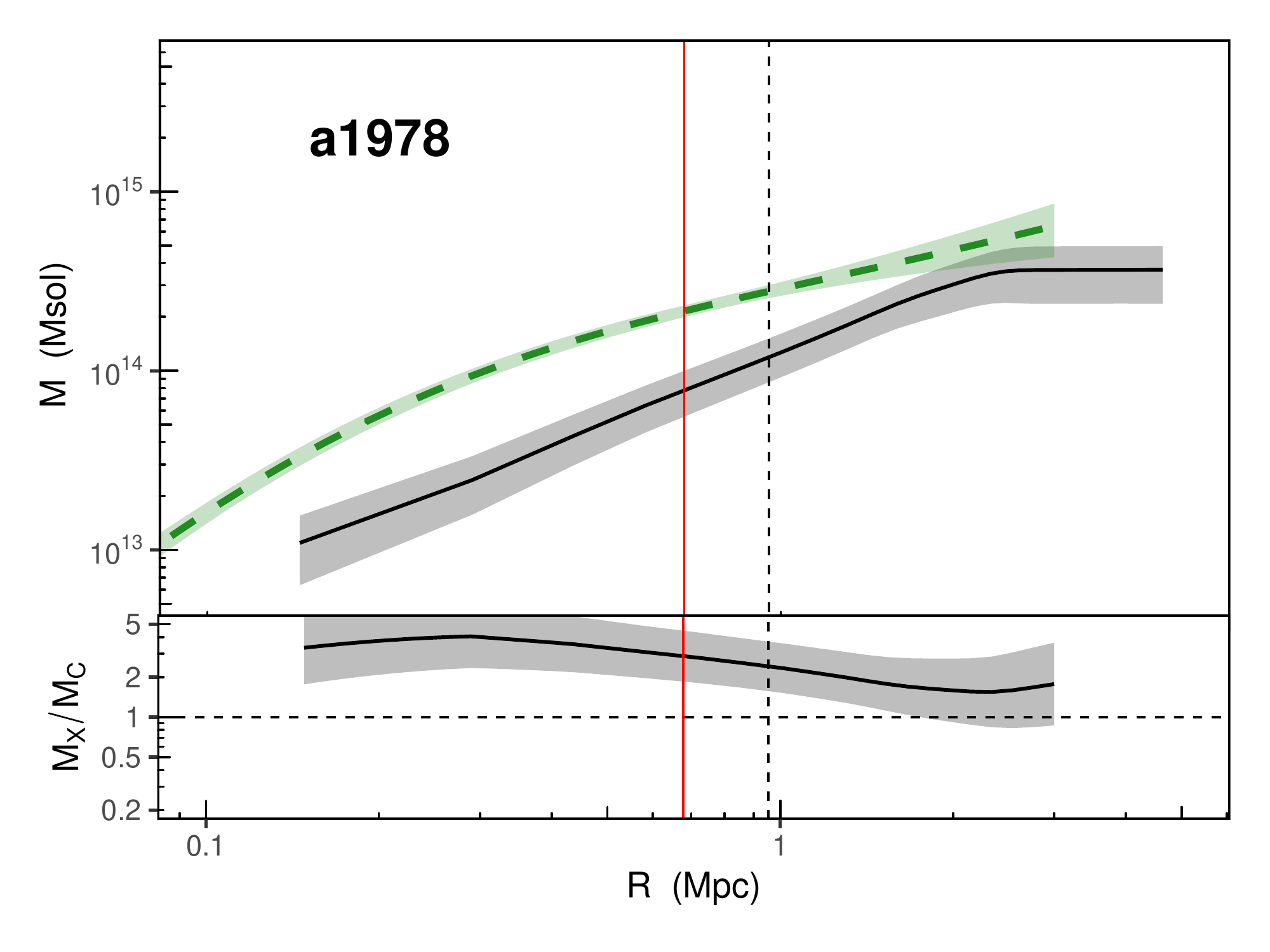}}
    \caption{\emph{- continued}}
  \end{figure*}

  \begin{figure*}
    \ContinuedFloat
    \centering
    \subfloat{\includegraphics[angle=0,width=200px]{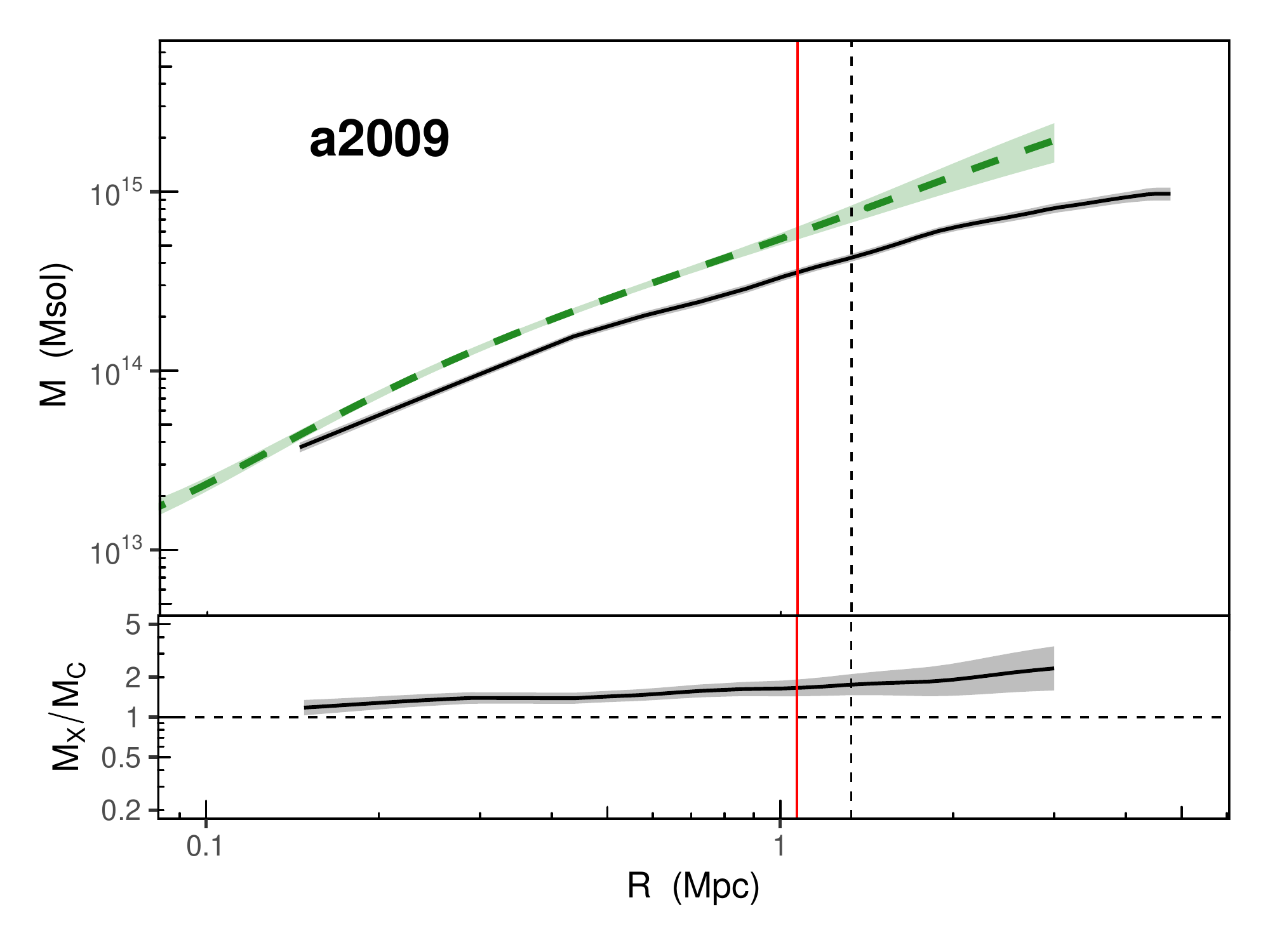}} \qquad
    \subfloat{\includegraphics[angle=0,width=200px]{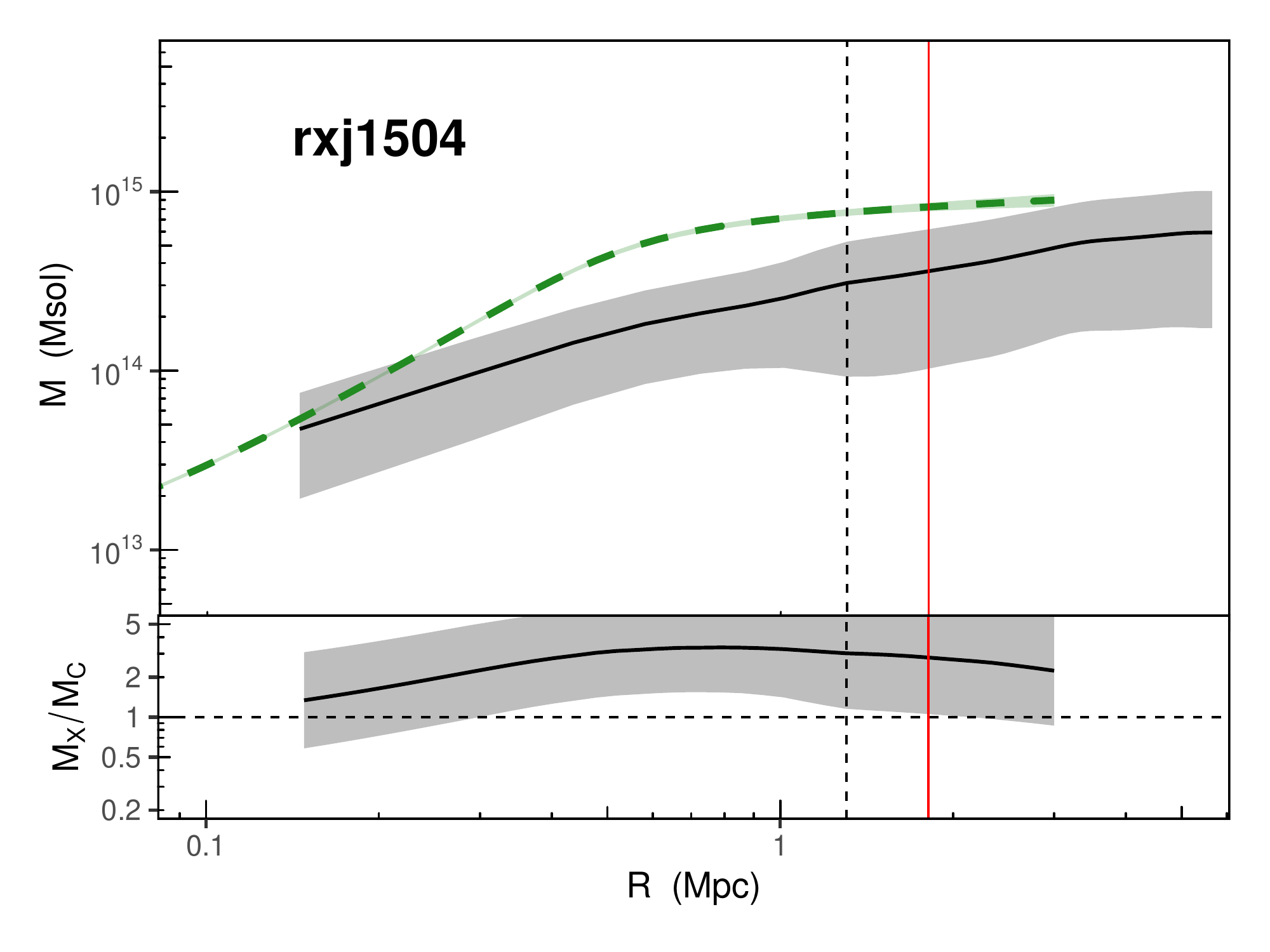}} \\
    \subfloat{\includegraphics[angle=0,width=200px]{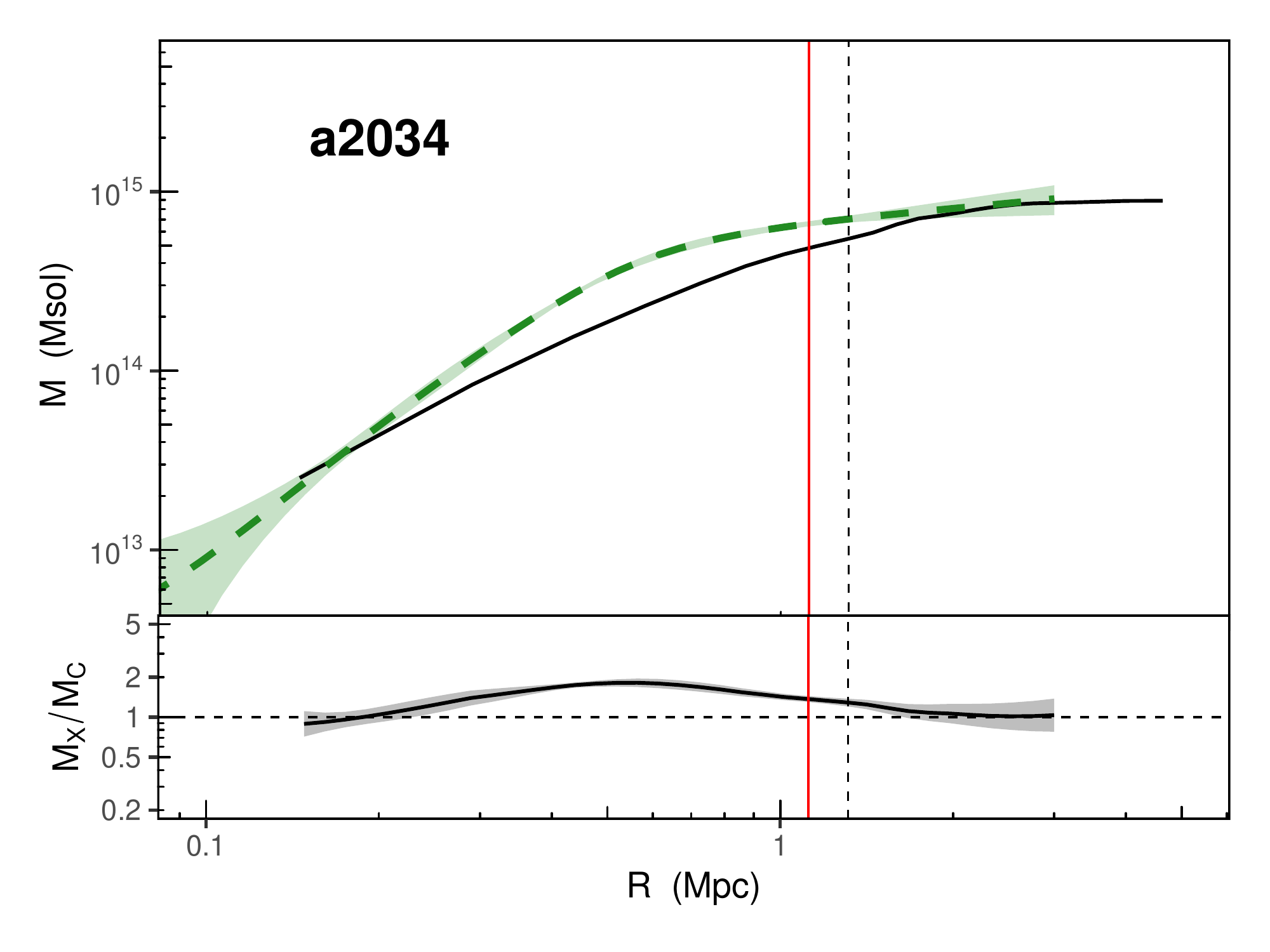}} \qquad
    \subfloat{\includegraphics[angle=0,width=200px]{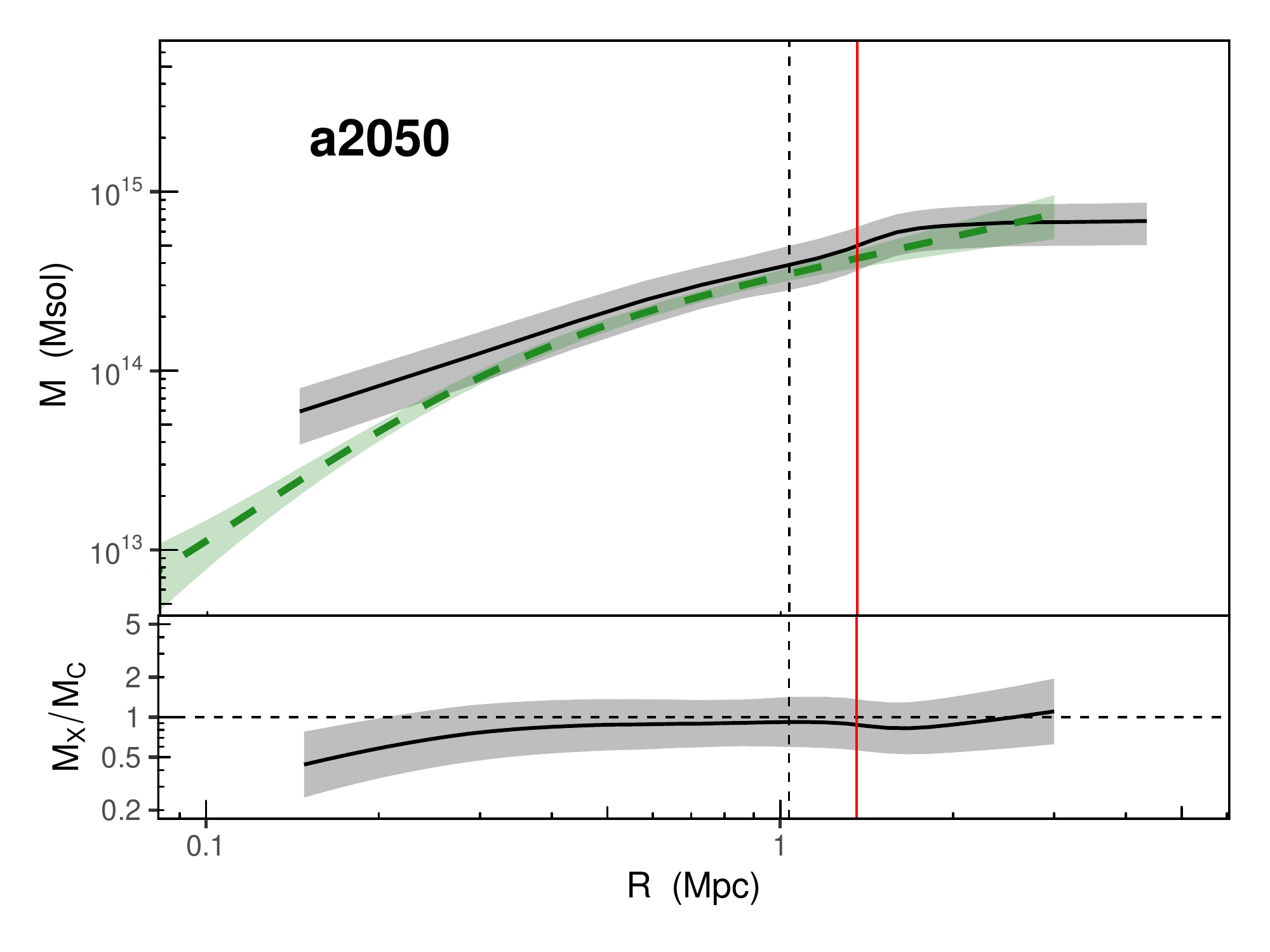}} \\
    \subfloat{\includegraphics[angle=0,width=200px]{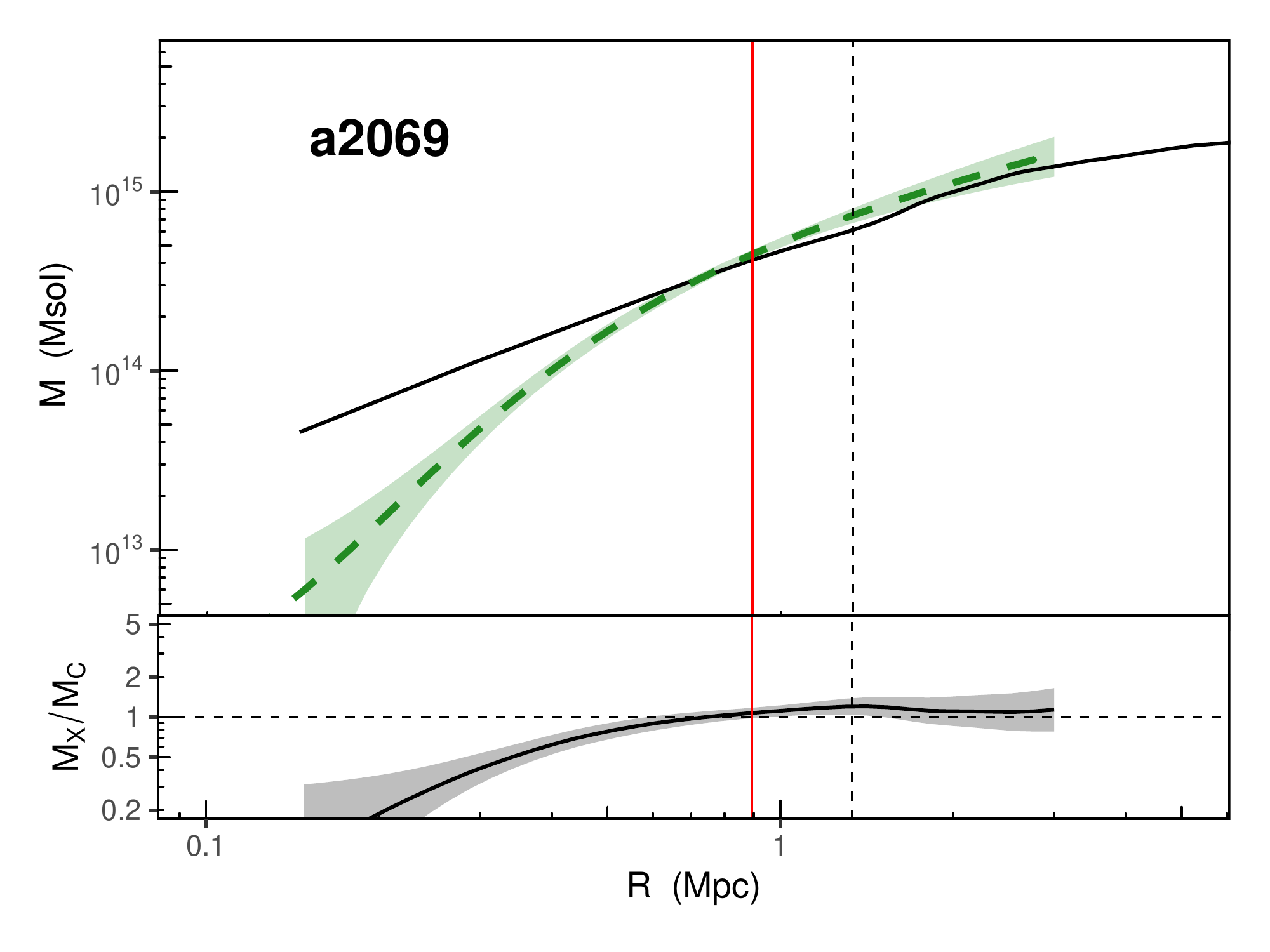}} \qquad
    \subfloat{\includegraphics[angle=0,width=200px]{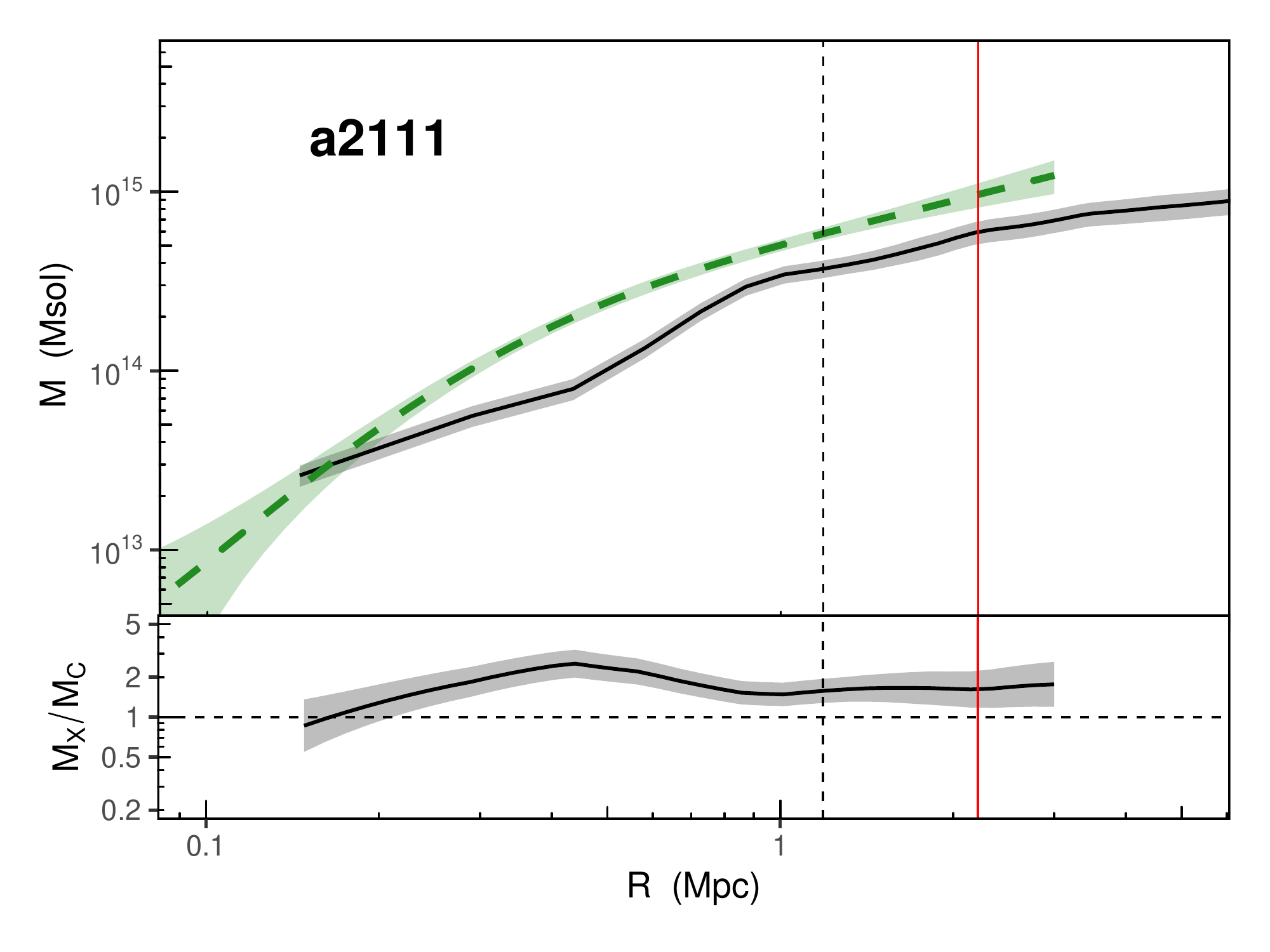}} \\
    \subfloat{\includegraphics[angle=0,width=200px]{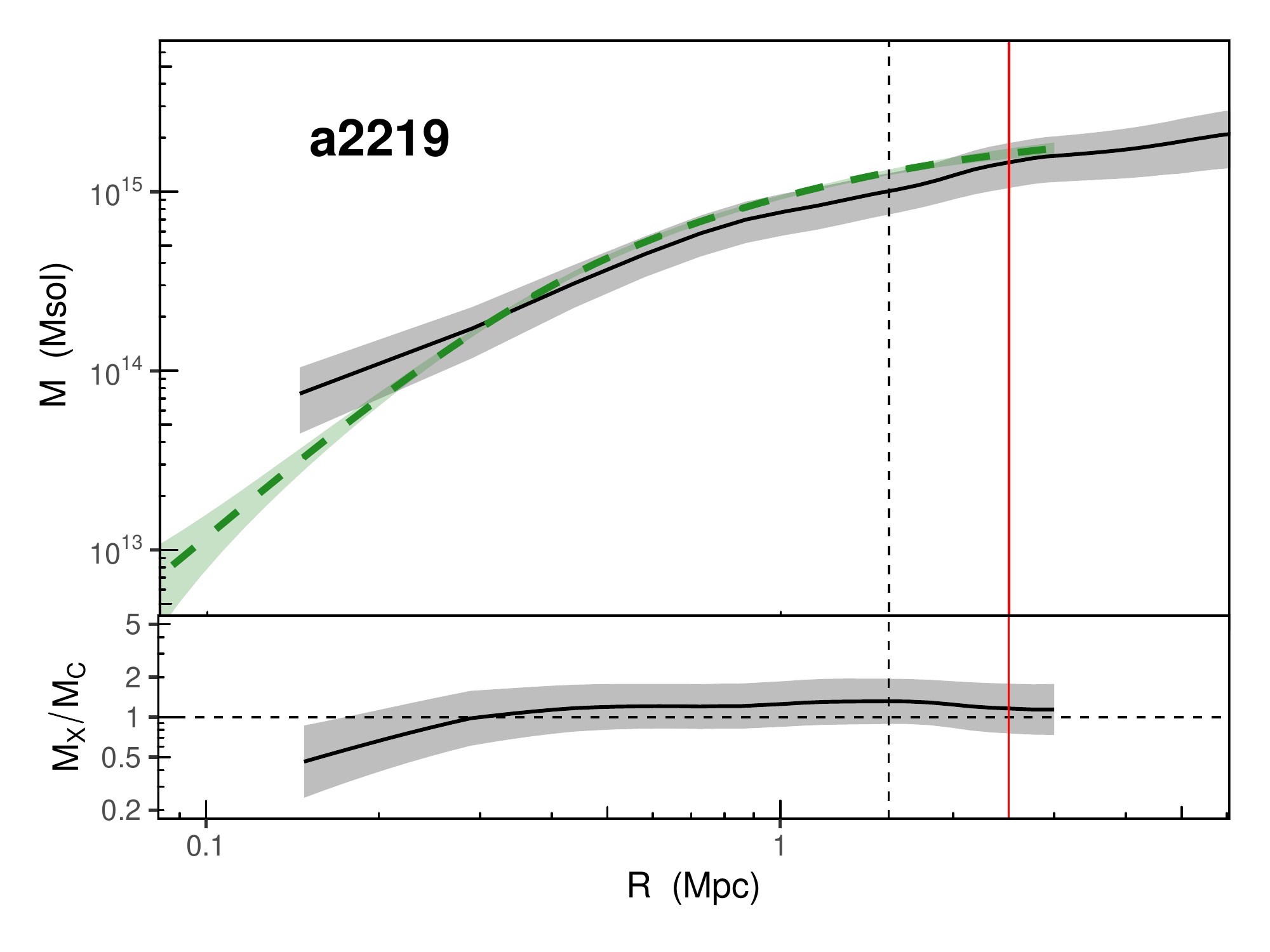}} \qquad
    \subfloat{\includegraphics[angle=0,width=200px]{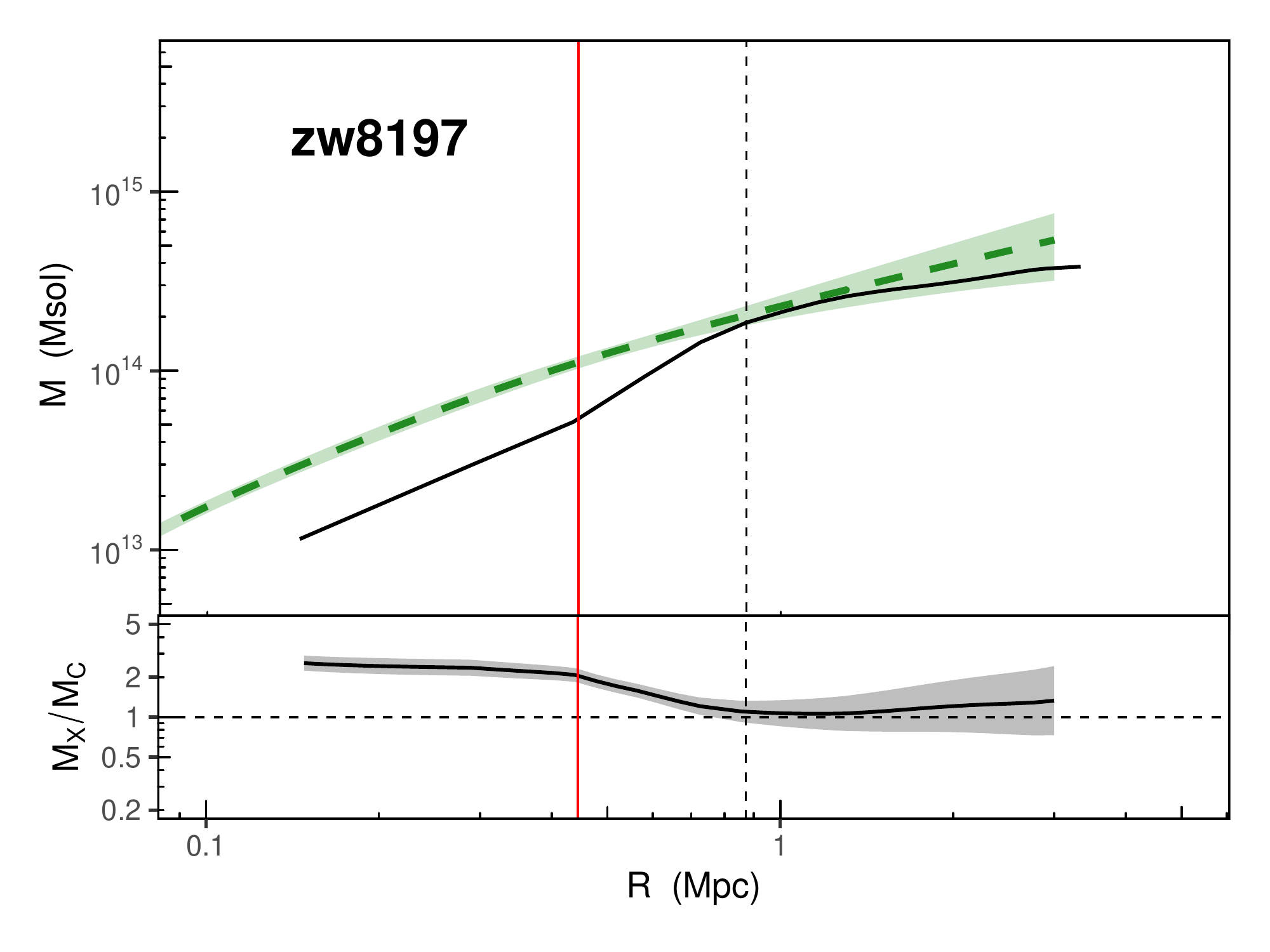}}
    \caption{\emph{- continued}}
  \end{figure*}

  \begin{figure*}
    \ContinuedFloat
    \centering
    \subfloat{\includegraphics[angle=0,width=200px]{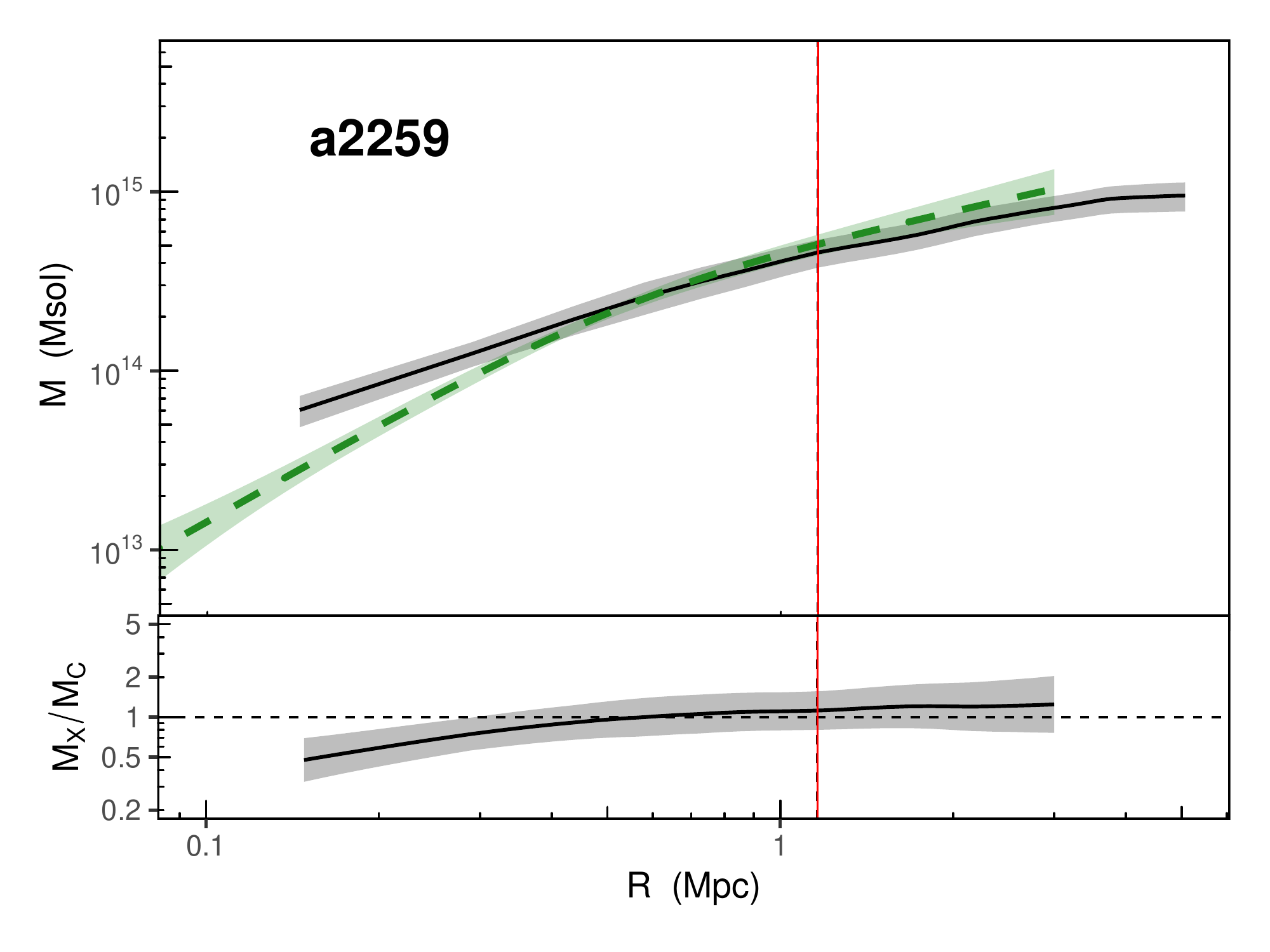}} \qquad
    \subfloat{\includegraphics[angle=0,width=200px]{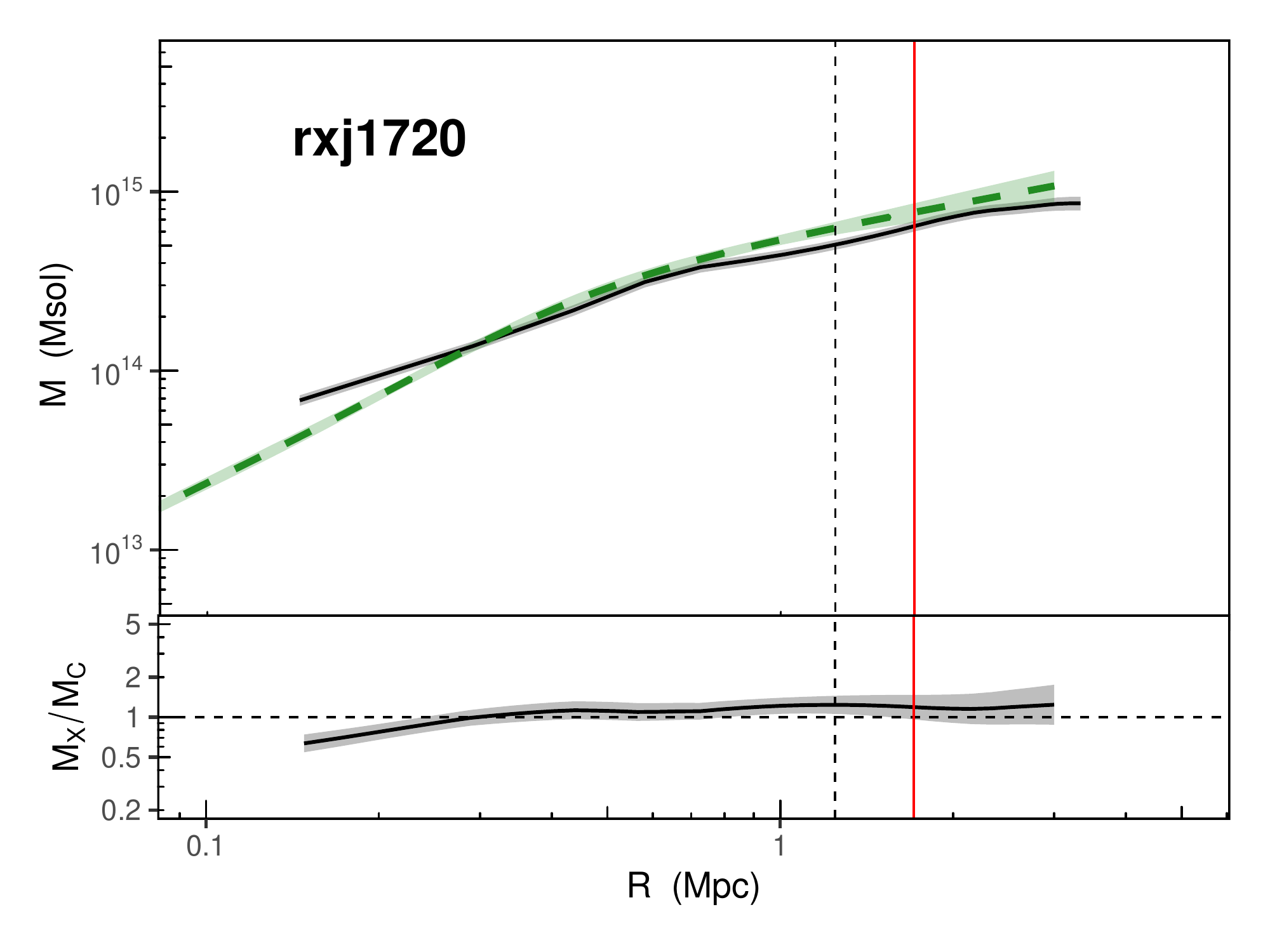}} \\
    \subfloat{\includegraphics[angle=0,width=200px]{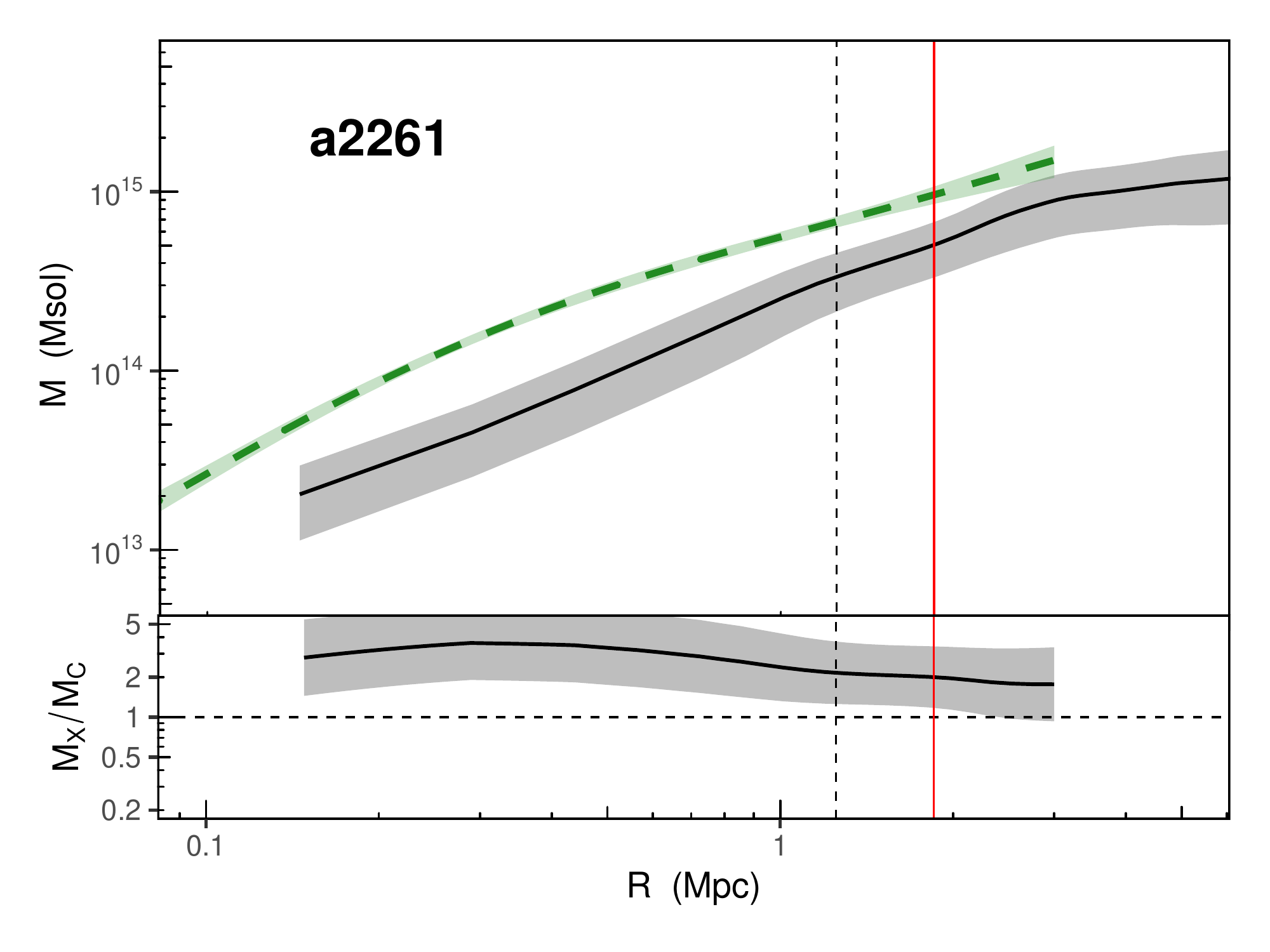}} \qquad
    \subfloat{\includegraphics[angle=0,width=200px]{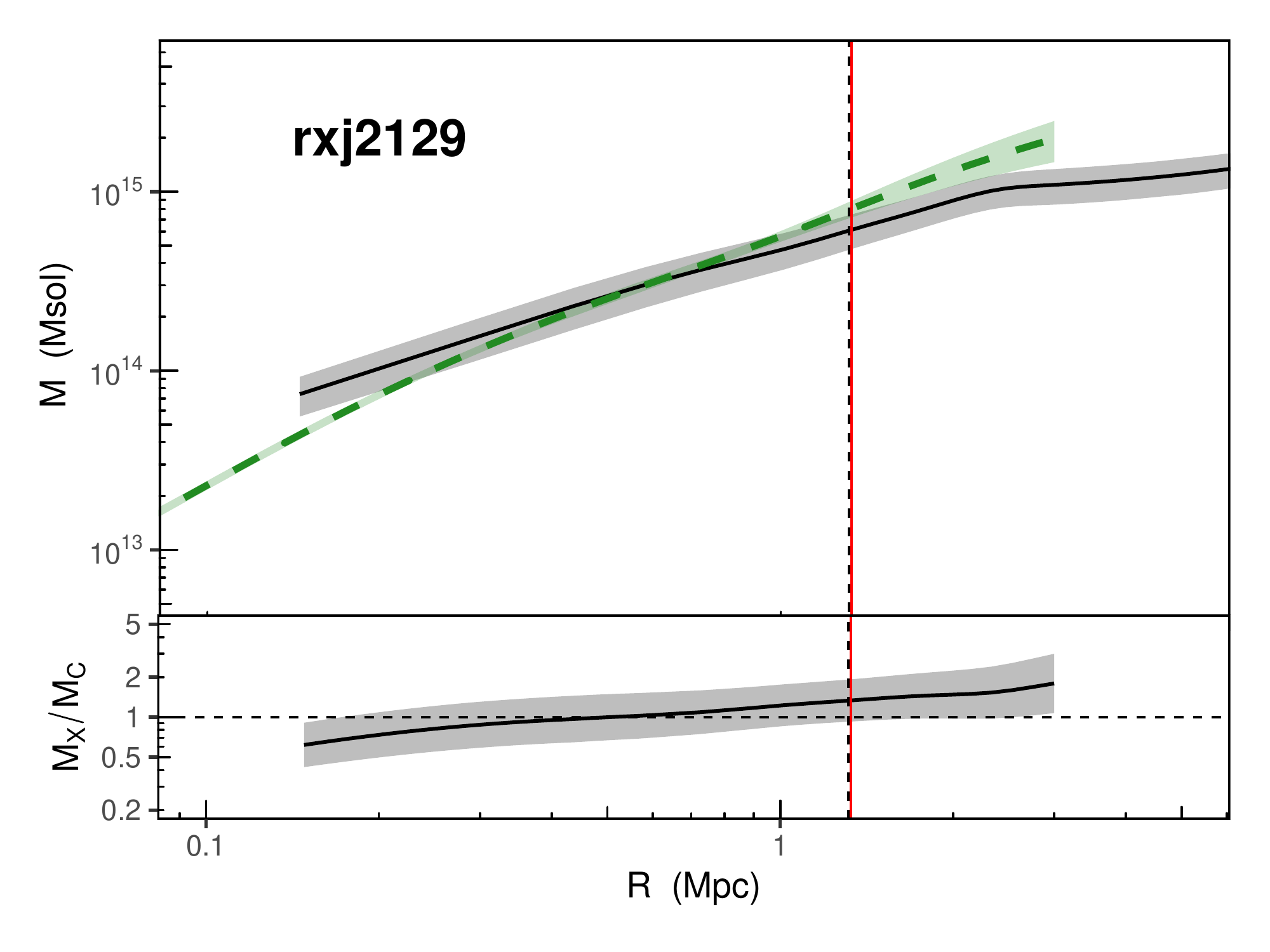}} \\
    \caption{\emph{- continued}}
  \end{figure*}

\end{appendix}

    \end{document}

%% file: newcom.tex
\usepackage{amsmath}

\newcommand{\rf}{\ensuremath{R_{\text{500}}}}

\newcommand{\fgas}{\ensuremath{f_{\text{gas}}}}

\newcommand{\gta}{\,\rlap{\raise 0.4ex\hbox{$>$}}{\lower 0.6ex\hbox{$\sim$}}\,}
\newcommand{\lta}{\,\rlap{\raise 0.4ex\hbox{$<$}}{\lower 0.6ex\hbox{$\sim$}}\,}

\newcommand{\cm}{\mbox{\ensuremath{\text{~cm}}}}

\newcommand{\kpc}{\mbox{\ensuremath{\text{~kpc}}}}

\newcommand{\s}{\mbox{\ensuremath{\text{~s}}}}

\newcommand{\keV}{\mbox{\ensuremath{\text{~keV}}}}

\newcommand{\erg}{\mbox{\ensuremath{\text{~erg}}}}

\newcommand{\degree}{\ensuremath{\mathrm{^\circ}}}
\newcommand{\arcm}{\ensuremath{\mathrm{^\prime}}}
\newcommand{\arcs}{\arcm\hskip -0.1em\arcm}

\newcommand{\pcmsq}{\ensuremath{\cm^{-2}}}

\newcommand{\ps}{\ensuremath{\s^{-1}}}

\newcommand{\flux}{\ensuremath{\erg \ps \pcmsq}}

\newcommand{\apropto}{\mathrel{\vcenter{
  \offinterlineskip\halign{\hfil$##$\cr
    \propto\cr\noalign{\kern2pt}\sim\cr\noalign{\kern-2pt}}}}}